\documentclass[a4paper,12pt]{article}
\pdfoutput=1

\usepackage{jheppub}
\usepackage[utf8]{inputenc}
\usepackage[english]{babel}

\usepackage{amsfonts}
\usepackage{mathrsfs}
\usepackage[mathcal]{euscript}
\usepackage{subfig}
\usepackage[section]{placeins}

\newcommand{\be}{\begin{equation}}
\newcommand{\ee}{\end{equation}}
\newcommand{\benn}{\begin{equation*}}
\newcommand{\eenn}{\end{equation*}}

\newcommand{\skipline}{\vspace{\baselineskip}}

\usepackage{mathtools}
\DeclarePairedDelimiter\corrfunc{\langle}{\rangle}
\DeclarePairedDelimiter\bra{\langle}{\rvert}
\DeclarePairedDelimiter\ket{\lvert}{\rangle}
\DeclarePairedDelimiterX\braket[2]{\langle}{\rangle}{#1 \delimsize\vert #2}
\DeclareMathOperator{\re}{Re}
\DeclareMathOperator{\im}{Im}

\newcommand{\eps}{\varepsilon}
\newcommand{\zbar}{\bar{z}}
\newcommand{\wbar}{\overline{w}}
\newcommand{\Tbar}{\overline{T}}

\newcommand{\Odag}{O^{\dagger}}
\newcommand{\om}{\omega}
\newcommand{\sbar}{\bar{\sigma}}
\newcommand{\Ecal}{\mathcal{E}}
\newcommand{\Qcal}{\mathcal{Q}}
\newcommand{\Ocal}{\mathcal{O}}

\newcommand{\largetime}{\underset{t\,\to +\infty}{\approx}}
\newcommand{\largex}{\underset{x\,\to +\infty}{\approx}}

\title{From locality to irregularity: Introducing local quenches in massive scalar field theory}

\author{Dmitry S. Ageev,}
\author{Aleksandr I. Belokon}
\author{and Vasilii V. Pushkarev}

\affiliation{Steklov Mathematical Institute, Russian Academy of Sciences,\\ Gubkin str. 8, 119991 Moscow, Russia}

\emailAdd{ageev@mi-ras.ru}
\emailAdd{belokon@mi-ras.ru}
\emailAdd{pushkarev@mi-ras.ru}

\abstract{In this paper, we initiate the study of operator local quenches in non-conformal field theories. We consider the dynamics of excited local states in massive scalar field theory in an arbitrary spacetime dimension and generalize the well-known two-dimensional CFT results. We derive the energy density, $U(1)$-charge density and $\phi^2(x)$-condensate post-quench dynamics, and identify different regimes of their evolution depending on the values of the field mass and the quench regularization parameter. For local quenches in higher-dimensional free massless scalar theories, we reproduce the structure of the available holographic results. We also investigate the local quenches in massive scalar field theory on a cylinder and show that they cause an erratic and chaotic-like evolution of observables with a complicated localization/delocalization pattern.}

\begin{document}
\maketitle
\newpage

\section{Introduction and summary}

Quantum dynamics of non-equilibrium quantum systems is an important area of research that brings together many seemingly unrelated topics --- condensed matter theory, quantum gravity, thermalization, quantum entanglement and chaos~\cite{Liu:2018crr, Zaanen:2015oix, Hartnoll:2016apf, Blake:2022uyo, Faulkner:2022mlp, Mezei:2018jco}. To study the dynamical properties of quantum systems, it is convenient to perturb the system in some way and probe the dynamics after this perturbation (which is referred to as a quantum quench). For example, what is called the ``global quench'' corresponds to a global perturbation of the system --- a change in a coupling constant or a state, as well as a global energy injection. In the AdS/CFT correspondence, much attention has been paid to this process and its close relationship with thermalization and, as a result, with the formation of black holes in recent years~\cite{Danielsson:1999fa, Balasubramanian:2010ce, Balasubramanian:2011ur, Ageev:2017wet}. In quantum field theory, the number of exactly and analytically solvable models is quite restricted. Important examples include two-dimensional conformal field theories (CFTs)~\cite{Calabrese:2006rx, Calabrese:2007rg, Calabrese:2016xau} and free field theories~\cite{Buchel:2013gba, Das:2014jna, Das:2014hqa, Das:2015jka}.

In the local quench, the initial perturbation of a theory is localized and its subsequent evolution is studied by examining correlation functions. In contrast to the global quench, this topic is much less explored despite its obvious applications in physics. The local quench has been the subject of study in the context of condensed matter theory~\cite{Calabrese:2007mtj, cmtloc3, cmtloc1, cmtloc2, cmtloc4, e23020220}, holographic duality~\cite{Nozaki:2013wia}, the information paradox~\cite{Agon:2020fqs, Bianchi:2022ulu} and high-energy physics~\cite{Zhang:2021hra}. The problem of the local quenches naturally arises in von Neumann theory of measurements when a local projective measurement is considered in a distributed quantum system. Such a measurement leads to appearance of decoherence waves~\cite{Katsnelson:2000, Katsnelson:local, Katsnelson:2016}. Localized excited states have been studied in different contexts in~\cite{Nozaki:2014hna, He:2014mwa, Astaneh:2014fga, Caputa:2014vaa, Asplund:2014coa, Caputa:2014eta, Guo:2015uwa, Caputa:2015waa, Caputa:2015tua, Chen:2015usa, Ageev:2015qbz, Ageev:2015ykq, Rangamani:2015agy, Caputa:2015qbk, Franchini2015universal, Franchini:2016ivt, He:2017lrg, Guo:2018lqq, Ageev:2018nye, Ageev:2018msv, Shimaji:2018czt, Apolo:2018oqv, Ageev:2018tpd, Caputa:2019avh, He:2019vzf, Bhattacharyya:2019ifi, He:2020qcs, Zenoni:2021iiv, Suzuki:2022xwv}. Most of the known studied examples of the local quenches are related to two-dimensional CFT~\cite{Nozaki:2013wia, Caputa:2014vaa, Asplund:2014coa, Caputa:2019avh}. The local quench introduced in~\cite{Calabrese:2007mtj} is called the geometric quench, which has been partially generalized in~\cite{Doyon:2014qsa} to a higher-dimensional case. This quench protocol involves joining two different theories with a boundary at some time moment. Another setup is called the operator local quench, where a localized excited state is prepared by inserting some local operator into the path-integral that prepares the state~\cite{Nozaki:2013wia, Asplund:2014coa, Caputa:2014vaa, Nozaki:2014uaa}. To the best of our knowledge, this type of quenches has been explored in full generality only in the context of two-dimensional CFT. 

\skipline

In this paper, we initiate the study of the operator local quenches beyond situations where conformal symmetry plays a decisive role. Namely, in this paper we focus on free massive scalar field theories in arbitrary dimension. We study the dynamics of different composite operators, including the renormalized scalar condensate corresponding to the operator $\phi^2(x)$ and the energy density, as well as the charge density for complex scalar field theories. In two-dimensional CFT, these correlators can be studied using conformal symmetry. The mass term, however, spoils conformal invariance, so we calculate all correlators in a straightforward manner via Wick's theorem.

\skipline

In two-dimensional CFT, the energy density dynamics following the local quench by a primary operator is well-known~\cite{Caputa:2014vaa}. In massless two-dimensional Klein-Gordon theory, the simplest choice of the quenching operator corresponds to the field derivative $\partial\phi$, which is chiral. We reproduce by  straightforward calculations the answers derived previously using conformal symmetry. The generalization of this protocol to the massive two-dimensional field theory allows to choose $\phi$ as the quenching operator (i.e. without any derivatives).

\skipline

Let us summarize our conclusions regarding the dynamics of the energy and charge densities after the local quench in massive, higher-dimensional and finite-volume theories.


\subsection*{Local quenches in two-dimensional massive theories}

The addition of the mass term to two-dimensional bosonic free CFT leads to substantial change in the quench dynamics. The explicit space and time dependence of the energy density evolution is expressed in terms of the Bessel functions. At large times with fixed spatial coordinates, we find the scalings of the energy density following local quenches by different operators in two-dimensional massive theory
\benn
    \begin{aligned}
        \corrfunc{\Ecal(t, x)}_{\partial\phi} & \largetime \frac{\alpha}{t} + \frac{2 \alpha x}{t^{2}} + O\left(t^{-3}\right), \quad \alpha = \frac{\pi m}{e^{2\eps m}K_2(2\eps m)}, \\
        \corrfunc{\Ecal(t, x)}_{\phi} & \largetime \frac{\beta}{t} + O\left(t^{-3}\right), \quad \beta = \frac{\pi m}{e^{2\eps m}K_0(2\eps m)},
    \end{aligned}
\eenn
where the subscript $\phi$ or $\partial\phi$ corresponds to the type of the quenching operator. The mass scale $m$ and the quench regularization parameter $\eps$ are contained in the universal constants $\alpha$ and $\beta$, which define the leading-order expressions for the time evolution.

The large-distance asymptotics with fixed time coordinates, which are power-law in the two-dimensional conformal case, now acquire exponential corrections
\benn
    \begin{aligned}
        \corrfunc{\Ecal(t, x)}_{\partial\phi} & \largex \frac{\alpha e^{2\eps m}}{x}\,e^{-2mx} e^{\frac{m}{x}\left(t^2 - \eps^2\right)} + O\left(x^{-2}e^{-2mx} e^{\frac{m}{x}\left(t^2 - \eps^2\right)}\right), \\
        \corrfunc{\Ecal(t, x)}_{\phi} & \largex \frac{\beta e^{2\eps m}}{x}\,\,e^{-2mx} e^{\frac{m}{x}\left(t^2 - \eps^2\right)} + O\left(x^{-2}e^{-2mx} e^{\frac{m}{x}\left(t^2 - \eps^2\right)}\right).
    \end{aligned}
\eenn
The fact that observables do not vanish outside the lightcone  is due to the non-locality of the initial perturbation and specifics of the quench protocol corresponding to finite\footnote{For $\varepsilon \rightarrow 
 0$, the massless stress-energy tensor is localized on the lighcone.} $\varepsilon$.

We should emphasize that the local $\partial\phi$-quench in massive theory smoothly interpolates into the massless case. To see how the introduction of $m$ governs the evolution of the energy density at large times and distances, let us give the asymptotics taken first in the limit $m \to 0$ and then in the limit of large time/distance,
\benn
   \begin{aligned}
        \corrfunc{\Ecal(t, x)}_{\partial\phi} & \underset{\substack{m\,\to\,0 \\ t\,\to +\infty}}{\approx} \frac{4\eps^2}{t^4} + 4m^2\eps^2\left(\frac{6x^2 - \eps^2}{t^4} + \frac{3x}{t^3} + \frac{1}{t^2}\right) + O\left(\frac{m^2}{t^5}\right), \\
        \corrfunc{\Ecal(t, x)}_{\partial\phi} & \underset{\substack{m\,\to\,0 \\ x\,\to +\infty}}{\approx} \frac{4\eps^2}{x^4} - 4m^2\eps^2\left(\frac{3t^2 - \eps^2}{x^4} + \frac{t}{x^3}\right) + O\left(\frac{m^2}{x^5}\right).
    \end{aligned}
\eenn 
The leading-order contribution coincides with the CFT result, which starts at $t^{-4}$. In turn, the $m \to 0$ limit of the local $\phi$-quench in $d = 2$ is ill-defined due to the IR-divergence $1/\beta \sim \ln m$ coming from the two-point function of the field $\phi$.

The presence of scales $m$ and $\eps$ in massive theories leads to different regimes of the energy propagation. In the case of the local quench by the field operator $\phi$, we observe the following energy evolution regimes:
\begin{itemize}
    \item For relatively small masses, the spatial dependence of the energy density has the form of a double-hill configuration of decreasing amplitude, which spreads during the evolution. Initially, the peaks propagate near the lightcone going away from it after some time. 
    \item For masses larger than some critical value, the energy density spreads as a single-maximum localized configuration. In this case, the maximum of the configuration stays at the quenching point all the time, with its absolute value decreases.
    \item If the mass is exactly equal to the critical value, we observe, instead of the maximum, a stretching large plateau with decreasing amplitude and two ``fronts'' leaving the quench point.
\end{itemize}

Similar dynamics can be seen in the case of the $\partial\phi$-quench with the only difference in the presence of a kind of ``chirality'', which is expected here, because the partial derivative with respect to the lightcone coordinate ``erases'' the energy flow in the opposite lightcone direction.

\subsection*{One-point function in momentum space}

It is of interest to investigate how the two-dimensional energy density one-point function behaves in the momentum space.

In two-dimensional CFT, the Fourier image is calculable analytically, acquiring the form of a combination of a delta-function and an exponential damping factor. The delta-function selects the modes corresponding to the propagation along the lightcone, and the exponential suppression arises due to the presence of the dimensional quench regularization parameter~$\eps$. 

For the massive quenches, the answer cannot be found analytically, so we rely on numerical results. The picture for the $\phi$-quench goes as follows. The increase of mass leads to localization of modes near the momentum space origin and, simultaneously, to the growth of the energy density between the lightcone parts $\om = \pm k$. At the critical mass, the density is described by a triangular-shaped configuration exponentially suppressed from the origin to the edges. For larger masses, the energy density modes are concentrated in the narrow dumbbell-like configuration located approximately between the lines $\om \approx \pm \, s \cdot k$, where $s < 1$ is a parameter, which slowly decreases with growing~$m$.

\subsection*{Higher-dimensional and complex free field theories}

We generalize the obtained results to higher dimensions and field theories with $U(1)$ charge. In higher-dimensional theories, the post-quench dynamics has the form of a spherically-symmetric wave front diverging from the quenching point. The large-time and large-distance asymptotics are given by
\benn
    \begin{aligned}
        \corrfunc{\Ecal(t, x^i)}_{\phi,\,d} & \largetime \frac{\gamma}{t^{d - 1}} + O\left(t^{-d - 1}\right), \\
        \corrfunc{\Ecal(t, x^i)}_{\phi,\,d} & \underset{\rho\,\to +\infty}{\approx} \frac{\gamma e^{2\eps m}}{\rho^{d - 1}} \, e^{-2m\rho} \, e^{\frac{m}{\rho}(t^2 - \eps^2)} + O\left(\rho^{-d}e^{-2m\rho} \, e^{\frac{m}{\rho}(t^2 - \eps^2)}\right),
    \end{aligned}
\eenn
where
\benn
    \gamma = \frac{\pi^{2-\frac{d}{2}}\eps^{\frac{d}{2} - 1}m^{\frac{d}{2}}}{e^{2\eps m}K_{\frac{d}{2} - 1}(2\eps m)}, \quad d > 2,
\eenn
and $\rho$ is the spatial distance from the quenching point. These leading-order expressions are universal in the sense that they also cover the corresponding two-dimensional answers.

The massless higher-dimensional version is free of the mentioned IR-divergence of the $\phi$-propagator, and the energy density after the $\phi$-quench has a simple rational form
\benn
    \begin{aligned}
        \corrfunc{\Ecal(t, x^i)}_{\phi,\,d}\Big|_{m\,=\,0} = \frac{(d - 2)\Gamma\left(\frac{d}{2}\right)}{\pi^{\frac{d}{2} - 1}} \cdot \frac{(2\eps)^{d - 2}\left(\eps^2 + t^2 + \rho^2\right)}{\left[\left(\rho^2 - t^2\right)^2 + 2\eps^2\left(\rho^2 + t^2\right) + \eps^4\right]^\frac{d}{2}}, \quad d > 2.
    \end{aligned}
\eenn

We obtain that the higher-dimensional generalization of the answer for the energy density dynamics following the $\partial\phi$-quench reproduces the known holographic results of~\cite{Nozaki:2013wia} up to a constant multiplier.

\skipline

Complex scalar field theories (both massive and massless) are especially important, because they enable the study of finite chemical potential effects and interactions with gauge fields. It seems that the local quenches in theories with a charge have not been thoroughly explored (see the studies of the holographic local quenches with a finite chemical potential in~\cite{Krikun:2019wyi, Ageev:2020acl}). The post-quench charge density dynamics is driven by the same constant $\beta$ and shares the same regimes (double-hill/plateau/single-hill) as the energy density. The massless limit of the charge density after the $\partial\phi$-quench in $d 
= 2$ is well-defined and reads
\benn
    \corrfunc{\Qcal(t, x)}_{\partial\phi}\Big|_{m\,=\,0} = \frac{16\eps^3}{((x - t)^2 + \eps^2)^3}.
\eenn
The charge density following the $\phi$-quench in $d$ dimensions is described by
\benn
    \corrfunc{\Qcal(t, x^i)}_{\phi,\,d}\Big|_{m\,=\,0} = \frac{2\Gamma\left(\frac{d}{2}\right)}{\pi^{\frac{d}{2} - 1}}\left[\frac{\eps - it}{\left((\eps - it)^2 + \rho^2\right)^{\frac{d}{2}}} + \frac{\eps + it}{\left((\eps + it)^2 + \rho^2\right)^{\frac{d}{2}}}\right], \quad d > 2.
\eenn

\subsection*{Finite-volume dynamics for massive theories}

In two-dimensional CFT, the analytical expression for the energy dynamics can be derived not only for the local quench of the vacuum state in flat space, but also on a cylinder. For a general primary operator, the localized perturbations wind around the cylinder interfering when meeting at some spacetime point. In massive theory, after introduction of a non-trivial geometry, besides the regularization parameter and mass, the scale defined by the circumference of the cylinder comes into play:
\begin{itemize}
    \item For small masses (compared to the circumference of the cylinder), we observe a combination of oscillations caused by winding of the perturbation around the cylinder and  modulations caused by non-zero mass. These modulations, for small masses, manifest themselves as a kind of spreading of a localized energy soliton all over the cylinder with subsequent revivals. It is worth noticing that the \mbox{$\partial\phi$-quench} of the massive theory on a cylinder smoothly contains in its massless limit the well-known CFT result for the local quench on a cylinder. As in the flat space, ``chirality" manifests here as helical winding of the perturbation along one diagonal of the lightcone around the cylinder. This behaviour persists until $m/L \sim 1$, and then the regime of the propagation changes. 
    \item Increase of mass leads to a complicated and seemingly chaotic picture of interference and localization/delocalization pattern of the energy density dynamics. For the $\partial\phi$-quench, we observe that a single localized initial perturbation spreads over the cylinder and then evolves with an erratic structure combined of decays and revivals of the energy localizations.
\end{itemize}

It is worth noticing that by ``chaotic picture'' we mean not the exponential growth of the out-of-time-ordered correlators (OTOC), but the erratic behaviour of the energy revivals and oscillations. The notion of the chaotic behaviour in this sense has been studied in quantum mechanics, scattering amplitudes, string theory and two-dimensional CFT, see, for example,~\cite{gutz2, gutz1, Rosenhaus:2020tmv, Ageev:2020qox, Gross:2021gsj, Jepsen:2021rhs}. It would be interesting to extend this understanding somehow, and we leave this for future research.

\section{Generalities and CFT\texorpdfstring{$_2$}{2} warm-up}
\label{sec:generalities}
\subsection*{Generalities}

In the operator local quench protocol, the quenched state $\ket{\Psi(t)}$ is produced by the insertion of a local operator $O$ at the spacetime point $(t_0, x_0)$ as follows~\cite{Nozaki:2014hna, Asplund:2014coa, Caputa:2014vaa, Nozaki:2014uaa}
\be
    \ket{\Psi(t)} = \mathcal{N}_{O} \cdot e^{-iH(t - t_0)} \cdot e^{-\eps H} O(t_0, x_0)\ket{0}.
    \label{eq:operator_insertion}
\ee
Here, $\mathcal{N}_O$ is a normalization factor, which provides the unit norm of the state; the parameter $\eps$ is an infinitesimal damping factor, which makes the corresponding Euclidean path-integral convergent (in Lorentzian case, this preserves a finite norm of the state) and serves as a regularization parameter of the UV degrees of freedom.

The local quench can be interpreted as a local excitation in a quantum system that might be tracked through the evolution of some observable. After the local quench, the evolution of the observable defined by a local operator $\Ocal$ on the state~\eqref{eq:operator_insertion} is given by
\be
    \corrfunc{\Ocal(t, x)}_{O} = \frac{\bra{\Psi} \Ocal(t, x) \ket{\Psi}}{\braket{\Psi}{\Psi}}.
    \label{eq:quench_def}
\ee
In this paper, we focus on a particular case, in which the quenching operator is chosen to be an operator creating a single field perturbation. For convenience, let us choose the spacetime point of the quenching operator as $(t_0, x_0) = (0, 0)$. The presence of the damping factor in~\eqref{eq:operator_insertion} effectively shifts $t_0 \to t_0 - i\eps$\footnote{The bra-state defined as the conjugate to~\eqref{eq:operator_insertion} is shifted in the opposite direction, i.e., $t_0 \to t_0 + i\eps$, see~\eqref{eq:our_setting}.} and from a physical point of view, this defines a local excitation with the characteristic length $2\eps$. Though this shift is complex, it has to do only with the preparation of the initial state, while the observables are real-valued.

Given this, the evolution of the observable $\Ocal$ after the single-point local quench reduces to a three-point correlation function and is defined by the following expression
\be 
    \corrfunc{\Ocal(t, x)}_{O} = \frac{\bra{0}\Odag(i\eps, 0)\Ocal(t, x)O(-i\eps, 0)\ket{0}}{\bra{0}\Odag(i\eps, 0)O(-i\eps, 0)\ket{0}}.
    \label{eq:our_setting}
\ee
This is a general rule: the $n$-point correlation function after the operator quench at $m$ points is effectively given by the $(n + m)$-point correlator. 

\skipline

Since we are interested in the real-time dynamics of correlation functions, given a Euclidean correlator, we perform Wick rotation as $\tau \to it$
\be
    \bra{0}\Odag(i\eps, 0) \Ocal(t, x) O(-i\eps, 0)\ket{0}_L = \bra{0}\Odag(\eps, 0) \Ocal(\tau, x) O(-\eps, 0)\ket{0}_E\Big\rvert_{\tau\,\to\,it}.
\ee
This naive rule $\tau \to it$ for one-point correlation functions in the local quench state for finite $\eps$ is widely used in the literature~\cite{Caputa:2014eta, Caputa:2014vaa, Shimaji:2018czt, Caputa:2019avh, Bhattacharyya:2019ifi}, and we will also follow it in all calculations. In general, when dealing with a real-time dynamics of correlators involving more than one operator $\Ocal$, one has to take care about operator ordering. This might be subtle, especially given the use of complex time in the state definition. We briefly comment on this in appendix~\ref{appendix:cont}.

\subsection*{Local quench in CFT\texorpdfstring{$_2$}{2} from Ward identities}
\label{sec:CFT_scalar_quench}

In two-dimensional CFT, this setup is exactly solvable up to a wide range of conditions due to the virtue of the conformal symmetry~\cite{Caputa:2014vaa}. Throughout this work, one of the main observables in~\eqref{eq:our_setting} will be the energy density $\Ecal(t,x)$, which is universally important in all physical setups. Let us review how the evolution of the energy density after a single-point operator quench produced by a primary operator $O$ with conformal dimensions $(h, \bar{h})$ (which we denote as $O_{(h, \bar{h})}$) can be derived using conformal Ward identities\footnote{Recall that in the theory of a free massless boson, $\partial_z\phi(z, \zbar)$ is a primary operator with conformal dimensions $(1, 0)$ and $\partial_{\zbar}\phi(z, \zbar)$ is a primary operator with conformal dimensions $(0, 1).$}.

\skipline

The two-point function of a primary operator $O$ with conformal dimensions $(h, \bar{h})$ is fixed by conformal symmetries
\be
    \corrfunc{O(z_0, \zbar_0)O(z_1, \zbar_1)} = \frac{1}{(z_1 - z_0)^{2h}(\zbar_1 - \zbar_0)^{2\bar{h}}},
    \label{eq:CFT_prop}
\ee
where $(z, \zbar)$ are holomorphic (or lightcone) coordinates related to Euclidean coordinates ($\tau, x$) as $z = x + i\tau$, $\zbar = x - i\tau$. Ward identity for a primary operator $O$ with conformal dimensions $(h, \bar{h})$ and the holomorphic part of the stress-energy tensor $T(z)$ has the form
\be
    \corrfunc{T(z)O_0 \ldots O_N} = \sum\limits_{k = 0}^N\left(\frac{h}{(z - z_k)^2} + \frac{1}{z - z_k}\partial_{w_k}\right)\corrfunc{O_0 \ldots O_N} + \text{reg}(z),
    \label{eq:Ward_identities}
\ee
with the analogous identity holding for $\Tbar(\zbar)$ (here we use the notation $O_i \equiv O(z_i, \zbar_i)$). Therefore, the three-point function necessary to calculate the energy density evolution according to \eqref{eq:our_setting} can be written as
\be
    \begin{aligned}
        & \frac{\corrfunc{O(z_0, \zbar_0) \left(T(z) + \Tbar(\zbar)\right) O(z_1, \zbar_1)}}{\corrfunc{O(z_0, \zbar_0)O(z_1, \zbar_1)}} = \frac{D\left[\corrfunc{O_0 O_1}\right]}{\corrfunc{O_0 O_1}} = \\
        & = \frac{h(z_1 - z_0)^2}{(z_1 - z)^2(z_0 - z)^2} + \frac{\bar{h}(\zbar_1 - \zbar_0)^2}{(\zbar_1 - \zbar)^2(\zbar_0 - \zbar)^2},
    \label{eq:CFT_quench_arb_h}
    \end{aligned}
\ee
where the differential operator $D$ follows directly from~\eqref{eq:Ward_identities} and has the form
\be
    \begin{aligned}
        D & = \frac{h}{(z - z_1)^2} + \frac{h}{(z - z_0)^2} + \frac{\bar{h}}{(\zbar - \zbar_1)^2} + \frac{\bar{h}}{(\zbar - \zbar_0)^2} + \\
        & + \frac{1}{z - z_1}\partial_{z_1} + \frac{1}{z - z_0}\partial_{z_0} + \frac{1}{\zbar - \zbar_1}\partial_{\zbar_1} + \frac{1}{\zbar - \zbar_0}\partial_{\zbar_0}.
    \end{aligned}
\ee
Transforming the holomorphic coordinates to Euclidean and performing Wick rotation $\tau \to it$ with the substitution of points $t_0 = -i\eps$, $t_1 = i\eps$ in~\eqref{eq:CFT_quench_arb_h}, the energy density evolution after the local quench by a primary operator $O_{(1,0)}$ is given~by
\be
    \corrfunc{\Ecal(t, x)}_{O_{(1, 0)}} = - \corrfunc{T(t, x) + \Tbar(t, x)}_{O_{(1, 0)}} = \frac{4\eps^2}{\left((x - t)^2 + \eps^2\right)^2}.
    \label{eq:CFT_result}
\ee
We choose this specific conformal dimension to match with the results in the next sections. Note that the energy density is chiral for this specific choice of the quenching operator, and generically, the perturbation propagates along both parts of the lightcone as it can be seen from~\eqref{eq:CFT_quench_arb_h}. The structure of this formula is in accordance with the holographic result in~\cite{Nozaki:2013wia}.

\subsection*{Straightforward calculation}

In the previous section, we briefly reviewed how the well-known expression \eqref{eq:CFT_result} describing the evolution after the local quench in two-dimensional CFT can be derived by use of conformal symmetries. Now let us derive \eqref{eq:CFT_result} \textit{straightforwardly} using the Euclidean coordinate-space propagator and Wick's theorem.

\skipline

The action for a free massless scalar field theory in two-dimensional Euclidean spacetime reads
\be
    S = \frac{A}{2}\int d\tau\, dx\left((\partial_\tau\phi)^2 + (\partial_x\phi)^2\right),
    \label{eq:massless_action}
\ee
where $A$ is an arbitrary constant, which we are going to maintain undefined temporarily.

By definition, the energy is given by
\be
    E = \int dx\,\Ecal(\tau, x),
\ee
where $\Ecal(\tau, x)$ denotes the energy density. It can be calculated from the $\tau\tau$-component of the stress-energy tensor
\be
    T_{\alpha\beta} = \frac{4\pi}{\sqrt{|g|}}\frac{\delta S}{\delta g^{\alpha\beta}},\quad \alpha, \beta = \tau, x,
    \label{eq:stress-energy_tensor}
\ee
where we choose the normalization factor in such a form for further convenience. Hence, $\Ecal(\tau, x)$ is given by
\be
    \Ecal(\tau, x) = \pi A\left(-(\partial_\tau\phi)^2 + (\partial_x\phi)^2\right) = 2\pi A\,\partial\phi(z, \zbar)\partial\phi(z, \zbar) + 2\pi A\,\bar{\partial}\phi(z, \zbar)\,\bar{\partial}\phi(z, \zbar).
\ee

After splitting $\Ecal$ into the holomorphic and antiholomorphic parts, it takes the form
\be
    \Ecal(z, \zbar) = -(T(z) + \Tbar(\zbar)),
    \label{eq:CFT_energy-momentum_tensor}
\ee
where
\begin{gather}
    T(z) \equiv -2\pi A\,\partial\phi(z, \zbar)\,\partial\phi(z, \zbar), \\
    \Tbar(\zbar) \equiv -2\pi A\,\bar{\partial}\phi(z, \zbar)\,\bar{\partial}\phi(z, \zbar).
\end{gather}

Since the action of the massless scalar field~\eqref{eq:massless_action} involves only derivative terms, it has a global symmetry under an arbitrary constant shift of the field value $\phi \to \phi + \text{const}$. The physical $n$-point functions should be restricted to those that respect this symmetry, and for correlators constructed of shift-invariant operators, the IR-divergences cancel out.

For technical purposes, we define a formal expression for the ``two-point correlation function'' of the operator $\phi$
\be
    \corrfunc{\phi(z_1, \zbar_1)\phi(z_0, \zbar_0)} = -\frac{1}{4\pi A}\ln\left[(z_1 - z_0)(\zbar_1 - \zbar_0)\right].
    \label{eq:CFT_propagator}
\ee
It is defined up to an arbitrary constant, which can be considered as the IR-divergence of the massive field in the limit $m \to 0$ (since this two-point function is not shift-invariant). In what follows, we always consider the massless limit of a correlator, if it is shift-invariant as $m \to 0$.

Taking the derivatives of~\eqref{eq:CFT_propagator}, we obtain the two-point function\footnote{Note that~\eqref{eq:CFT_prop} and~\eqref{eq:sec2twop} differ in signs. In the CFT derivation of~\eqref{eq:CFT2}, this sign cancels out from numerator and denominator in~\eqref{eq:CFT_quench_arb_h}, and is, therefore, irrelevant. However, it is crucial for the straightforward calculation of~\eqref{eq:CFT2}, because in the end, it gives a positive energy density, as well as the Casimir effect in a finite volume, which is also determined by the transformation law of the CFT stress-energy tensor~\eqref{eq:Tz_transform}.} of the operator $\partial\phi$ without IR-divergences
\be
    \corrfunc{\partial\phi(z_1, \zbar_1)\partial\phi(z_0, \zbar_0)} = \partial_{z_1}\partial_{z_0}\corrfunc{\phi(z_1, \zbar_1)\phi(z_0, \zbar_0)} = -\frac{1}{4\pi A}\frac{1}{(z_1 - z_0)^2}.
    \label{eq:sec2twop}
\ee
This result coincides with the CFT two-point function~\eqref{eq:CFT_prop} up to a sign, with $h = 1$, $\bar{h} = 0$ if we choose the norm $A = 1/(4\pi)$.

The correlator corresponding to the evolution of the energy density after the local quench by the operator $\partial\phi$ is
\be
    \corrfunc{\Ecal(z, \bar z)}_{\partial\phi} = \frac{\bra{\partial\phi(z_1, \zbar_1)} {\cal E}(z,\zbar) \ket{\partial\phi(z_0, \zbar_0)}}{\corrfunc{\partial\phi(z_1, \zbar_1)\partial\phi(z_0, \zbar_0)}},
    \label{eq:Edphi}
\ee
and can be derived using~\eqref{eq:sec2twop} and Wick's theorem.

Energy density defined by~\eqref{eq:CFT_energy-momentum_tensor} contain composite operators\footnote{In what follows, by a composite operator we mean an operator, which contains a multiplication of several field operators taken in the same spacetime point.}. The numerator of~\eqref{eq:Edphi} can be calculated as follows
\be
    \begin{aligned}
        & \bra{\partial\phi(z_1, \zbar_1)} T(z) \ket{\partial\phi(z_0, \zbar_0)} = -2\pi A\lim_{\substack{w \to z \\ \wbar \to \zbar}}\partial_{z_1}\partial_{z_0}\partial_w\partial_z\corrfunc{\phi(z_1, \zbar_1)\phi(w,\wbar)\phi(z, \zbar)\phi(z_0, \zbar_0)} = \\
        & = -\frac{1}{8\pi A}\lim_{w \to z}\left[\frac{1}{(w - z_0)^2(z - z_1)^2} + \frac{1}{(w - z_1)^2(z - z_0)^2} + \frac{1}{(w - z)^2(z_0 - z_1)^2}\right] = \\
        & = -\frac{1}{4\pi A}\frac{1}{(z - z_0)^2(z - z_1)^2} - \frac{1}{8\pi A}\lim_{w \to z}\left[\frac{1}{(w - z)^2(z_0 - z_1)^2}\right].
        \label{eq:CFT_divergences}
    \end{aligned}
\ee
The last term in this expression is divergent, so we use the point-splitting regularization: $w - z = \wbar - \zbar \equiv \delta$, and consider the limit $\delta \to 0$. To obtain a finite answer for expressions of the form $\bra{\Psi(x_1)}\Ocal(x)^2\ket{\Psi(x_2)}$, we exploit the subtraction scheme defined as
\be
    \begin{aligned}
    & \bra{\Psi(x_1)}\Ocal(x)^2\ket{\Psi(x_2)}\Big|_\text{finite} = \\
    & = \lim\limits_{y\,\to\,x}\big[\bra{\Psi(x_1)} \Ocal(x)\Ocal(y) \ket{\Psi(x_2)} - \braket{\Psi(x_1)}{\Psi(x_2)}\bra{0}\Ocal(x)\Ocal(y)\ket{0}\big].
    \label{eq:sub_plain}
    \end{aligned}
\ee
In this way, taking into account the definition~\eqref{eq:CFT_energy-momentum_tensor}, we reproduce the answer~\eqref{eq:CFT_quench_arb_h} obtained earlier
\be
    \frac{\bra{\partial_{z_1}\phi(z_1, \zbar_1)} \Ecal(z,\zbar) \ket{\partial_{z_0}\phi(z_0, \zbar_0)}}{\corrfunc{\partial_{z_1}\phi(z_1, \zbar_1)\partial_{z_0}\phi(z_0, \zbar_0)}} = -\frac{(z_1 - z_0)^2}{(z - z_1)^2(z - z_0)^2},
    \label{eq:z_quench}
\ee
which leads to \eqref{eq:CFT_result} after the appropriate choice of $z, z_0$ and $z_1$
\be 
    \corrfunc{\Ecal(t, x)}_{\partial\phi} = \frac{4\eps^2}{\left((x - t)^2 + \eps^2\right)^2}.
    \label{eq:CFT2}
\ee

\begin{figure}[t]
    \centering
	\subfloat[]{\includegraphics[width=0.4\textwidth]{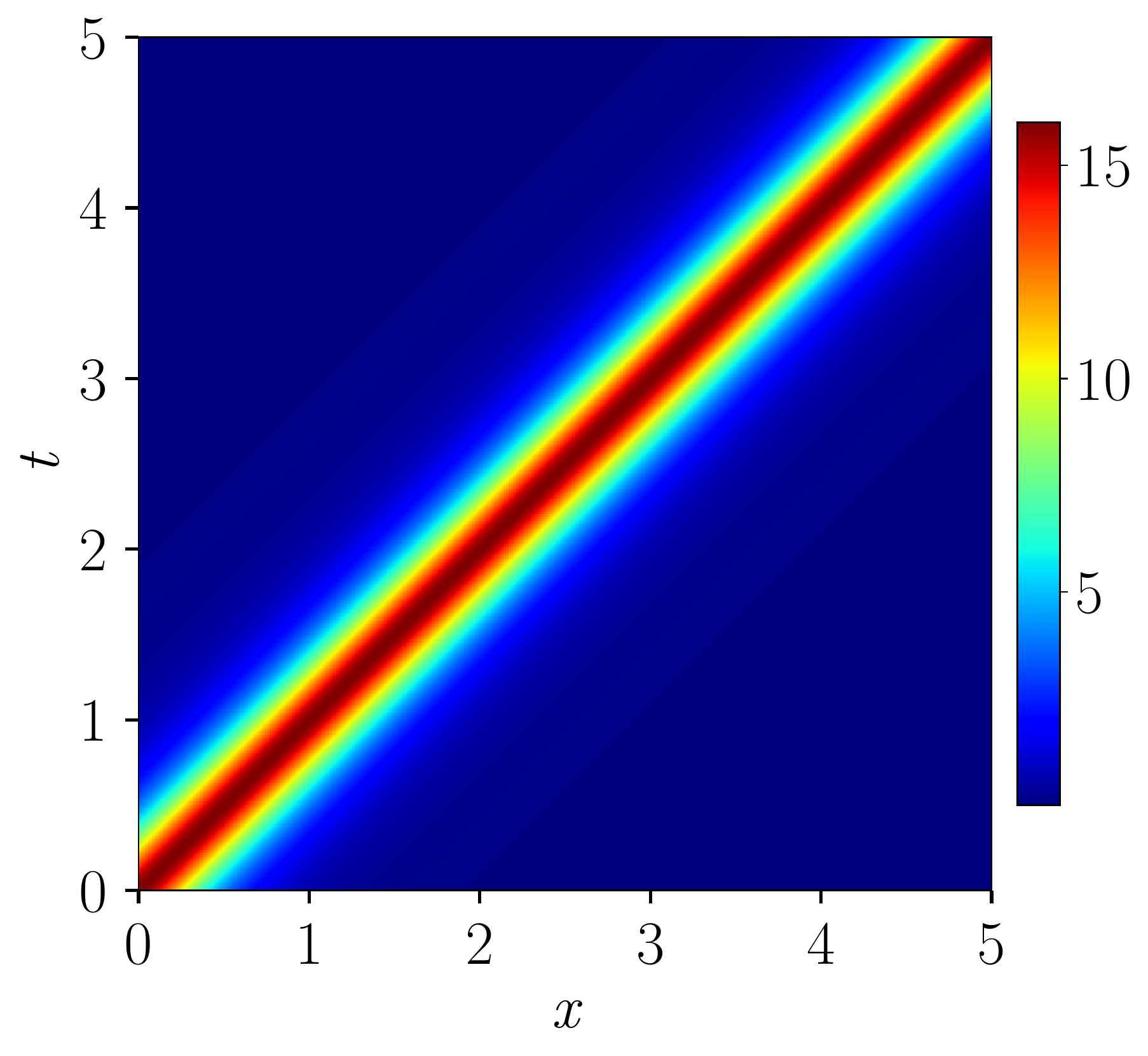}}
	\hspace{0.05\textwidth}
	\subfloat[]{\includegraphics[width=0.45\textwidth]{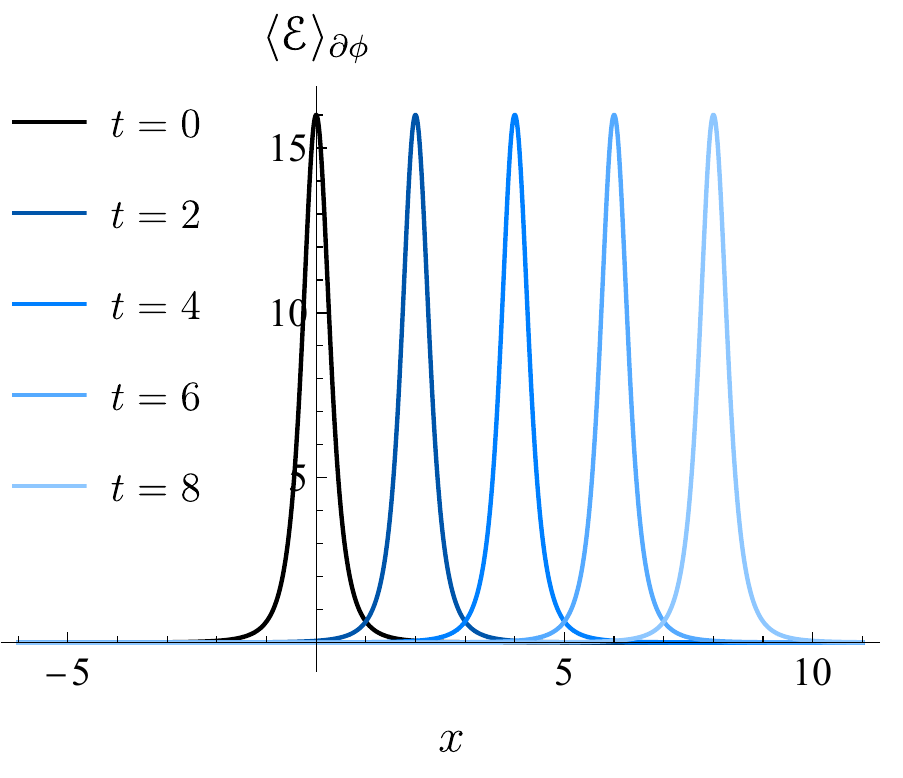}}
	\caption{\textit{Left:} Energy density evolution following the local quench by the operator $\partial\phi$ in two-dimensional CFT~\eqref{eq:CFT2} with $\eps = 0.5$. The soliton-like perturbation of the characteristic width $2\eps$ propagates along the lightcone $x = t$ and does not dilute with time. \textit{Right:} The same evolution; each line in the figure corresponds to a configuration that the perturbation has at a particular time moment.}
	\label{fig:CFT}
\end{figure}

The energy perturbation corresponding to~\eqref{eq:CFT2} has a localized soliton-like shape and propagates along the lightcone (see figure~\ref{fig:CFT}). The chirality of the operator $\partial\phi$ manifests itself explicitly in the specific direction, in which this perturbation propagates. The parameters of the configuration such as the amplitude and the width are determined by the value of $\eps$. The total energy obtained by integrating~\eqref{eq:CFT2} over~$x$ is given by
\be 
    E = \frac{2\pi}{\eps},
\ee 
and shows the divergence for $\eps \to 0$.

It is interesting to study the energy one-point correlation function corresponding to~\eqref{eq:CFT2} in momentum space $(\om,\,k)$. Performing the Fourier transformation, we obtain the following expression
\be
    \begin{aligned}
        \corrfunc{\Ecal(\om, k)}_{\partial\phi} = & -2\pi e^{\eps \om} \om\,\theta(-\om)\delta(\om + k) + \frac{2\pi}{\eps} e^{\eps \om} \,\theta(-\om)\delta(\om + k) + \\
        & + 2\pi e^{-\eps \om} \om\,\theta(\om)\delta(\om + k) + \frac{2\pi}{\eps} e^{-\eps \om}\,\theta(\om)\delta(\om + k),
    \end{aligned}
    \label{eq:FT_CFT}
\ee
which explicitly shows the localization of modes along the lightcone with the exponential suppression corresponding to larger $\om$.

\subsection*{Local quench in finite volume theory}
\label{sec:cylinder}

An important physical setup with a wide range of applications in lower-dimensional quantum systems is an out-of-equilibrium theory defined in a finite volume, for example, on a cylinder. Massless scalar field on a cylinder satisfies the periodic boundary condition: $\phi(t, x + L) = \phi(t, x)$. Correlators in two-dimensional CFT on a cylinder with coordinates $(\sigma,\sbar)$ can be obtained from those on a plane $(z,\zbar)$ by applying the following conformal map
\be
    z \to \sigma: \quad z = \exp\left(-{\frac{2i\pi\sigma}{L}}\right).
    \label{eq:plane_to_cyl_map}
\ee
It is straightforward to obtain the propagator on a cylinder using the flat-space result for the two-point function of primary operators~\eqref{eq:CFT_prop} and their transformation rule
\be
    \begin{aligned}
        & \corrfunc{\mathcal{O}_1(\sigma_1, \sbar_1) \ldots \mathcal{O}_n(\sigma_n, \sbar_n)} = \\
        & = \prod^n_{i\,=\,1}\left(\frac{dz_i}{d\sigma_i}\right)^{h} \prod^n_{i\,=\,1}\left(\frac{d\zbar_i}{d\sbar_i}\right)^{\bar{h}} \left[\corrfunc{\mathcal{O}_1(z_1, \zbar_1) \ldots \mathcal{O}_n(z_n, \zbar_n)}\Big\rvert_{\begin{subarray}{l} z_i\,\to\,\sigma_i \\ \zbar_i\,\to\,\sbar_i \end{subarray}}\right].
    \end{aligned}
\ee
For the choice of mapping~\eqref{eq:plane_to_cyl_map}, the two-point function on a cylinder takes the form
\be
    \begin{aligned}
        & \corrfunc{\mathcal{O}(\sigma_1, \sbar_1)\mathcal{O}(\sigma_0, \sbar_0)} = \\
        & = \left(\frac{2\pi}{L}\right)^{2(h + \bar{h})} \left(2\sin\left[\frac{\pi(\sigma_1 - \sigma_0)}{L}\right]\right)^{-2h}\left(2\sin\left[\frac{\pi(\sbar_1 - \sbar_0)}{L}\right]\right)^{-2h},
    \end{aligned}
\ee
where we choose holomorphic coordinates $(\sigma, \sbar)$ on a cylinder as $\sigma = x + i\tau$ and $\sbar = x - i\tau$. The transformation rule of the holomorphic part of the energy-momentum tensor under local conformal transformations reads
\be
    T(\sigma) = \left(\frac{dz}{d\sigma}\right)^{2}T(z(\sigma)) + \frac{c}{12}(Sz)(\sigma),
    \label{eq:Tz_transform}
\ee
where $(Sz)(\sigma)$ is the Schwartzian derivative of $z(\sigma)$ with respect to $\sigma$, and $c$ is the central charge.

Applying the map~\eqref{eq:plane_to_cyl_map} to~\eqref{eq:CFT_quench_arb_h} and taking into account the transformation of the stress-energy tensor~\eqref{eq:Tz_transform}, we obtain the energy density evolution on a cylinder after the local quench by a primary operator $O_{(h, \bar{h})}$
\be
    \begin{aligned}
        \corrfunc{\Ecal(\sigma, \sbar)}_{O_{(h, \bar{h})}} & = -\frac{\pi^2 c}{3L^2} - \frac{\pi^2h}{L^2} \cdot \frac{\sin^2\left[\frac{\pi(\sigma_0 - \sigma_1)}{L}\right]}{\sin^2\left[\frac{\pi(\sigma - \sigma_0)}{L}\right] \sin^2\left[\frac{\pi(\sigma - \sigma_1)}{L}\right]} - \\
        & - \frac{\pi^2\bar{h}}{L^2} \cdot\frac{\sin^2\left[\frac{\pi(\sbar_0 - \sbar_1)}{L}\right]}{\sin^2\left[\frac{\pi(\sbar - \sbar_0)}{L}\right]\sin^2\left[\frac{\pi(\sbar - \sbar_1)}{L}\right]},
    \end{aligned}
\ee
which for the conformal dimensions $h = 1$, $\bar{h} = 0$ and the central charge $c = 1$ gives
\be
    \corrfunc{\Ecal(t, x)}_{O_{(1, 0)}} = -\frac{\pi^2}{3L^2} + \frac{4\pi^2}{L^2} \cdot \frac{\sinh^2\left[\frac{2\pi\eps}{L}\right]}{\left(\cos\left[\frac{2\pi(x - t)}{L}\right] - \cosh\left[\frac{2\pi\eps}{L}\right]\right)^2}.
    \label{eq:CFT_quench_cylinder}
\ee
In contrast to the operator local quench in two-dimensional thermal CFT~\cite{Caputa:2014eta}, the constant term in~\eqref{eq:CFT_quench_cylinder} is negative. This term denotes the vacuum energy, and its negativity is known as the Casimir effect, which arises due to a finite volume of the cylinder. From the CFT point of view, the explanation is as follows. On a plane, the vacuum expectation value of $T(z)$ is zero. However, mapping it to a cylinder, we obtain a non-zero value of the Schwartzian derivative, which for the energy density~$\Ecal$ takes a negative value. The second dynamical term describes the evolution of the excited quenched state, whose energy is clearly positive.

\skipline

This derivation fully relies on conformal symmetries. It can also be obtained straightforwardly using Wick's theorem. The two-point function for free massless scalar field on a cylinder can be formally obtained as the Green's function of the Klein-Gordon operator with imposed periodic boundary conditions (see appendix~\ref{appendix:cyl} for details). The reason why this Green's function is not well-defined is that it contains an IR divergence, just like the two-point function on a plane~\eqref{eq:CFT_propagator}. Taking derivatives with respect to holomorphic coordinates, we eliminate this divergence and obtain a well-defined two-point function of the field operator $\partial\phi$ (refer to~\eqref{eq:app:massless_dphi_2point_cyl})
\be
    \corrfunc{\partial\phi(\sigma_1, \sbar_1)\partial\phi(\sigma_0, \sbar_0)} = - \frac{1}{4\pi A}\left(\frac{\pi}{L}\right)^2\sin^{-2}\left[\frac{\pi(\sigma_1 - \sigma_0)}{L}\right].
\ee
Here $A$ is the normalization factor of the action. Using this expression along with Wick's theorem, we get that the dynamics of the holomorphic part of the energy density obeys
\be
    \begin{aligned}
        & \bra{\partial\phi(\sigma_1, \sbar_1)} T(\sigma) \ket{\partial\phi(\sigma_0, \sbar_0)} = \\
        & = -2\pi A\lim_{\substack{\xi\,\to\,\sigma \\ \bar{\xi}\,\to\,\sbar}}\partial_{\sigma_1}\partial_{\sigma_0}\partial_{\xi}\partial_{\sigma}\corrfunc{\phi(\sigma_1, \sbar_1)\phi(\xi, \bar{\xi})\phi(\sigma, \sbar)\phi(\sigma_0, \sbar_0)} = \\
        & = -\frac{\pi^3}{4 A L^4}\sin^{-2}\left[\frac{\pi(\sigma - \sigma_0)}{L}\right]\sin^{-2}\left[\frac{\pi(\sigma_1 - \sigma)}{L}\right] - \\
        & -\frac{\pi^3}{8 A L^4} \lim_{\substack{\xi\,\to\,\sigma \\ \bar{\xi}\,\to\,\sbar}}\left(\sin^{-2}\left[\frac{\pi(\sigma_1 - \sigma_0)}{L}\right] \sin^{-2}\left[\frac{\pi(\xi - \sigma)}{L}\right]\right).
    \end{aligned}
\ee
To eliminate divergences in the last term, we use the following subtraction scheme
\be
    \frac{\bra{\Psi(x_1)}\Ocal(x)^2\ket{\Psi(x_2)}}{\braket{\Psi(x_1)}{\Psi(x_2)}}\Bigg|_\text{finite} = \lim\limits_{x\,\to\,y}\left[\frac{\bra{\Psi(x_1)} \Ocal(x)\Ocal(y) \ket{\Psi(x_2)}}{\braket{\Psi(x_1)}{\Psi(x_2)}} - \bra{0}\Ocal(x)\Ocal(y)\ket{0}_{\text{flat}}\right],
    \label{eq:sub_cyl}
\ee
i.e., we subtract the two-point function for the flat space. Such a choice of subtraction keeps the constant term intact. For the total energy-momentum tensor $T(\sigma) + \Tbar(\sbar)$, this constant is the same as that one, which arises in the derivation based on Ward identities, and reflects the fact that the energy-momentum tensor transforms anomalously under conformal transformations. One should notice that it is more correct to use the covariant subtraction scheme on a cylinder~\cite{Wald:1984rg}. However, we would like to leave this for future research and stay with a naive subtraction scheme, which, nevertheless, gives a correct insight into the quantum dynamics.

\begin{figure}
    \centering
	\includegraphics[width=0.8\textwidth]{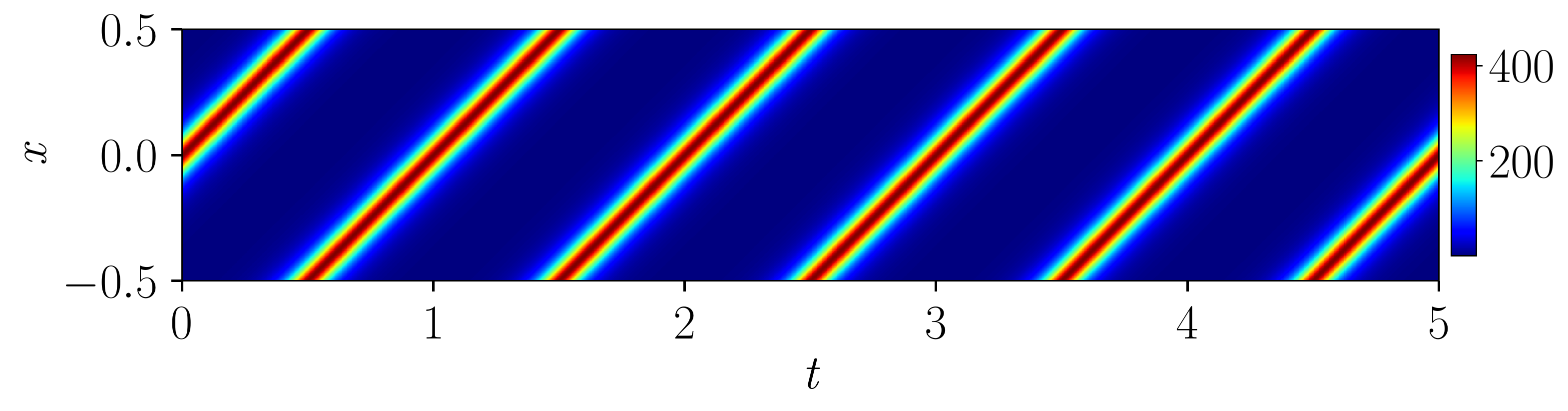}
    \caption{Energy density evolution after the local quench by the operator $\partial\phi$ in two-dimensional CFT on a cylinder~\eqref{eq:CFT_quench_cylinder}. The parameters are $\eps = 0.1$ and $L = 1$.}
	\label{fig:CFT_quench_cylinder}
\end{figure}

Finally, we arrive at the same result that was obtained as a consequence of conformal symmetries~\eqref{eq:CFT_quench_cylinder}, $\corrfunc{\Ecal(t, x)}_{\partial\phi} = \corrfunc{\Ecal(t, x)}_{O_{(1, 0)}}$. The evolution of the energy density after the local quench on a cylinder has the form of a localized perturbation freely winding around the cylinder (figure~\ref{fig:CFT_quench_cylinder}).

\section{Local quenches in massive scalar field theory}
\label{sec:mass_deform}
\subsection{Two-dimensional massive scalar field}
\subsubsection*{Local quench by operator $\partial\phi$}

Now let us turn to different types of the local quenches in the theory of a free massive scalar field in two dimensions. In massive scalar field theory, the quench by the field operator $\phi$ is well-defined in contrast to the massless case (it does not possess IR-divergences). However, to compare with the two-dimensional CFT result it is reasonable to start with the $\partial\phi$-quench and explore how the presence of mass $m$ affects the energy evolution.

\skipline

The Euclidean action of a free massive scalar field theory has the form
\be
    S = \frac{1}{8\pi}\int d\tau\,dx\left((\partial_\tau\phi)^2 + (\partial_x\phi)^2 + m^2\phi^2\right),
    \label{eq:massive_theory}
\ee
with the energy density calculated using~\eqref{eq:stress-energy_tensor}
\be
    \Ecal(\tau, x) = \frac{1}{4}\left(-(\partial_\tau\phi)^2 + (\partial_x\phi)^2 + m^2\phi^2\right).
\ee
In holomorphic coordinates, it can be rewritten as
\be
    \Ecal(z, \zbar) = \frac{1}{2}\left(\partial\phi(z, \zbar)
    \right)^2 + \frac{1}{2}\left(\bar{\partial}\phi(z, \zbar)\right)^2 + \frac{1}{4}m^2\phi^2(z, \zbar).
\ee
matching the CFT expression~\eqref{eq:CFT_energy-momentum_tensor} in the massless limit.

The two-point correlation function corresponding to the massive scalar field theory is given by the following expression
\be
    \corrfunc{\phi(z_1, \zbar_1)\phi(z_0, \zbar_0)} = 2K_0\left(m \sqrt{(z_1 - z_0)(\zbar_1 - \zbar_0)}\right),
    \label{eq:mass_2d_propagator}
\ee
which is the Green's function of the corresponding Klein-Gordon differential operator. Hence, the correlator of $\partial\phi$-operators has the form
\be
    \corrfunc{\partial_{z_1}\phi(z_1,\zbar_1)\partial_{z_0}\phi(z_0,\zbar_0)} = -\frac{m^2}{2}\cdot\frac{\zbar_1 - \zbar_0}{z_1 - z_0} \cdot K_2\left(m\sqrt{(z_1 - z_0)(\zbar_1 - \zbar_0)}\right).
    \label{eq:2dim-two-point}
\ee
Repeating the same steps as in the straightforward calculation in the previous section, we obtain the evolution of the energy density after the $\partial\phi$-quench
\be
    \begin{aligned} 
        & \corrfunc{\Ecal(t, x)}_{\partial\phi} = \frac{m^2}{2K_2(2\eps m)}\left[\frac{2}{\eps^2 + (t - x)^2}\left|\sqrt{(\eps - it)^2 + x^2}\,K_1\left(m\sqrt{(\eps - it)^2 + x^2}\right)\right|^2\right. + \\
        & + \left.\frac{\eps^2 + (t + x)^2}{\eps^2 + (t - x)^2}\left|K_2\left(m\sqrt{(\eps - it)^2 + x^2}\right)\right|^2 + \left|K_0\left(m\sqrt{(\eps - it)^2 + x^2}\right)\right|^2\right].
    \end{aligned}
    \label{eq:CFT_mass_deformed}
\ee
The divergent and the constant terms eliminated by the appropriate subtraction procedure (as in the massless case considered previously, \eqref{eq:sub_plain}) have the form
\begin{gather}
    {\cal C}_{\partial\phi} = \frac{m^2}{2}\left[\frac{1}{2} - \gamma_E - \ln\left(\frac{m}{2}\right)\right], \\
    {\cal D}_{\partial\phi} = - \lim_{\delta\,\to\,0}\left[\frac{1}{\delta^2} + \frac{m^2}{2}\ln\delta + O(\delta)\right],
    \label{eq:CFT_mass_deformed_divergences}
\end{gather}
where $\gamma_E$ denotes the Euler's constant, and $\delta$ is the point-splitting regularization parameter (see the comment after~\eqref{eq:CFT_divergences}). Also, one can observe the presence of the logarithmic divergence, which disappears when $m = 0$.

\skipline

The first thing worth noticing about the expression~\eqref{eq:CFT_mass_deformed} is that the energy is conserved, i.e., one can check numerically that the spatial integral of~\eqref{eq:CFT_mass_deformed} remains constant during the evolution. 

To get some intuition on how the mass $m$ changes the behaviour of the energy density, let us study different asymptotic regimes, which can be straightforwardly obtained from~\eqref{eq:CFT_mass_deformed}. At large times with fixed spatial coordinate, the energy density one-point function behaves~as
\be
    \corrfunc{\Ecal(t, x)}_{\partial\phi} \largetime \frac{\alpha}{t} + \frac{2\alpha x}{t^2} + O\left(t^{-3}\right),
    \label{eq:dphi_large_times}
\ee
where we have introduced the constant $\alpha$
\be
    \alpha = \frac{\pi m}{e^{2\eps m}K_2(2\eps m)}.
    \label{eq:alpha_def}
\ee
In contrast to the CFT answer~\eqref{eq:CFT2}, which starts at $O\left(t^{-4}\right)$, this expression starts directly from the $O\left(t^{-1}\right)$ term. One can see how additional modes included in the large-time dynamics are controlled by the constant $\alpha$: for small masses, $\alpha \sim \eps^2 m^3$, while for large masses, $\alpha$ depends on $m$ as $\alpha \sim \sqrt{\eps}\,m^{3/2}$.

For large spatial coordinate with fixed time, the asymptotic also changes already in the leading order in $x$ acquiring exponential suppression~\footnote{We have used that $K_n(x) \largex e^{-x} \sqrt{\frac{\pi}{2x}} + O\left(e^{1/x^2}x^{-3/2}\right).$
}
\be
    \corrfunc{\Ecal(t, x)}_{\partial\phi} \largex \frac{\alpha e^{2\eps m}}{x}\,e^{-2mx} e^{\frac{m}{x}\left(t^2 - \eps^2\right)} + O\left(x^{-2}e^{-2mx} e^{\frac{m}{x}\left(t^2 - \eps^2\right)}\right).
\ee
The limit of small mass yields the following large-time and distance asymptotics (first we expand in series with respect to mass and then with respect to time/distance)
\be
   \begin{aligned}
        \corrfunc{\Ecal(t, x)}_{\partial\phi} & \underset{\substack{m\,\to\,0 \\ t\,\to +\infty}}{\approx} \frac{4\eps^2}{t^4} + 4m^2\eps^2\left(\frac{6x^2 - \eps^2}{t^4} + \frac{3x}{t^3} + \frac{1}{t^2}\right) + O\left(\frac{m^2}{t^5}\right), \\
        \corrfunc{\Ecal(t, x)}_{\partial\phi} & \underset{\substack{m\,\to\,0 \\ x\,\to +\infty}}{\approx} \frac{4\eps^2}{x^4} - 4m^2\eps^2\left(\frac{3t^2 - \eps^2}{x^4} + \frac{t}{x^3}\right) + O\left(\frac{m^2}{x^5}\right).
    \end{aligned}
\ee
One can note that the observables are non-vanishing outside the lightcone. This feature is a consequence of the setup: the regulator $\eps$ spreads and smears the perturbation making it non-local with the initial characteristic width\footnote{Notice that the two-point correlation function of a massive relativistic scalar field is non-vanishing outside the lightcone.} $2\eps$.

\begin{figure}[ht]
    \centering
    \subfloat[$m = 1$]{\includegraphics[width=0.437\textwidth]{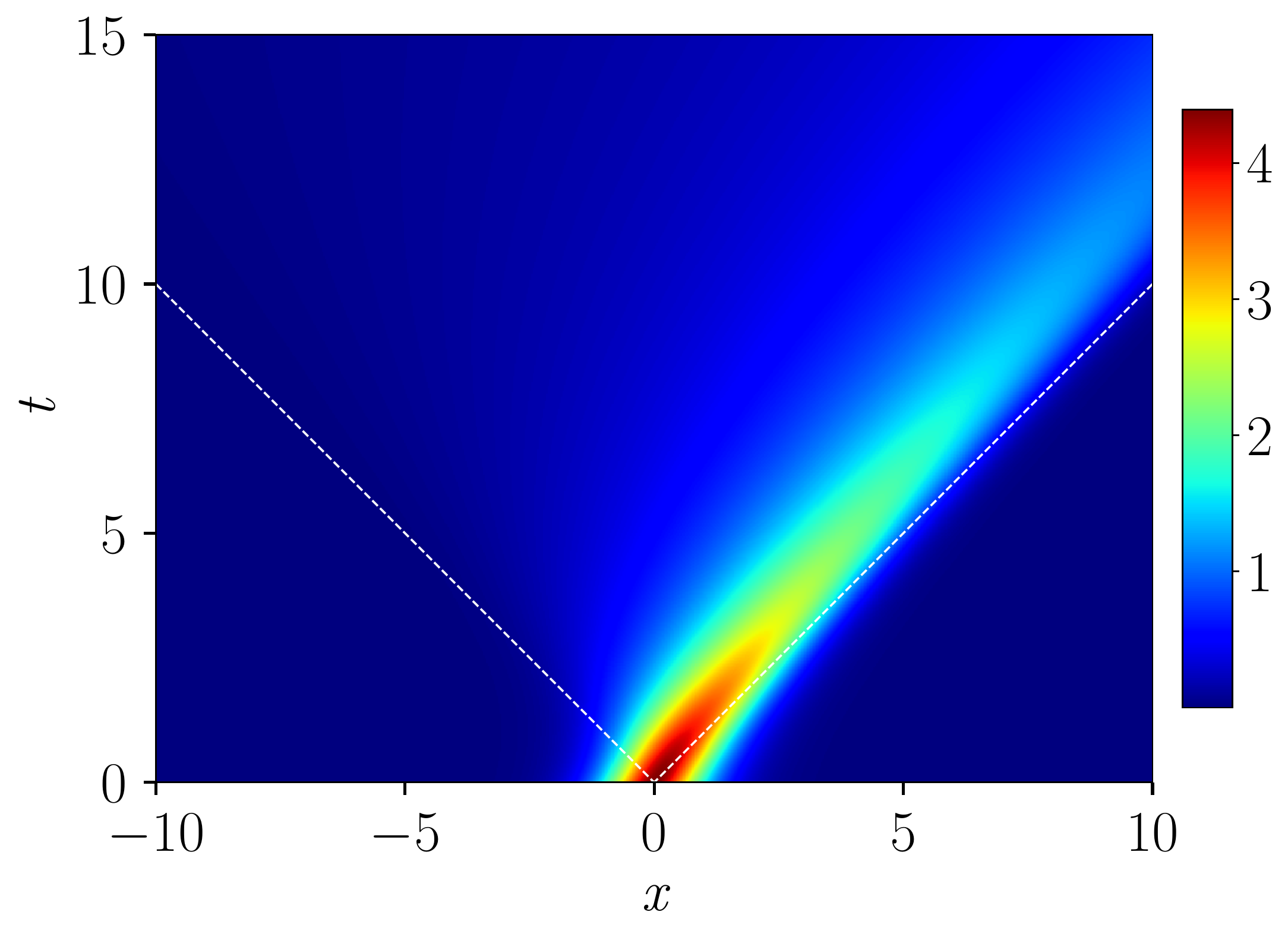}}
    \hspace{0.05\textwidth}
    \subfloat[$m = 15$]{\includegraphics[width=0.455\textwidth]{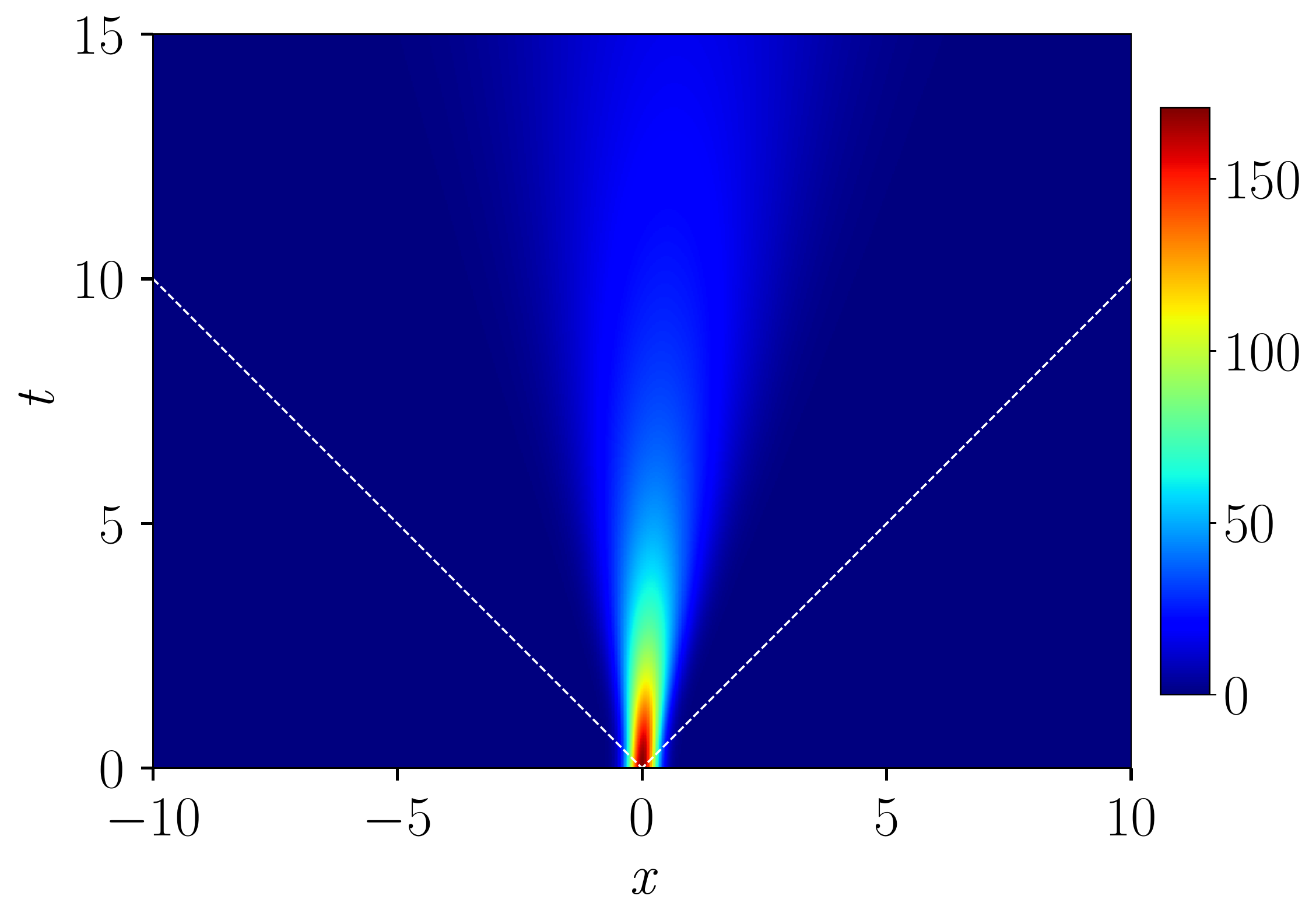}} \\
    \subfloat[$m = 1$]{\includegraphics[width=0.45\textwidth]{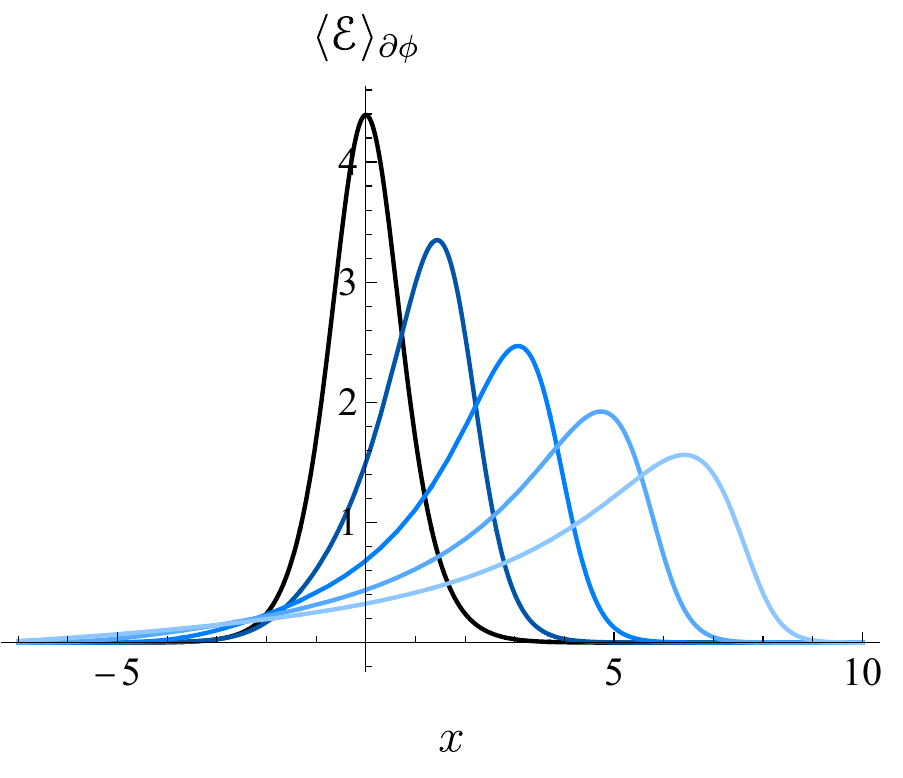}} \hspace{0.05\textwidth}
    \subfloat[$m = 15$]{\includegraphics[width=0.45\textwidth]{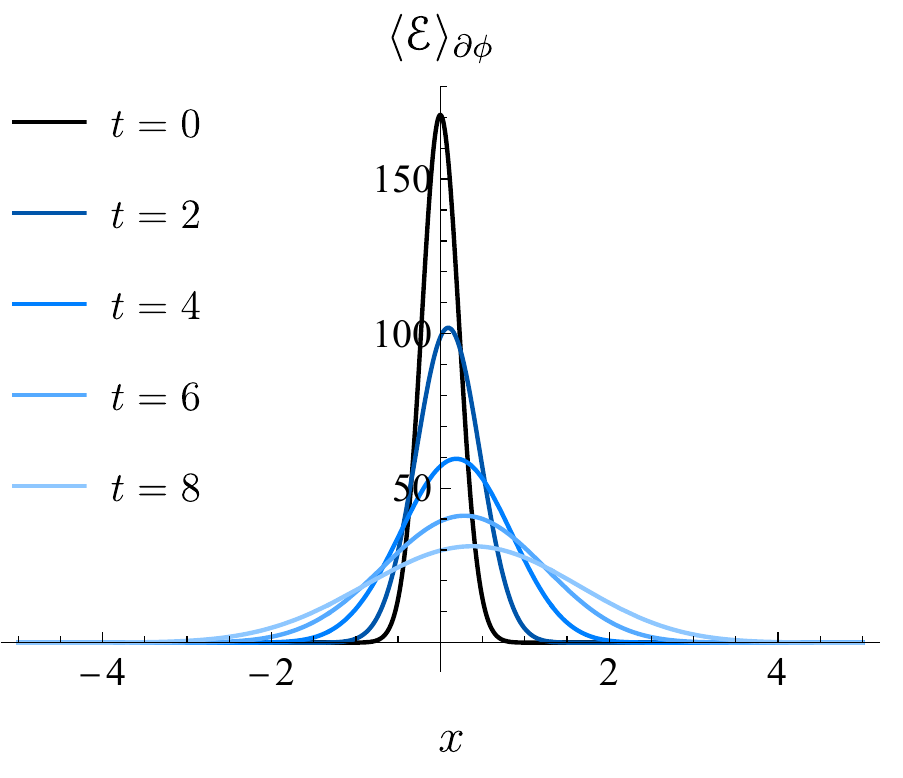}}
    \caption{\textit{Top}: Energy density evolution following the local  $\partial\phi$ in massive scalar field theory. The left figure corresponds to $m = 1$ and the right one to $m = 15$, and $\eps = 1.5$ is fixed for both figures. Dotted lines mark the lightcone. \textit{Bottom:} Spatial energy density distribution after the local $\partial\phi$ quench for fixed time moments. The left figure corresponds to $m = 1$ and the right one to $m = 15$, and $\eps = 1.5$ is fixed for both figures.}
    \label{fig:dphiE-massive}
\end{figure}

In figure~\ref{fig:dphiE-massive}, one can see how the introduction of the mass scale affects the quench dynamics. Due to the presence of mass, the perturbation ``rotates'' away from the lightcone and shows dissipation effects. Larger masses correspond to larger initial localization of the perturbation, i.e., for large masses the energy density evolution starts from a relatively narrow localized structure, which gradually dissipates in space.

\begin{figure}
	\centering
	\subfloat[massless (CFT)]{\includegraphics[width=0.325\textwidth]{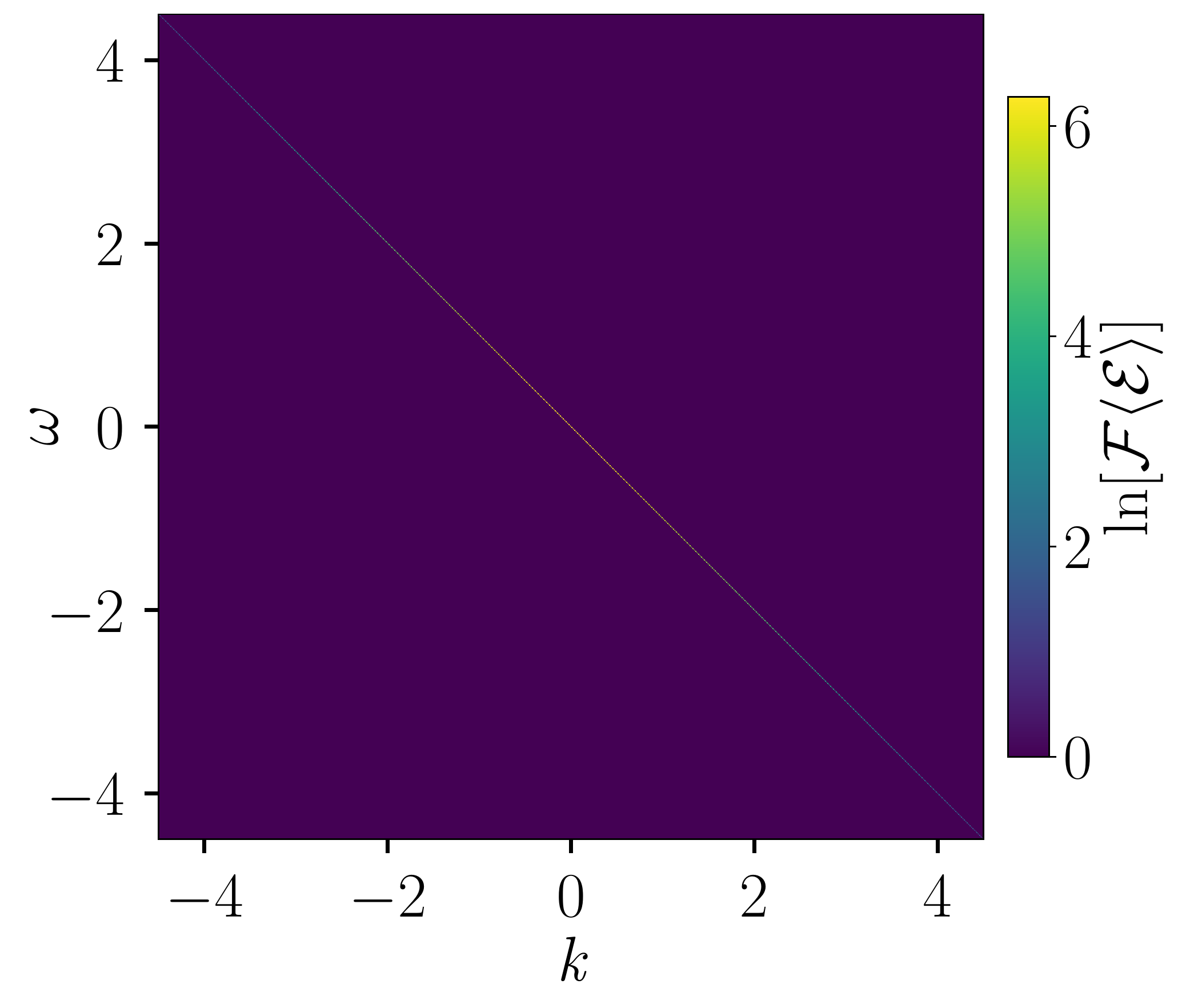}} \hfill
	\subfloat[$m = 0.1$]{\includegraphics[width=0.325\textwidth]{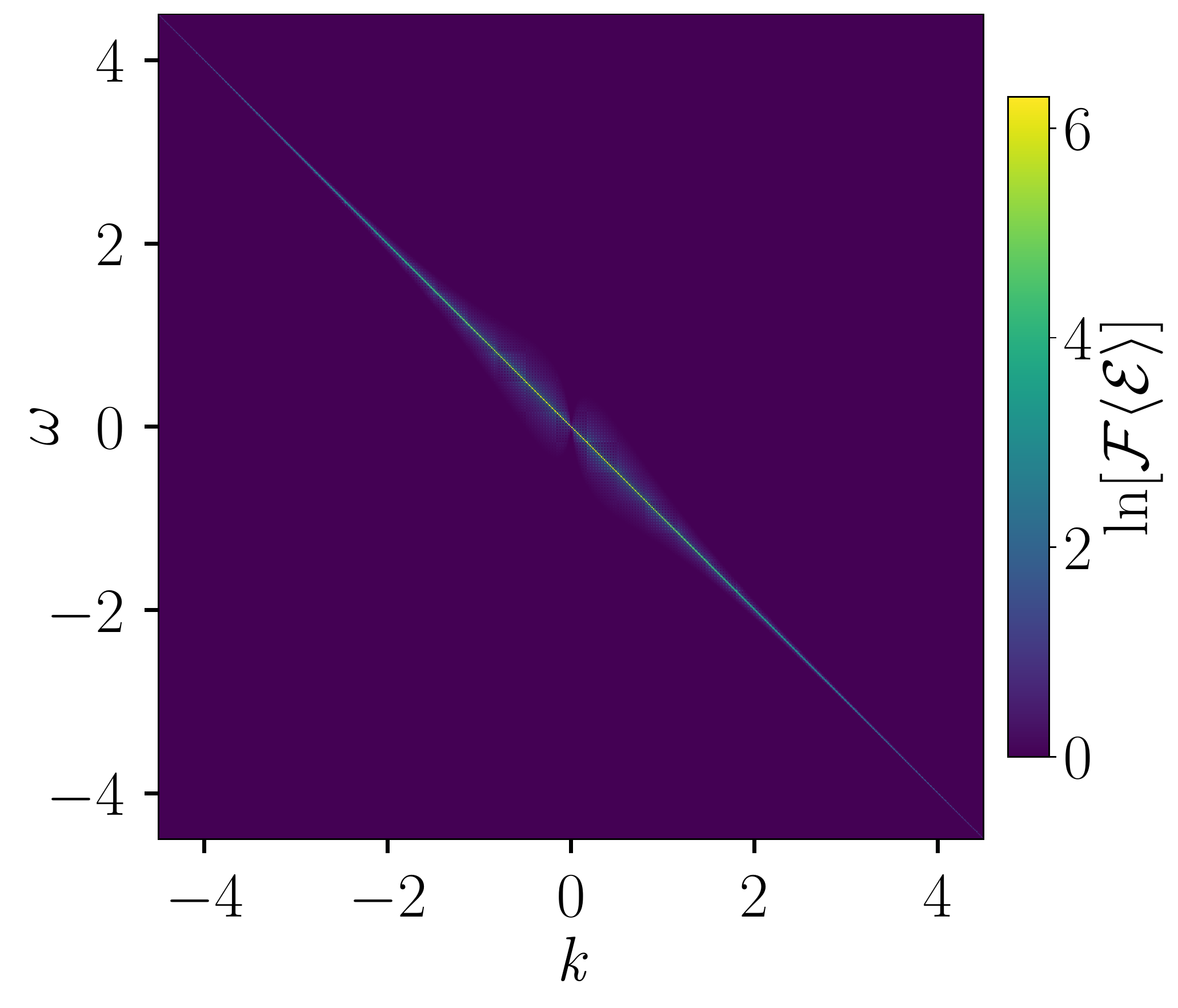}} \hfill
	\subfloat[$m = 0.5$]{\includegraphics[width=0.325\textwidth]{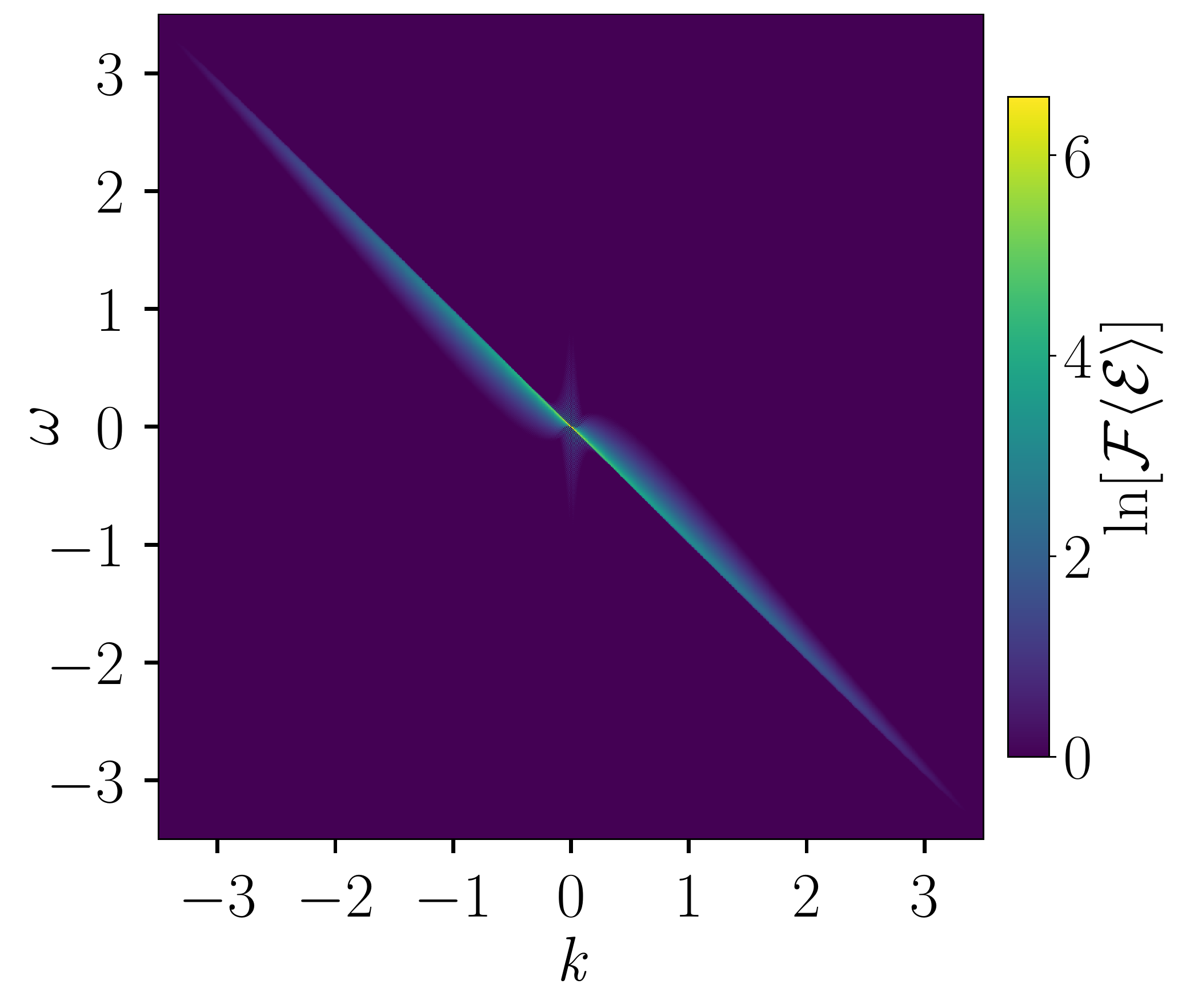}} \\
	\subfloat[$m = 1$]{\includegraphics[width=0.325\textwidth]{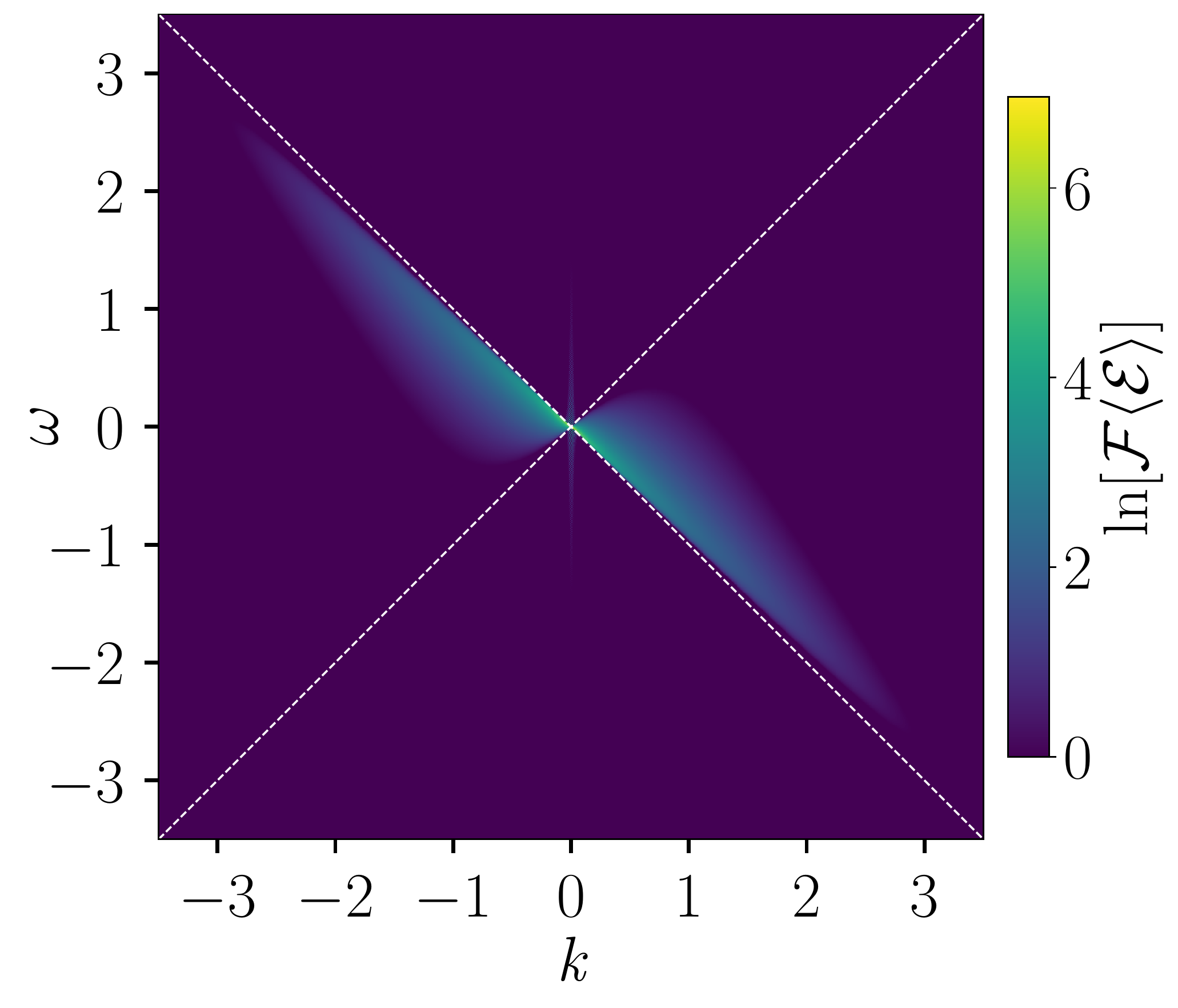}} \hfill
	\subfloat[$m = 3$]{\includegraphics[width=0.325\textwidth]{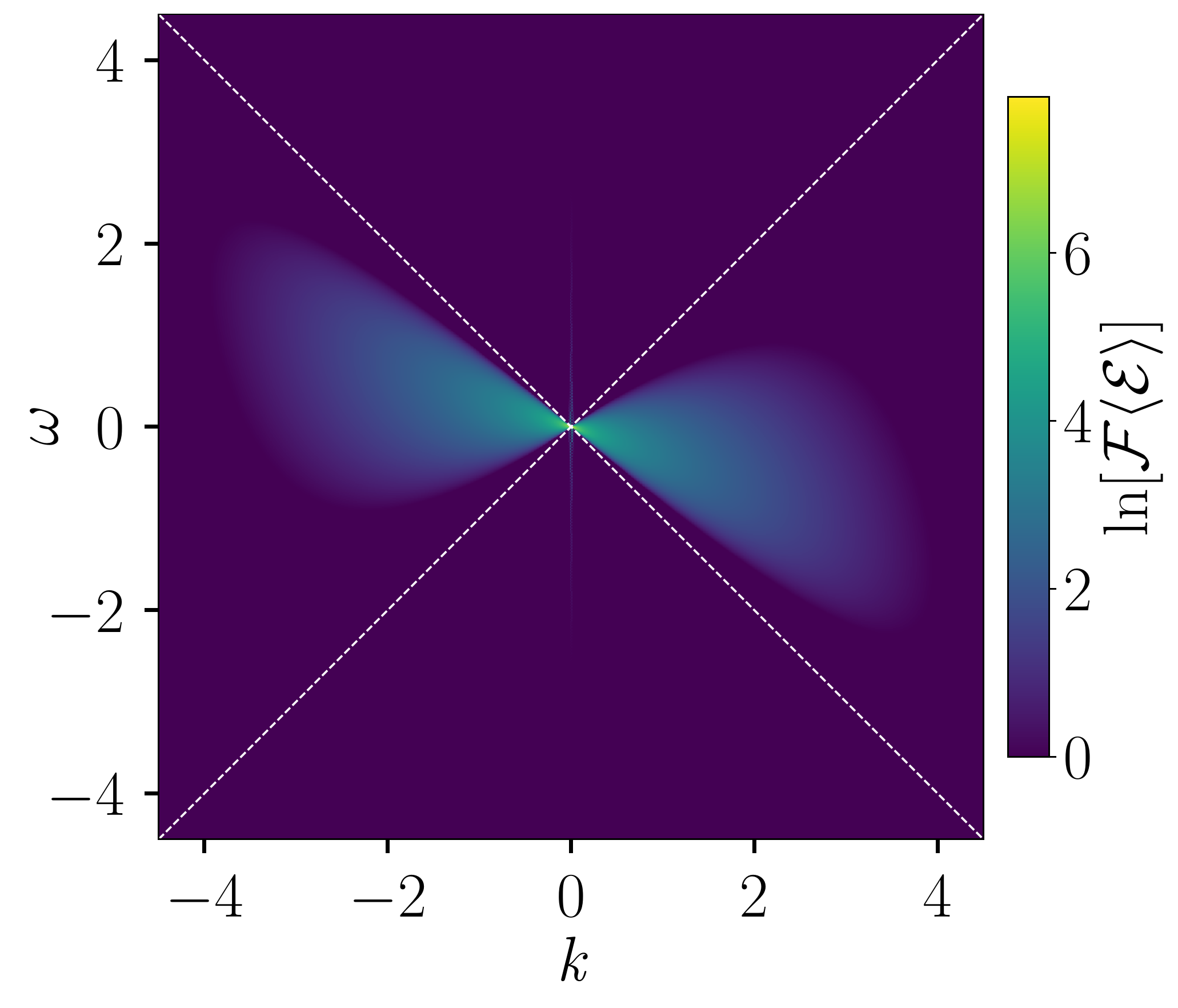}} \hfill
	\subfloat[$m = 20$]{\includegraphics[width=0.333\textwidth]{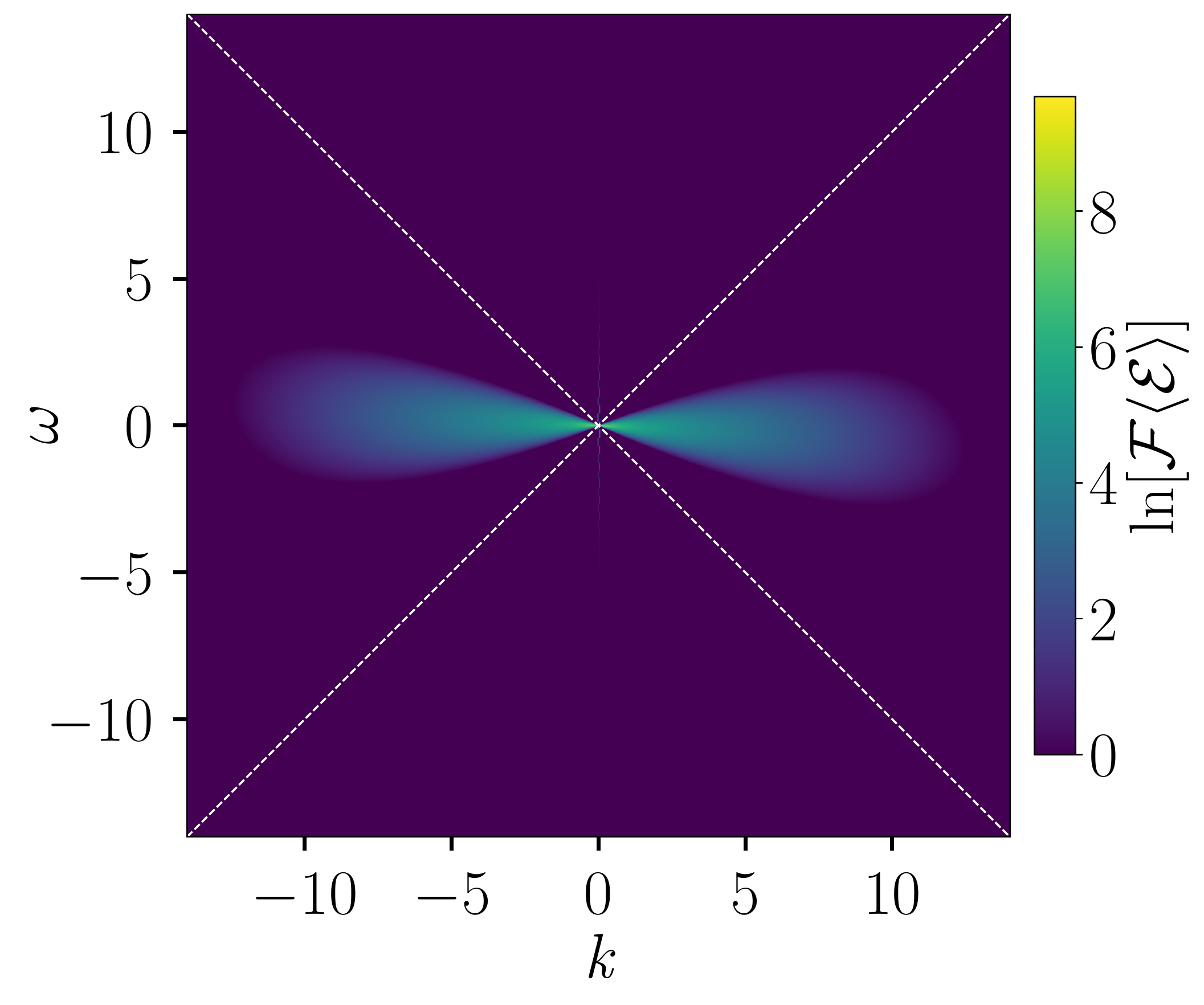}}
	\caption{Fourier images of the energy density evolution after the local quench by the operator $\partial\phi$ in massive scalar field theory, $\mathcal{F}\corrfunc{\Ecal}_{\partial\phi}$. Dotted lines in figures d), e) and f) mark the lightcone, $\om = \pm k$. Note that the images represent the real part of the Fourier transformation in logarithmic scale.}
	\label{fig:FT_dphiE-massive}
\end{figure}

To study distribution of modes in the energy density evolution after the $\partial\phi$-quench, we transform~\eqref{eq:CFT_mass_deformed} to momentum space (see figure~\ref{fig:FT_dphiE-massive}). We observe that in contrast to the massless case where the Fourier image of the energy density is proportional to $\delta(\om + k)$ (see~\eqref{eq:FT_CFT}), in the massive case, the configuration rolls off the lightcone part $\om = -k$ and gets more localized around $\om = 0$ as $m$ increases. With the further growth of mass, the configuration takes a dumbbell-like shape stretching along the line $\om = 0$ and dissipating rapidly for large values of $k$.

\skipline

Finally, let us consider the dynamics of the composite operator $\phi^2$, which for the massive field theory is one of the constituents of the energy density operator as well as the most simple example of a composite operator. It is the simplest non-vanishing one-point correlator whose post-quench dynamics after the subtraction has the form
\be
    \corrfunc{\phi^2(t, x)}_{\partial\phi} = \frac{4}{\left(\eps^2 + (t - x)^2\right)K_2(2\eps m)}\left|\sqrt{(\eps - it)^2 + x^2}K_1\left(m \sqrt{(\eps - it)^2 + x^2}\right)\right|^2.
\ee

\subsubsection*{Local quench by operator $\phi$}

In contrast to the operator $\partial\phi$, the local quench protocol based on the quenching operator $\phi$ does not have a nice implementation in two-dimensional massless bosonic field theory because the corresponding propagator is ill-defined. However, in massive field theory, there are no IR-divergences.

Straightforward calculation with the two-point function of the operator $\phi$~\eqref{eq:mass_2d_propagator} results in the following expression for the post-quench energy density evolution
\be
    \begin{aligned}
        \corrfunc{\Ecal(t, x)}_{\phi} = \frac{m^2}{K_0(2\eps m)}\left[\left(\eps^2 + t^2 + x^2\right)\left|\frac{K_1\left(m\sqrt{(\eps - it)^2 + x^2}\right)}{\sqrt{(\eps - it)^2 + x^2}}\right|^2\right. + \\ + \Bigg.\left|K_0\left(m\sqrt{(\eps - it)^2 + x^2}\right)\right|^2\Bigg].
    \label{eq:massive_quench}
    \end{aligned}
\ee
The structure of the divergent terms is the same as in the case of the $\partial\phi$-quench~\eqref{eq:CFT_mass_deformed_divergences}, $\mathcal{C}_{\phi} = \mathcal{C}_{\partial\phi}$ and $\mathcal{D}_{\phi} = \mathcal{D}_{\partial\phi}$. More generally, this is a consequence of the fact that the divergent terms come from the composite operator and not from the quenching one.

The large-time (at fixed spatial distance) and large-distance (at fixed time) asymptotics correspondingly have the form
\be
    \begin{aligned}
        \corrfunc{\Ecal(t, x)}_{\phi} & \largetime \frac{\beta}{t} + \left[\frac{1}{4m}\left(\eps - 2m\left(\eps^2 + x^2(2\eps m - 3)\right)\right) + \frac{1}{8m^2}\right]\frac{\beta}{t^3} + O\left(t^{-5}\right), \\
        \corrfunc{\Ecal(t, x)}_{\phi} & \largex \frac{\beta e^{2\eps m}}{x}\,\,e^{-2mx} e^{\frac{m}{x}\left(t^2 - \eps^2\right)} + O\left(x^{-2}e^{-2mx}\, e^{\frac{m}{x}\left(t^2 - \eps^2\right)}\right),
    \end{aligned}
    \label{eq:phi_asymptotics}
\ee
where the constant $\beta$ is defined as
\be 
    \beta = \frac{\pi m}{e^{2\eps m}K_0(2\eps m)}.
    \label{eq:beta_def}
\ee

\begin{figure}[t!]
    \centering
    \subfloat[$m = 0.1$]{\includegraphics[width=0.3355\textwidth]{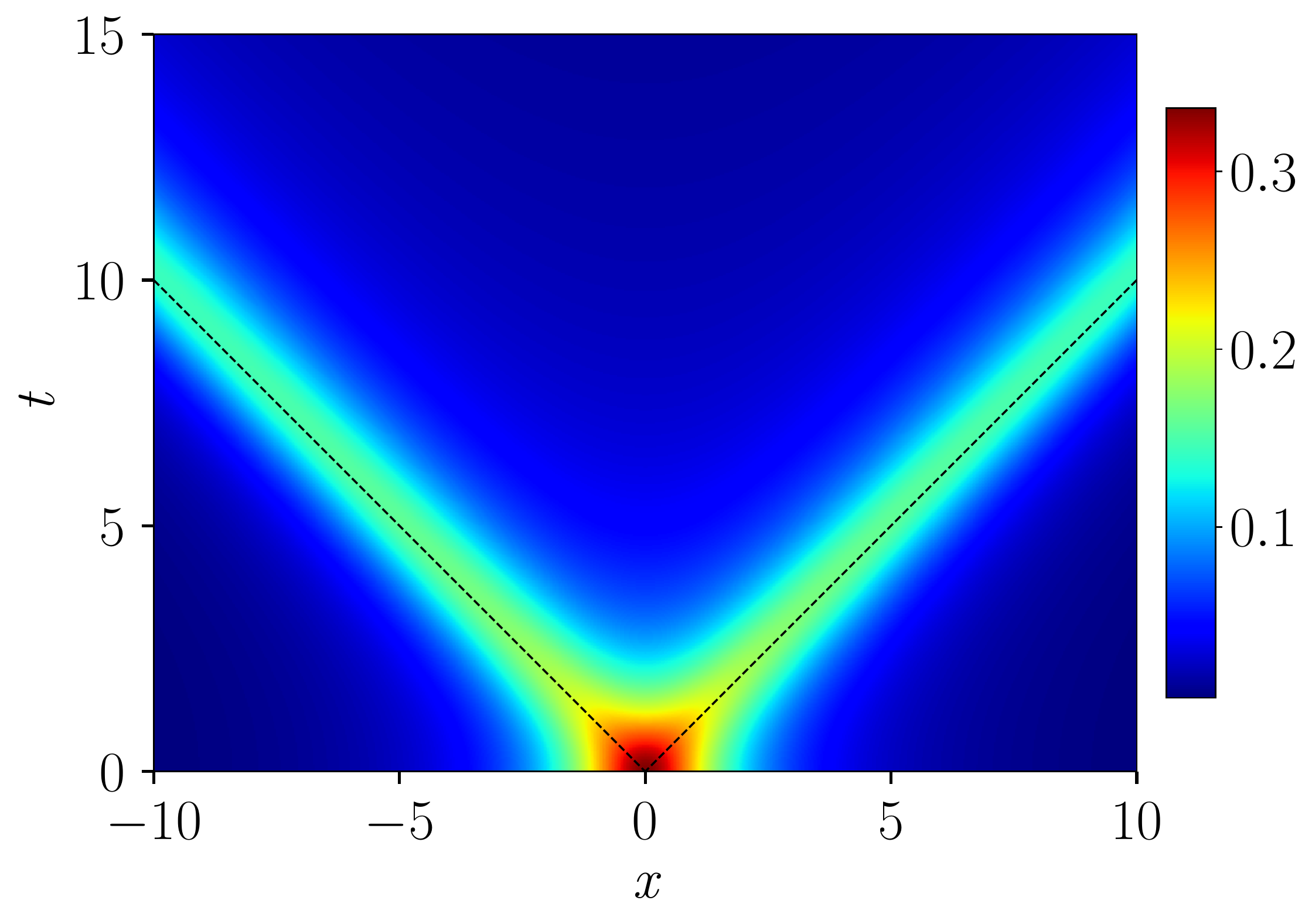}} \hfill
    \subfloat[$m = m_{\text{crit}} = 1$]{\includegraphics[width=0.325\textwidth]{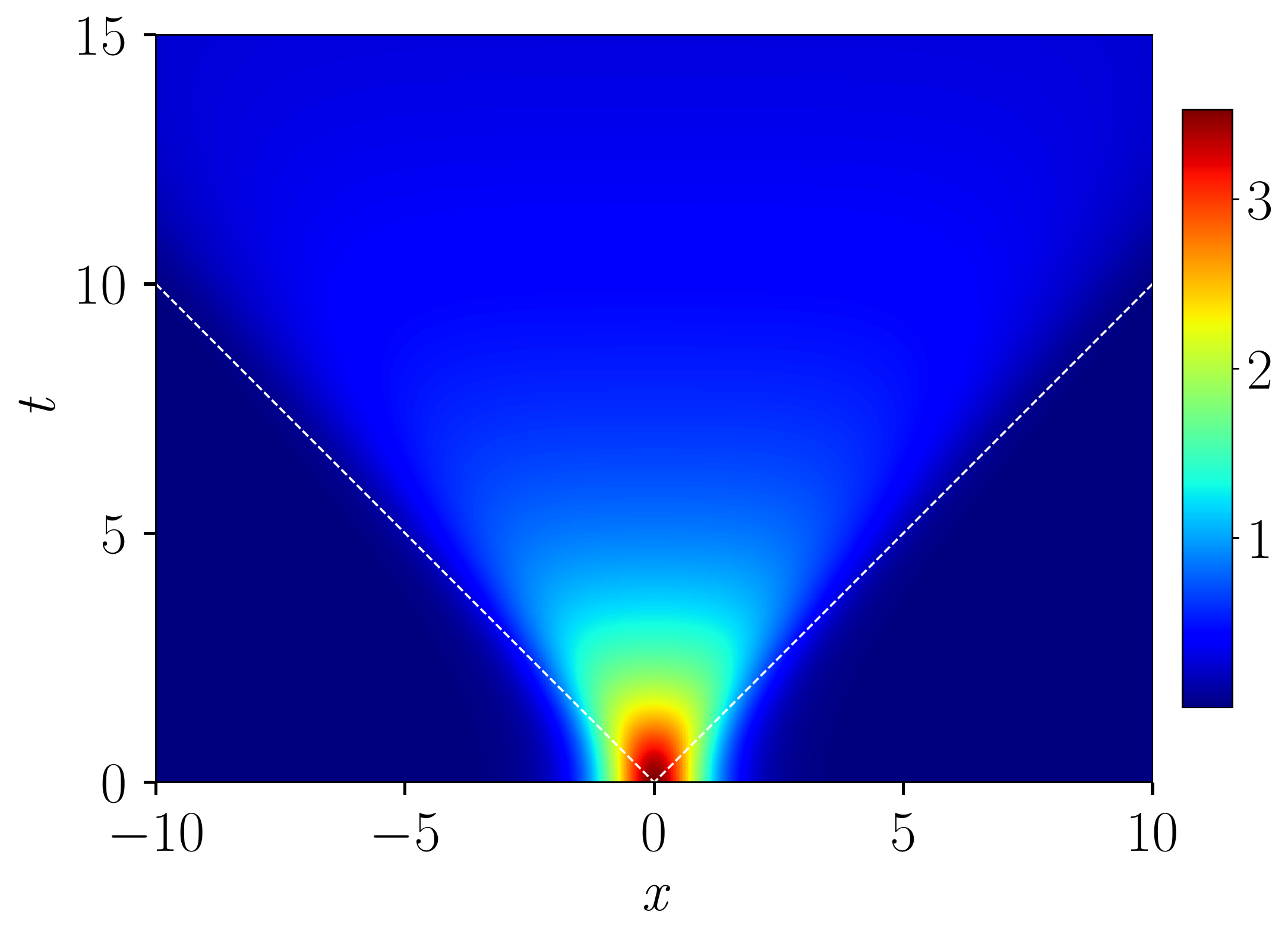}} \hfill
    \subfloat[$m = 5$]{\includegraphics[width=0.3316\textwidth]{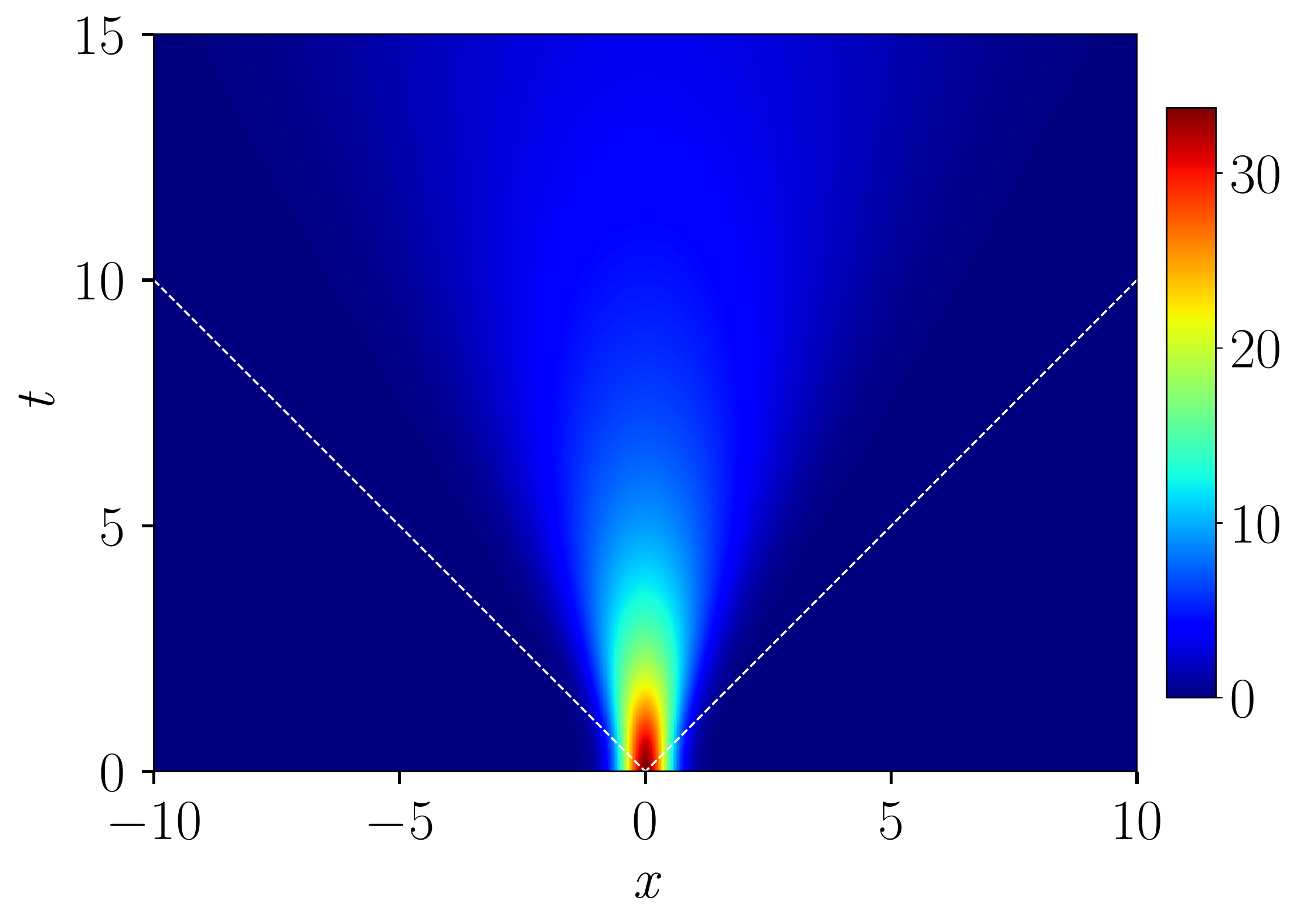}} \\
    \subfloat[$m = 0.1$]{\includegraphics[width=0.33\textwidth]{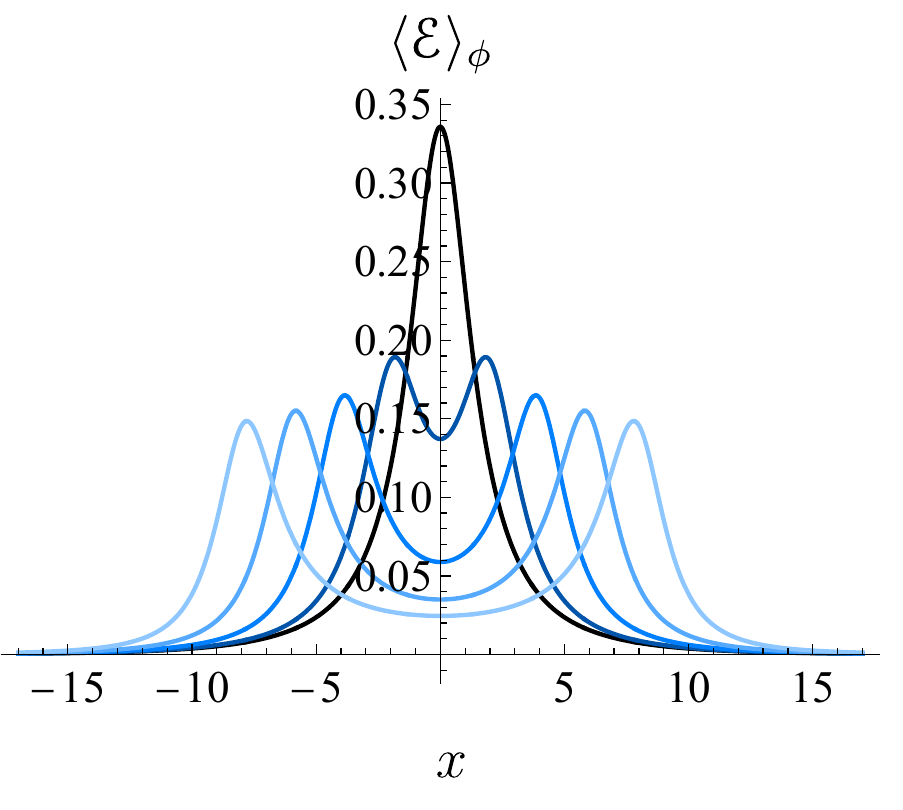}} \hfill
    \subfloat[$m = m_{\text{crit}} = 1$]{\includegraphics[width=0.33\textwidth]{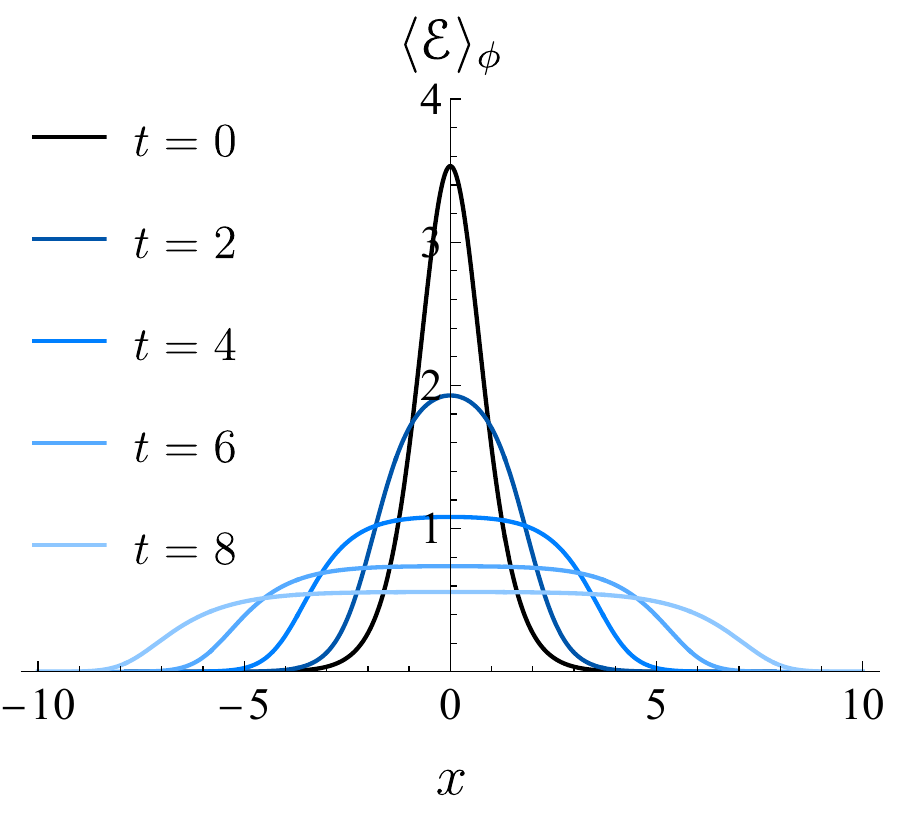}} \hfill
    \subfloat[$m = 5$]{\includegraphics[width=0.33\textwidth]{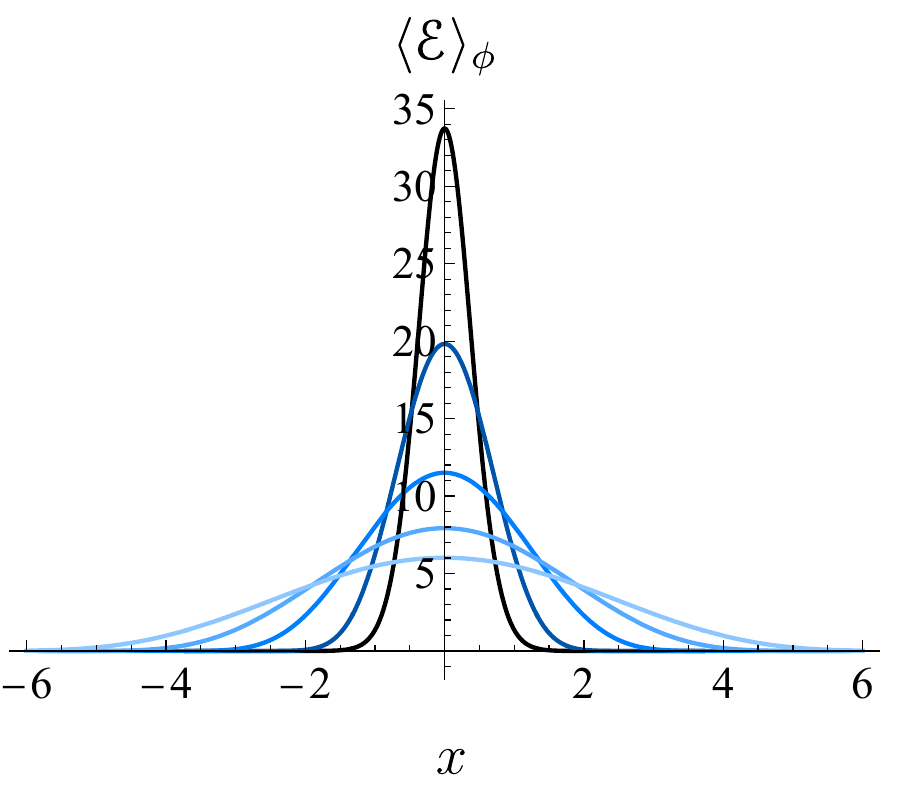}}
    \caption{\textit{Top:} Energy density evolution after the local quench by the operator $\phi$ in massive scalar field theory. The left figure corresponds to $m = 0.1$, the middle one to $m = m_{\text{crit}} = 1$ and the right one to $m = 5$; $\eps = 1.5$ is fixed for each figure. Dotted lines mark the lightcone. \textit{Bottom:} Spatial energy density distribution after the local $\phi$-quench for fixed time moments. The left figure corresponds to $m = 0.1$, the middle one to $m = m_{\text{crit}} = 1$ and the right one to $m = 5$; $\eps = 1.5$ is fixed for each figure.}
    \label{fig:PhiE-massive}
\end{figure}

In contrast to the $\partial\phi$-quench, the $\phi$-quench generates a perturbation that propagates along both sides of the lightcone (see figure~\ref{fig:PhiE-massive}). One can distinguish three different regimes of the energy density evolution following the quench. For small masses, the perturbation has a single-maximum shape, which after some time turns into a double-hill configuration with decreasing amplitude. For large masses, the configuration has a single-maximum shape for all times. In the critical regime, the perturbation propagates mostly as a flat plateau. It is possible to give a simple estimation for $m_{\text{crit}}$ taking into account that the single-maximum configuration is characterised by a negative value of the second derivative with respect to $x$ at $x = 0$, while the double-hill configuration has a positive value of the second derivative. In the critical case, the change in sign occurs at the time infinity. Hence, $m_{\text{crit}}$ is defined as the value of the mass $m$, at which the second derivative of the large-time asymptotic~\eqref{eq:phi_asymptotics} changes the sign
\be
    \frac{\partial^2}{\partial x^2}\corrfunc{\Ecal(t, x)}_{\phi}\Big|_{\begin{subarray}{l}x\,=\,0 \\ m\,=\,m_{\text{crit}}\end{subarray}} \largetime 0.
    \label{eq:crit_m}
\ee
In the leading order, this gives the following estimation 
\be 
    m_{\text{crit}} = \frac{3}{2\eps}.
    \label{eq:m_crit}
\ee 

\begin{figure}
	\centering
	\subfloat[massless]{\includegraphics[width=0.325\textwidth]{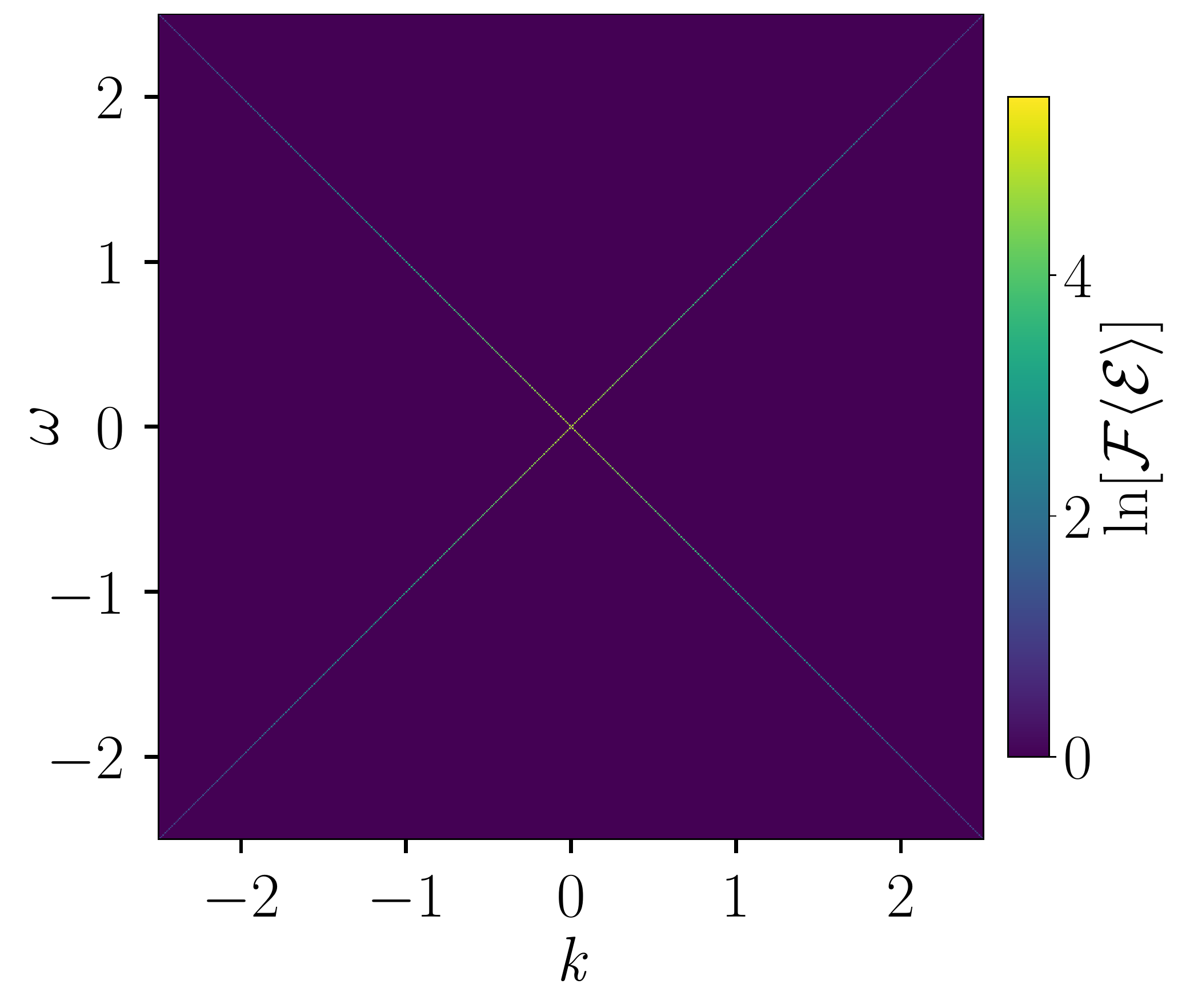}} \hfill
	\subfloat[$m = 0.1$]{\includegraphics[width=0.325\textwidth]{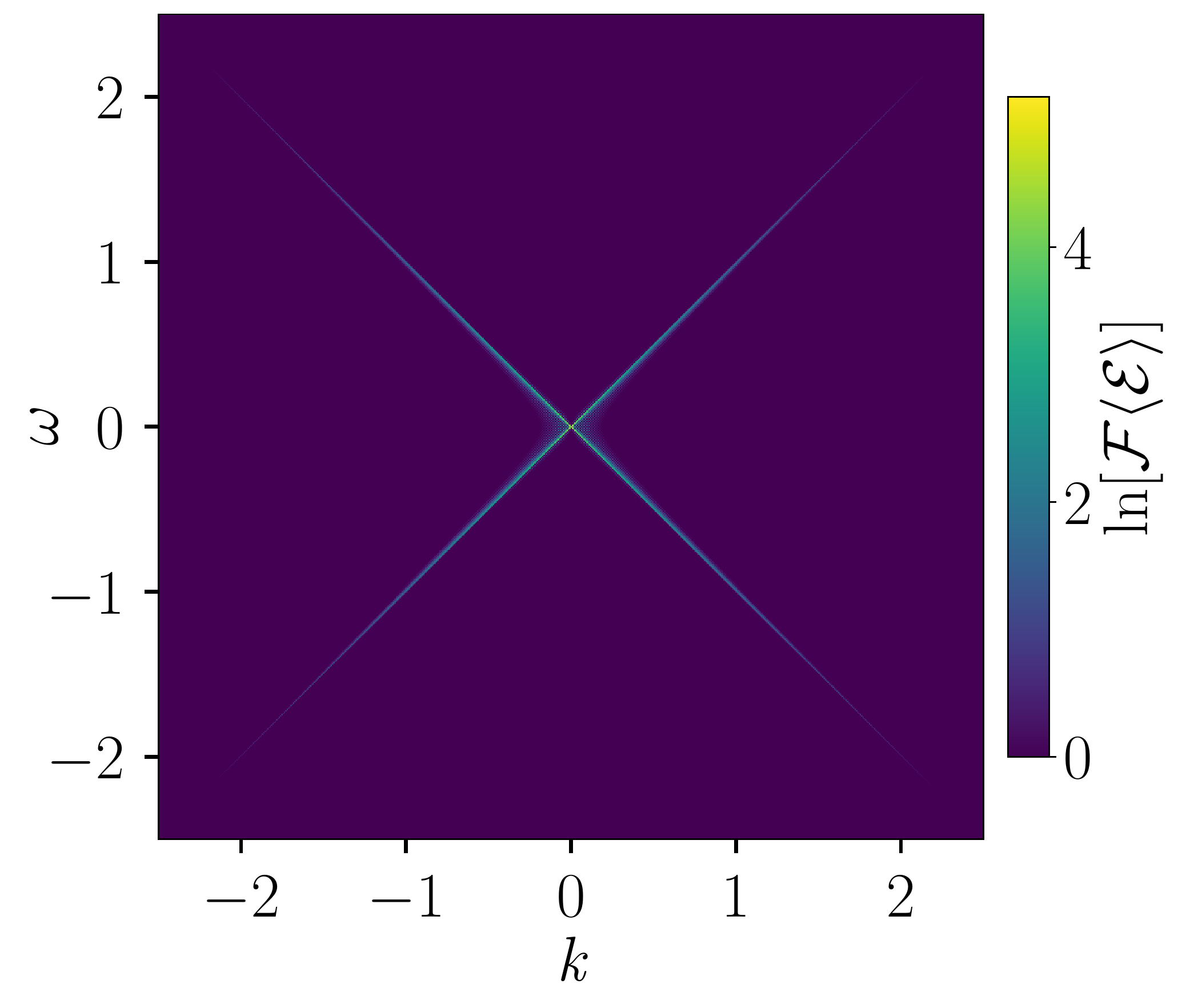}} \hfill
	\subfloat[$m = 0.5$]{\includegraphics[width=0.325\textwidth]{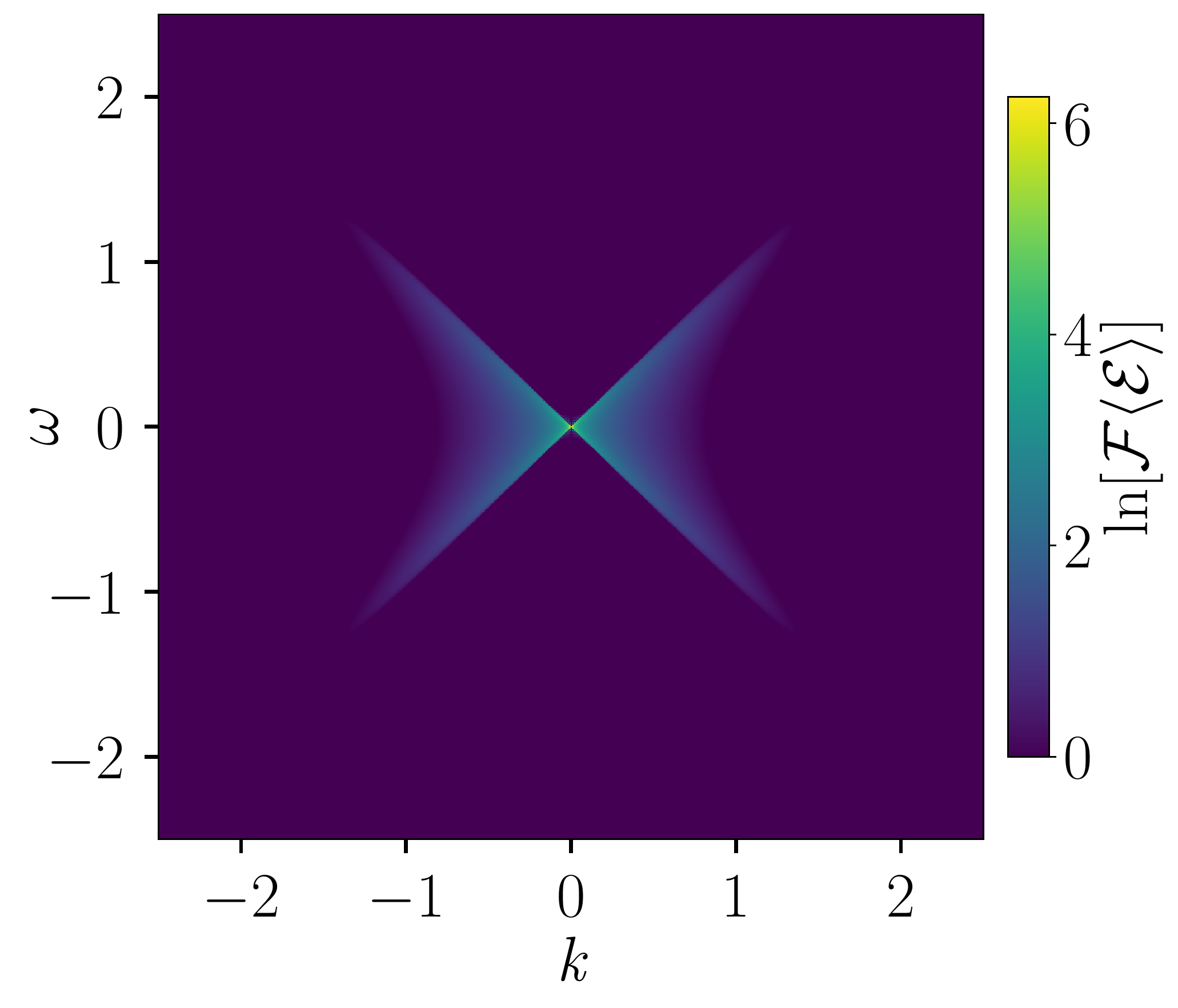}} \\
	\subfloat[$m = 1$]{\includegraphics[width=0.325\textwidth]{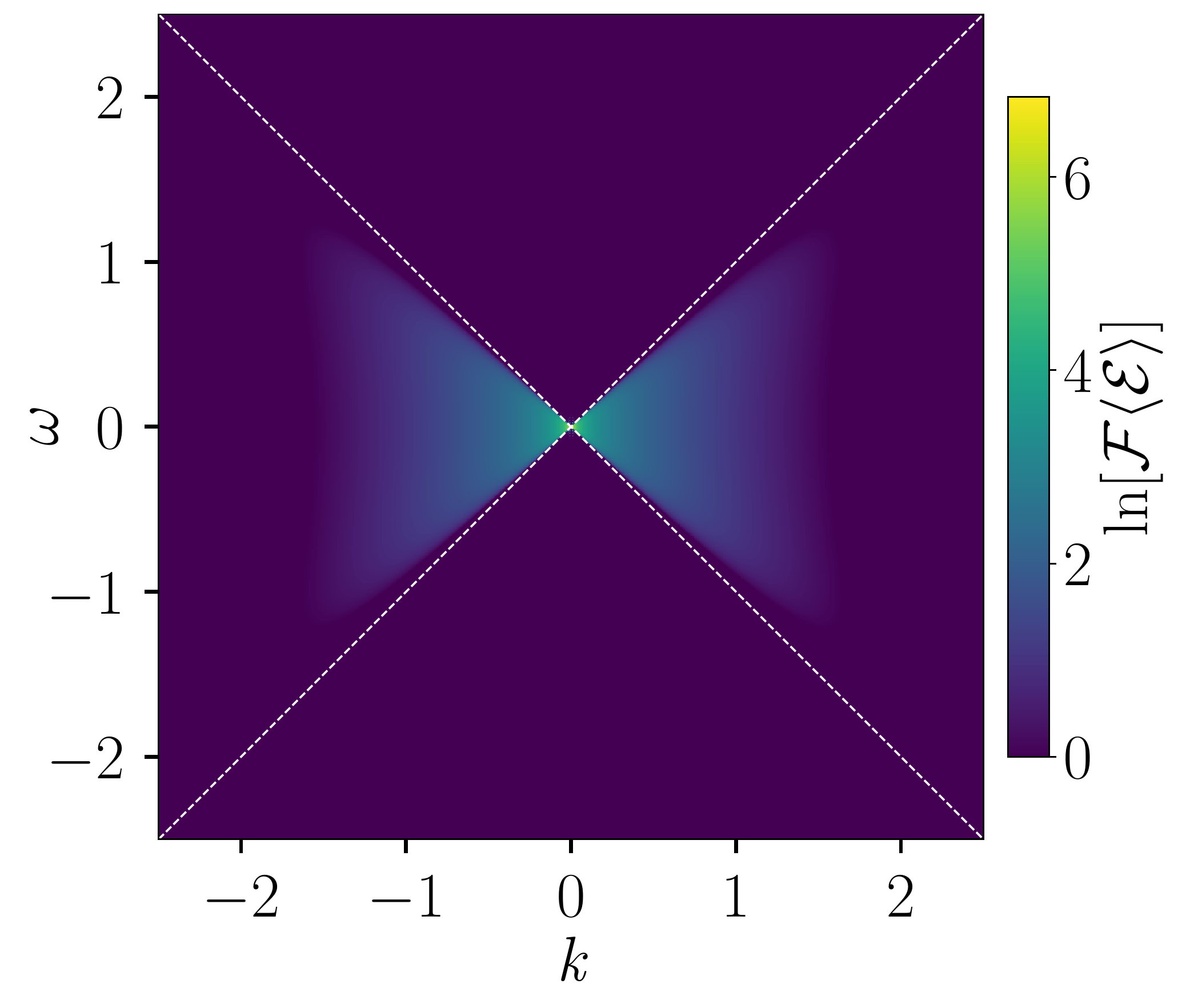}} \hfill
	\subfloat[$m = 3$]{\includegraphics[width=0.325\textwidth]{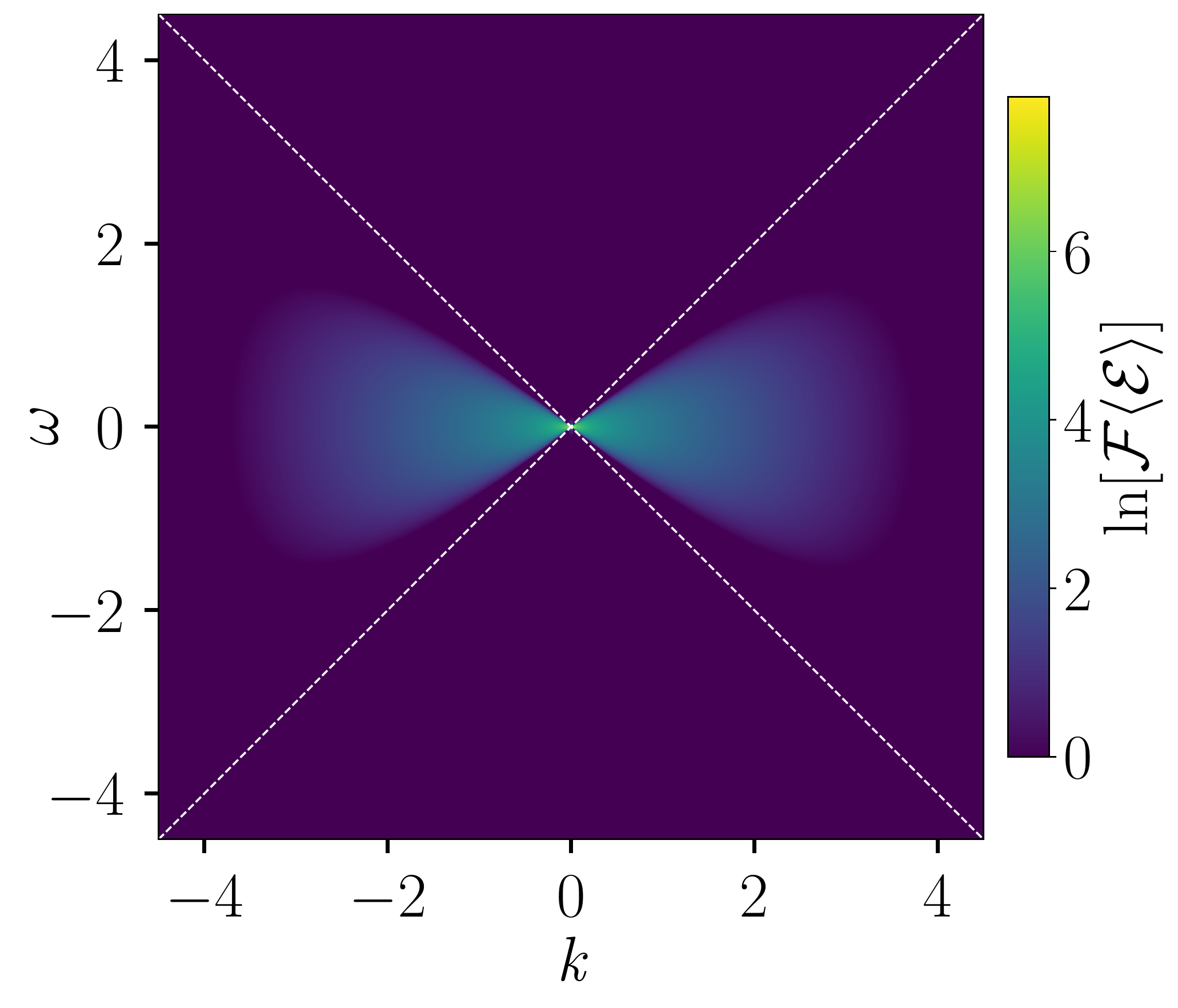}} \hfill
	\subfloat[$m = 20$]{\includegraphics[width=0.333\textwidth]{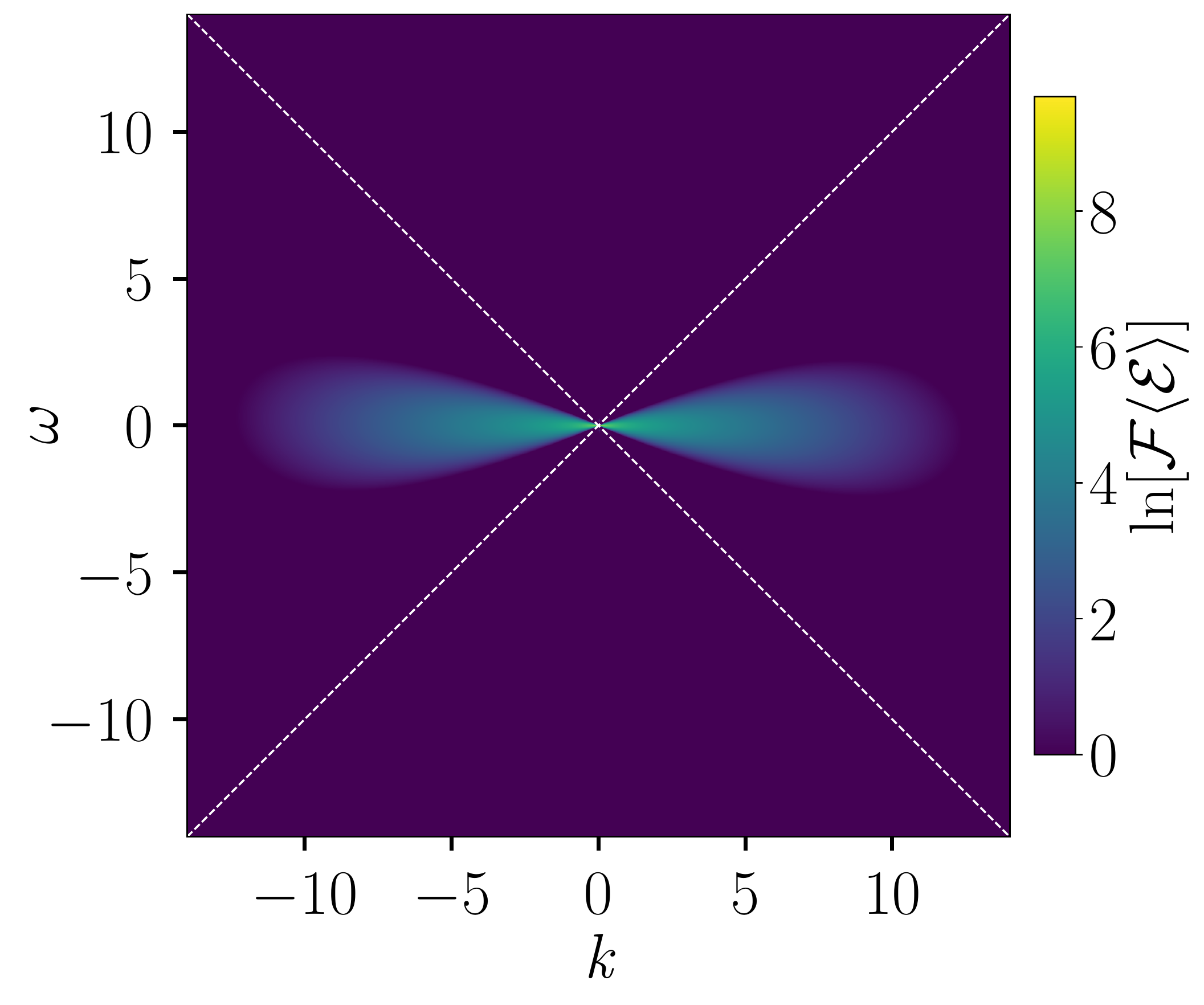}}
	\caption{Fourier images of the energy density evolution after the local quench by the operator $\phi$ in massive scalar field theory, $\mathcal{F}\corrfunc{\Ecal}_{\phi}$. There is a critical value of mass, $m_{\text{crit}} \sim 1$ for $\eps = 1.5$. Dotted lines in figures d), e) and f) mark the lines $\om = \pm k$. Note that the images represent the real part of the Fourier transformation in logarithmic scale.}
	\label{fig:FT_phiE-massive}
\end{figure}

It is interesting to observe how the evolution of the energy density is represented in momentum space. For small masses, the configuration slightly deviates from the lightcone $\om = \pm k$, forming a concave shape near the origin (see figure~\ref{fig:FT_phiE-massive}). At the critical mass $m_\text{crit}$~\eqref{eq:m_crit}, the energy density distribution $\corrfunc{\Ecal(\om, k)}_\phi$ has the form of a triangle. The amplitude rapidly decreases with increasing $k$. Further increase of mass leads to a localization around $\om = 0$ in a dumbbell-like shape.

\skipline

For completeness, let us also present the result (with the divergences subtracted) for the evolution of the $\phi^2$-operator after the quench by the operator~$\phi$
\be
    \corrfunc{\phi^2(t, x)}_{\phi} = \frac{4}{K_0(2\eps m)}\left|K_0\left(m\sqrt{(\eps - it)^2 + x^2}\right)\right|^2.
\ee

\subsection{Scalar field in \texorpdfstring{$d > 2$}{d > 2} dimensions}

Now let us turn our attention to the quantum dynamics of scalar field theories in \mbox{$d > 2$}. The $d = 2$ and $d > 2$ cases share a lot of similarities, however, higher-dimensional theories possess well-defined two-point functions of the operator $\phi$ in contrast to the $d = 2$ case. Higher-dimensional excitations of the vacuum by fermionic local operators as well as quantum entanglement of local operators in different types of CFT were also studied in~\cite{Nozaki:2014hna, Caputa:2014vaa, Caputa:2015qbk}. In terms of AdS/CFT correspondence, the local quench by a primary operator in $d = 2$ dimensions has a natural interpretation in terms of an infalling particle perturbing the AdS$_3$ bulk. The exact dual of $d > 2$ local quenches is not known by now. In~\cite{Nozaki:2013wia}, it was conjectured that the construction of a ``falling black hole'' of~\cite{Horowitz:1999gf, Nozaki:2013wia} can be considered as a higher-dimensional analogue of the local quench. In this construction, the source deforming the AdS$_d$ bulk is a heavy localized object (``the centre of a falling black hole'', see~\cite{Nozaki:2013wia}). The details of this construction as well as the exact dual seem to be unclear. It is interesting to compare these results at least with observables in simplest \mbox{$d$-dimensional} CFT (an example of which is a free massless scalar field theory). At the same time, however, studying only one type of observables (stress-energy tensor dynamics in our case) is not enough to establish exact holographic dual in its full generality.

\subsubsection*{Massless theory}

The Euclidean action of a free massless scalar field for the higher-dimensional case has the form
\be
    S = \frac{1}{8\pi}\int d^{\,d}x\left((\partial_\tau\phi)^2 + \partial\phi^i\partial\phi_i\right),
\ee
where the summation goes over the spatial indices, $i = 1, \ldots, d$, $d > 2$. The two-point function derived from the corresponding $d$-dimensional Klein-Gordon equation reads
\be
    \corrfunc{\phi(\tau_1, x_1^i)\phi(\tau_0, x_0^i)}_d = \frac{\Gamma\left(\frac{d}{2} - 1\right)}{\pi^{\frac{d}{2} - 1}} \cdot \frac{1}{R^{d - 2}} ,
\ee
where $R$ is the spacetime distance between points, $R = \sqrt{(\tau_1 - \tau_0)^2 + \Sigma_i (x_1^i - x_0^i)^2}$, and the energy density for this theory is given by
\be
    \Ecal(t, x^i) = \frac{1}{4}\left(-(\partial_{\tau}\phi)^2 + \partial^i\phi\partial_i\phi\right).
\ee

Generalization of lightcone coordinates in $d$ dimensions is given by
\be
    u = \rho + i\tau, \quad v = \rho - i\tau,
\ee
where $\rho = \sqrt{x^ix_i}$ is the spatial distance to the quenching point. Then, the two-point function of the field $\partial\phi$ takes the form
\be
    \corrfunc{\partial\phi(u_1, v_1)\partial\phi(u_0, v_0)}_d = \frac{\Gamma\left(\frac{d}{2} + 1\right)}{\pi^{\frac{d}{2} - 1}} \cdot \frac{(v_1 - v_0)^2}{\left[(u_1 - u_0)(v_1 - v_0)\right]^{\frac{d}{2} + 1}}.
\ee

Let us consider the operator local quench protocol~\eqref{eq:operator_insertion} with the operator $\partial\phi$. Applying Wick's theorem, we obtain that the expectation value of the $uu$-component of the stress-energy tensor in the quenched state is described by
\be
    \corrfunc{T_{uu}}_{\partial\phi,\,d}\Big|_{m\,=\,0} = \frac{\Gamma\left(\frac{d}{2} + 1\right)}{\pi^{\frac{d}{2} - 1}} \cdot \frac{\eps^2 + v^2}{\eps^2 + u^2} \cdot \frac{(2\eps)^d}{\left[(\eps^2 + u^2)(\eps^2 + v^2)\right]^{\frac{d}{2}}}.
    \label{eq:d_dim_dphi_quench_m_0}
\ee
For $d = 3$ and $d = 4$, this gives
\be
    \begin{aligned}
        & \corrfunc{T_{uu}}_{\partial\phi,\,3d}\Big|_{m\,=\,0} = \frac{6\eps^3}{\left(\eps^2 + u^2\right)^\frac{5}{2}\sqrt{\eps^2 + v^2}}, \\
        & \corrfunc{T_{uu}}_{\partial\phi,\,4d}\Big|_{m\,=\,0} = \frac{32\eps^4}{\pi\left(\eps^2 + u^2\right)^3\left(\eps^2 + v^2\right)}.
    \end{aligned}
\ee
As in $d = 2$, these formulae coincide up to a constant with the corresponding stress-energy tensor components of the holographic $d$-dimensional CFT, dual to a falling massive particle in AdS$_{d + 1}$ (see equations~(2.20), (2.21) in~\cite{Nozaki:2013wia}). It would be interesting to understand the relation between the constant multipliers as well as to add other insights on holographic correspondence, and we leave this for future investigation.

\skipline

For $\phi$-quench, we obtain that
\be
   \corrfunc{\Ecal(t, x^i)}_{\phi,\,d}\Big|_{m\,=\,0} = \frac{(d - 2)\Gamma\left(\frac{d}{2}\right)}{\pi^{\frac{d}{2} - 1}} \cdot \frac{(2\eps)^{d - 2}\left(\eps^2 + t^2 + \rho^2\right)}{\left[\left(\rho^2 - t^2\right)^2 + 2\eps^2\left(\rho^2 + t^2\right) + \eps^4\right]^\frac{d}{2}}.
    \label{eq:d_dim_phi_quench_m_0}
\ee
Note that in contrast to the massless case in two dimensions~\eqref{eq:CFT2}, in which the perturbation does not dissipate over time, in the case of $d > 2$, the amplitude of the perturbation decays.

The total energy calculated from~\eqref{eq:d_dim_phi_quench_m_0} is
\be
    E = \int d^{\,d - 1}x \, \corrfunc{\Ecal(t, x^i)}_{\phi,\,d}\Big|_{m\,=\,0} = \frac{(d - 2)\pi}{\eps}.
\ee

\subsubsection*{Massive theory}

Now let us consider massive $d$-dimensional scalar field theory
\be
    S = \frac{1}{8\pi}\int d^{\,d}x\left((\partial_\tau\phi)^2 + \partial\phi^i\partial\phi_i + m^2\phi^2\right),
\ee
with the energy density corresponding to this action
\be
    \Ecal(\tau, x^i) = \frac{1}{4}\left(-(\partial_{\tau}\phi)^2 + \partial^i\phi\partial_i\phi + m^2\phi^2\right).
\ee
The $d$-dimensional two-point function of the massive scalar field is given by
\be
    \corrfunc{\phi(\tau_1, x_1^i)\phi(\tau_0, x_0^i)}_d = 2\left(\frac{m}{2\pi}\right)^{\frac{d}{2} - 1} \cdot \frac{K_{\frac{d}{2} - 1}(mR)}{R^{\frac{d}{2} - 1}}.
\ee
It simplifies in the case of a half-integer order, explicitly revealing how the mass $m$ controls the exponential suppression. For example, the $d = 3$ and $d = 5$ cases read
\be
    \begin{aligned}
        & \corrfunc{\phi(\tau_1, x_1^i)\phi(\tau_0, x_0^i)}_{3d} = \frac{e^{-mR}}{R}, \\
        & \corrfunc{\phi(\tau_1, x_1^i)\phi(\tau_0, x_0^i)}_{5d} = \frac{m}{2\pi}\left(1 + \frac{1}{mR}\right)\frac{e^{-mR}}{R^2}.
    \end{aligned}
\ee
The energy density evolution following the $\phi$-quench is given by the expression
\be
    \begin{aligned}
        \corrfunc{\Ecal(t, x^i)}_{\phi,\,d} & = \frac{m^{\frac{d}{2} + 1}\eps^{\frac{d}{2} - 1}}{\pi^{\frac{d}{2} - 1}K_{\frac{d}{2} - 1}(2\eps m)}\left|\left((\eps - it)^2 + \rho^2\right)\right|^{\frac{d}{2}} \times \\
        & \times \left(\left(\eps^2 + t^2 + \rho^2\right)\left|K_{\frac{d}{2}}\left(m\sqrt{(\eps - it)^2 + \rho^2}\right)\right|^2\right. + \\
        & + \left.\left|\sqrt{(\eps - it)^2 + \rho^2}\,K_{\frac{d}{2} - 1}\left(m\sqrt{(\eps - it)^2 + \rho^2}\right)\right|^2\right).
    \end{aligned}
    \label{eq:massive_quench_ddim}
\ee

Let us briefly describe the structure of divergences of the composite operator $\Ecal$. The constant and the divergent terms now depend on the number of spacetime dimensions. The explicit series expansion in the regularization parameter $\delta \to 0$ is
\be
    \begin{aligned}
        & \mathcal{C}_{\phi,\,d} + \mathcal{D}_{\phi,\,d} = -\frac{m^d\Gamma\left(-\frac{d}{2}\right)}{2^d\pi^{\frac{d}{2} - 1}} + \\
        & + \lim_{\delta\,\to\,0}\left[\frac{m^2\Gamma\left(\frac{d}{2} - 1\right)}{2\pi^{\frac{d}{2} - 1}d^{\frac{d}{2}}} \cdot \frac{1}{\delta^{d - 2}} - \frac{m^4\Gamma\left(\frac{d}{2} - 2\right)}{8\pi^{\frac{d}{2} - 1}d^{\frac{d}{2} - 1}} \cdot \frac{1}{\delta^{d - 4}} + \ldots + O\left(\frac{1}{\delta^{d - 2n}}\right)\right], \quad n \in \mathbb{Z}.
    \end{aligned}
    \label{eq:d_dim_div}
\ee
For an odd number of dimensions, $d \equiv 2l + 1$, the $0^{\text{th}}$-order term is a finite constant with respect to $\delta$, while higher-order terms are divergent starting from $1/\delta^{d-2}$ and involving all the terms up to  $\sim 1/\delta^{d - 2n}$. For even $d \equiv 2l$, the $0^{\text{th}}$-order term is an infinite constant, since it has poles independent of $\delta$. Higher orders contribute as divergences: $\sim 1/\delta^{2(l - n)}$ with $l < n$. The last non-zero term $l = n$ (in the limit $\delta \to 0$) in the series~\eqref{eq:d_dim_div} contains a pole of the gamma-function $\Gamma\left(\frac{d}{2} - l\right)$ and a constant part for any $l > 2$.

Note that these divergences, however, do not influence the observables, since they do not contribute to the final answer after the subtraction procedure.

The large-time (at fixed spatial distance) and large-distance (at fixed time) asymptotics of the energy density dynamics given by~\eqref{eq:massive_quench_ddim} up to $O\left(t^{-d - 3}\right)$ and $O\left(\rho^{-d}e^{-2m\rho} \, e^{\frac{m}{\rho}(t^2 - \eps^2)}\right)$, respectively, read
\be
    \begin{aligned}
        & \corrfunc{\Ecal(t, x^i)}_{\phi,\,d} \largetime \frac{\gamma}{t^{d - 1}} + \\
        & + \Big[(2m\rho)^2(d + 1 - 2\eps m) + (d - 1)(d - 1 + (d - 1)2\eps m - (2\eps m)^2)\Big] \cdot \frac{\gamma}{8m^{2} t^{d + 1}},
   \end{aligned}
   \label{eq:phiDdim_large-t}
\ee
and
\be
    \corrfunc{\Ecal(t, x^i)}_{\phi,\,d}  \underset{\rho\,\to +\infty}{\approx} \frac{\gamma e^{2\eps m}}{\rho^{d - 1}} \, e^{-2m\rho} \, e^{\frac{m}{\rho}(t^2 - \eps^2)},
\ee
where
\be
    \gamma = \frac{\pi^{2 - \frac{d}{2}}\eps^{\frac{d}{2} - 1}m^{\frac{d}{2}}}{e^{2\eps m}K_{\frac{d}{2} - 1}(2\eps m)}, \quad d > 2.
    \label{eq:gamma_def}
\ee
Note that both series start not directly from the first order, but have the lowest non-vanishing order depending on $d$ (compare this with the two-dimensional result~\eqref{eq:phi_asymptotics}).

\begin{figure}
    \centering
    \subfloat[$t = 0$]{\includegraphics[width=0.307\textwidth]{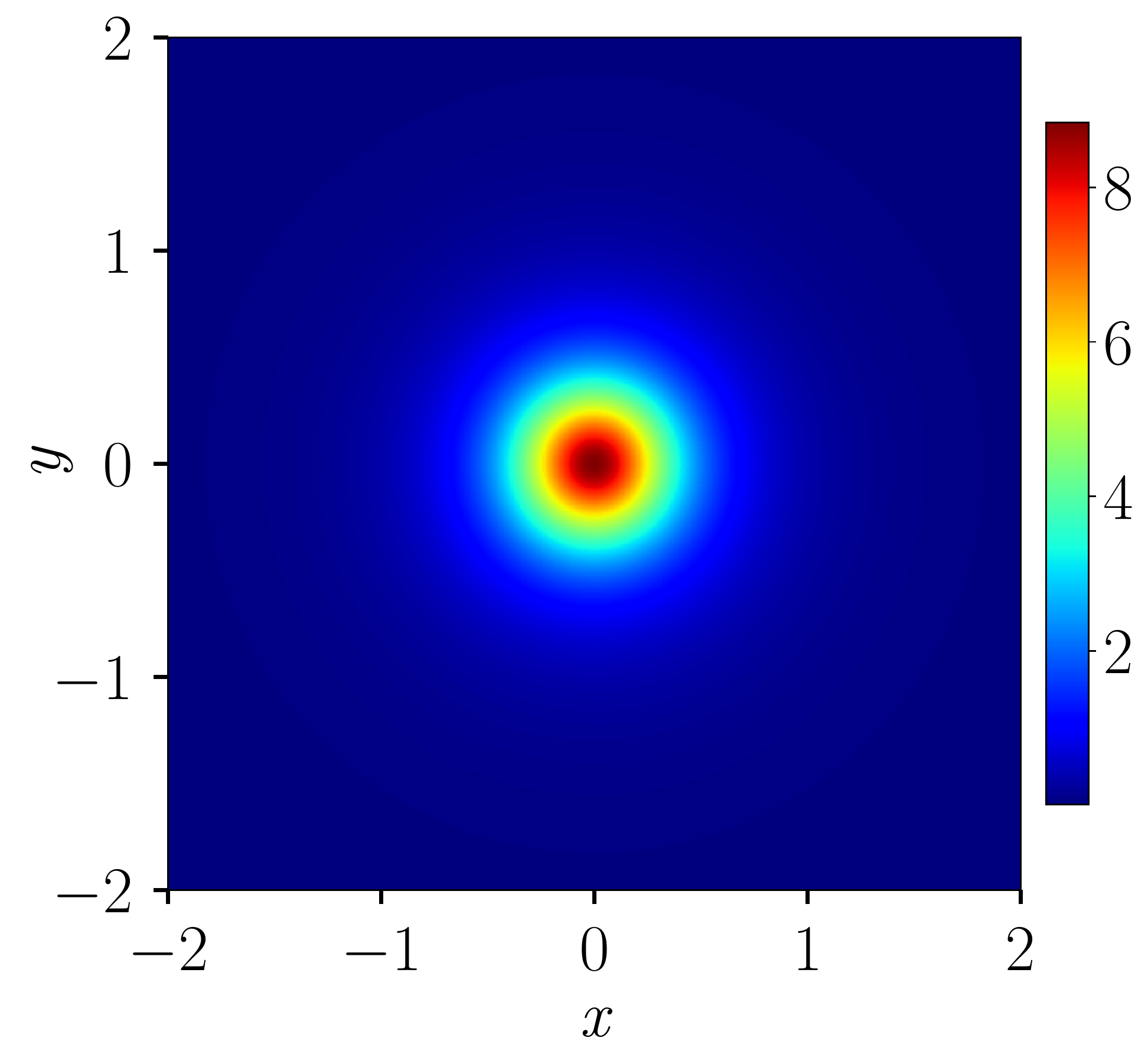}} \hfill
    \subfloat[$t = 0.5$]{\includegraphics[width=0.32\textwidth]{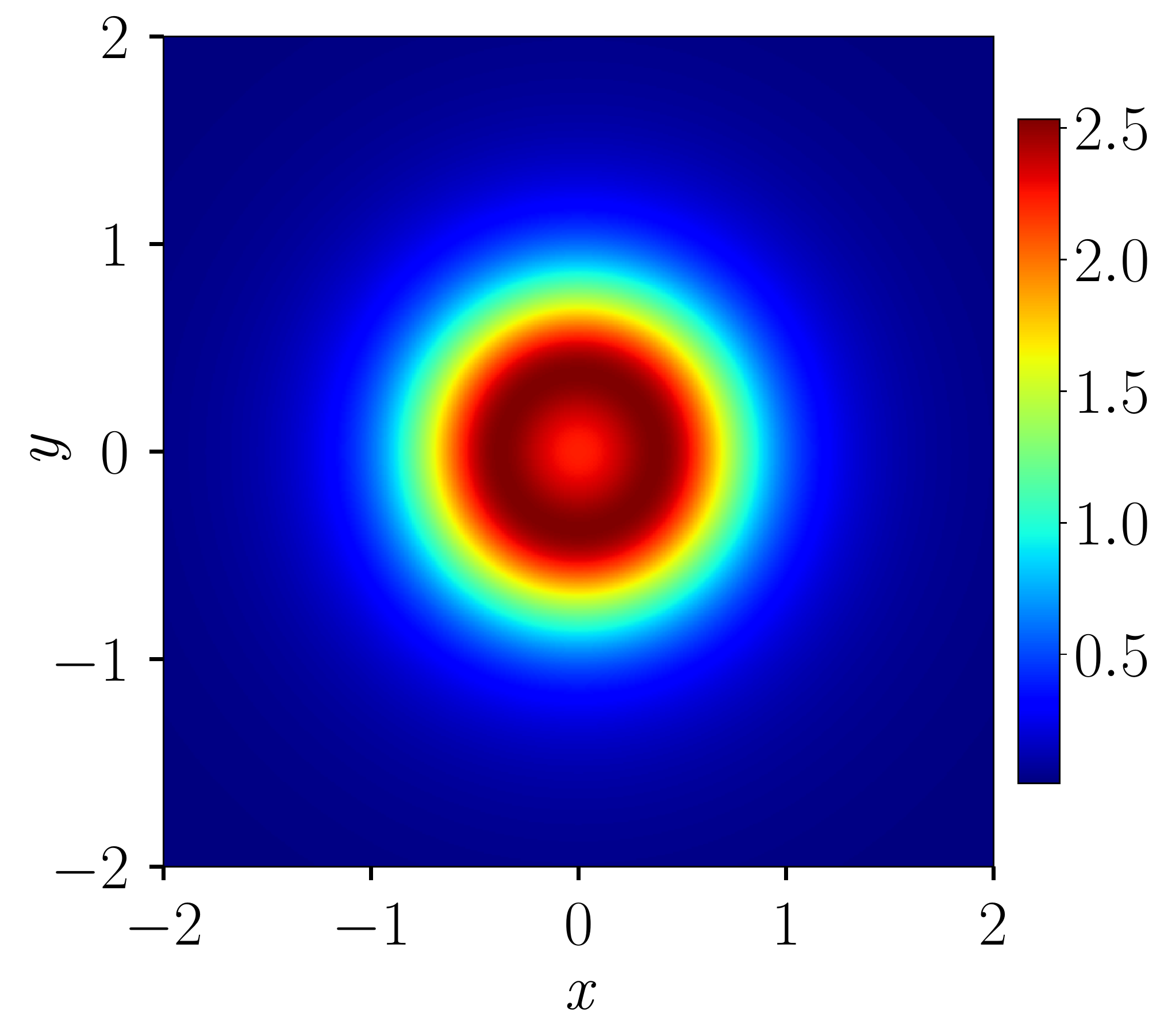}} \hfill
    \subfloat[$t = 1$]{\includegraphics[width=0.32\textwidth]{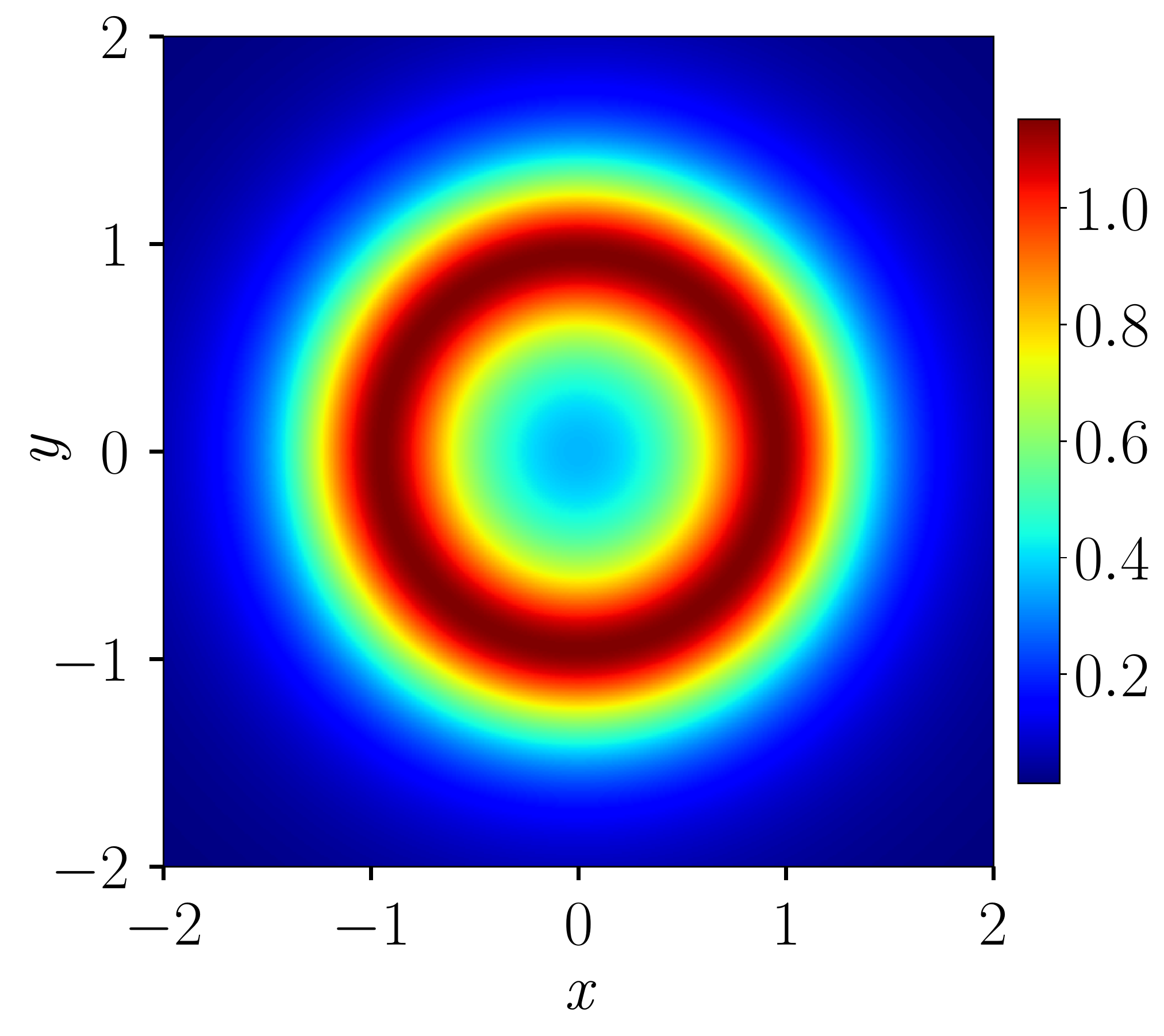}} \\
    \subfloat[$d = 3$]{\includegraphics[width=0.45\textwidth]{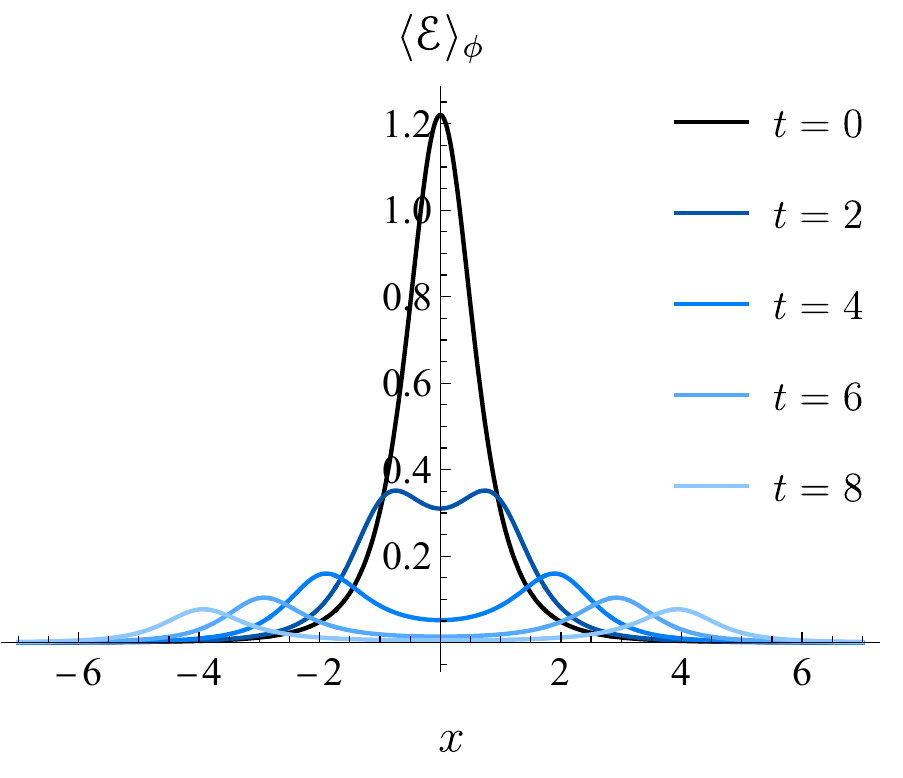}}
    \hspace{0.05\textwidth}
    \subfloat[$d = 4$]{\includegraphics[width=0.45\textwidth]{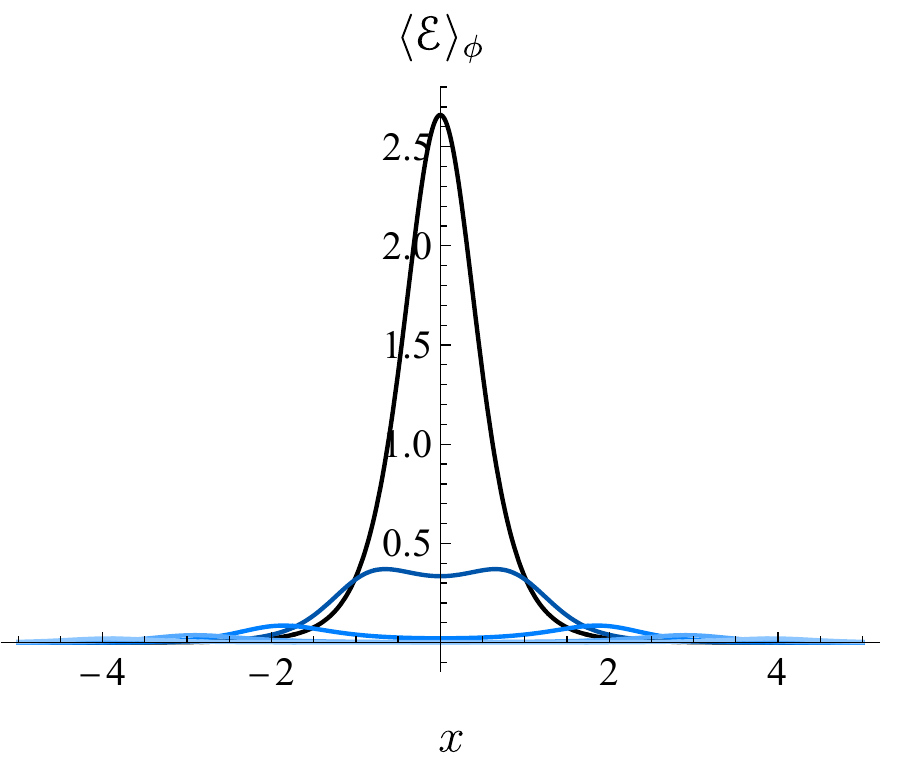}}
    \caption{\textit{Top:} Spatial dependence of the energy density for $t = 0$ (left), $t = 0.5$ (middle) and $t = 1$ (right), following the local quench by the operator $\phi$ in $d = 3$ theory. The parameters are $\eps = 1$ and $m = 0.1$. \textit{Bottom:} Spatial energy density distribution along the $x$-axis after the local quench by the operator $\phi$ in $d = 3$ (left) and $d = 4$ (right) theories correspondingly for fixed time moments; $\eps = 1$, $m = 0.1$.}
    \label{fig:massive_quench_3dim}
\end{figure}

Qualitatively, the perturbation (see figure~\ref{fig:massive_quench_3dim} for the $d = 3$ case) propagates as radially symmetric waves.

\skipline

The evolution of the $\phi^2$-operator after the quench by the operator~$\phi$ is given by
\be
    \corrfunc{\phi^2(t, x^i)}_{\phi,\,d}\Big|_{m\,=\,0} = \frac{4\Gamma\left(\frac{d}{2}\right)}{(d - 2)\pi^{\frac{d}{2} - 1}} \cdot \frac{(2\eps)^{d - 2}}{\left[\left(\rho^2 - t^2\right)^2 + 2\eps^2\left(\rho^2 + t^2\right) + \eps^4\right]^{\frac{d}{2} - 1}},
\ee
in the massless case, and by
\be
    \corrfunc{\phi^2(t, x^i)}_{\phi,\,d} = \frac{4(m\eps)^{\frac{d}{2} - 1}}{\pi^{\frac{d}{2} - 1} K_{\frac{d}{2} - 1}(2m\eps)} \cdot \frac{\left|K_{\frac{d}{2} - 1}(m\sqrt{(\eps - it)^2 + \rho^2}\right|^2}{\left|\sqrt{(\eps - it)^2 + \rho^2}\right|^{d - 2}},
\ee
in the massive.

\section{Local quenches in complex scalar field theory}

The developed technique allows to explore the influence of the local quench not only on the energy density, but on any other observable. In this section, we extend the operator quench protocol~\eqref{eq:operator_insertion} to the case of a complex scalar field and calculate the dynamics of the charge density operator $\Qcal(t, x)$ after the local quench. 

\skipline

The Euclidean action of a free massive complex scalar field in two-dimensions has the form
\be
    S = \frac{1}{8\pi}\int d\tau\,dx\left(\partial_\tau\phi\partial_\tau\phi^* + \partial_x\phi\partial_x\phi^* + m^2\phi\phi^*\right).
\ee
The theory possesses the conserved charge, which we define as
\be
    Q = \int dx\,\left(\phi^*{\partial_\tau\phi} - {\partial_\tau\phi}^*\phi\right).
\ee
We consider the excitation created by the field $\phi$ and its conjugate $\phi^*$ inserted according to the operator local quench protocol~\eqref{eq:operator_insertion} as in the non-charged case. The charge dynamics after the quench is described by the following correlation function
\be
    \corrfunc{\Qcal(t, x)}_{\phi} = \frac{\left\langle 0\left|\phi^{*}(i\eps, 0) \mathcal{Q}(t, x) \phi(-i\eps, 0)\right| 0\right\rangle}{\left\langle 0\left|\phi^{*}(i\eps, 0) \phi(-i\eps, 0)\right|0\right\rangle},
\ee
where the charge density operator is
\be
    \Qcal(\tau, x) = \phi^*\partial_\tau\phi - \partial_\tau\phi^*\phi.
    \label{eq:charge_density}
\ee
The regularization and subtraction procedures for this composite operator hold the same as in the previous sections.

The only non-vanishing two-point function of the theory includes both the field and its conjugate counterpart
\be
    \corrfunc{\phi(\tau_1, x_1)\phi^*(\tau_0, x_0)} = 2 K_0\left(m\sqrt{(\tau_1 - \tau_0)^2 + (x_1 - x_0)^2}\right).
\ee
Starting with this two-point function and following the usual steps of the calculation we find that the charge density $\Qcal$ after the local $\phi$-quench is given by
\be
    \begin{aligned}
        & \corrfunc{\Qcal(t, x)}_{\phi} = \frac{2m}{K_0(2\eps m)}\left[\frac{(\eps - it)K_0\left(m \sqrt{(\eps + it)^2 + x^2}\right)K_1\left(m\sqrt{(\eps - it)^2 + x^2}\right)}{\sqrt{(\eps - it)^2 + x^2}} + \text{c.c.}\right].
        \label{eq:charge_quench}
    \end{aligned}
\ee 

The point-splitting in the composite operator results in a divergent term of the form
\be
    \mathcal{D}^{\Qcal}_{\phi} = \lim_{\delta\,\to\,0}\left[\frac{2}{\delta} + O(\delta)\right].
\ee

The large-time (at fixed spatial distance) and large-distance (at fixed time) dynamics of the charge density is
\be
    \begin{aligned}
        \corrfunc{\Qcal(t, x)}_{\phi} & \largetime \frac{2}{m}\cdot\frac{\beta}{t} - \frac{(2\eps^2 m + 4\eps m^2 x^2 - \eps - 4mx^2)}{2m^2} \cdot \frac{\beta}{t^3} + O\left(t^{-5}\right), \\
        \corrfunc{\Qcal(t, x)}_{\phi} & \largex \frac{2\beta e^{2\eps m}\eps}{mx^2}\,\,e^{-2mx} e^{\frac{m}{x}\left(t^2 - \eps^2\right)} + O\left(x^{-3}e^{-2mx}\, e^{\frac{m}{x}\left(t^2 - \eps^2\right)}\right),
    \end{aligned}
    \label{eq:charge_asymptotics}
\ee
where $\beta$ is defined by the expression~\eqref{eq:beta_def}.

\begin{figure}[ht]
	\centering
	\subfloat[$m = 0.1$]{\includegraphics[width=0.455\textwidth]{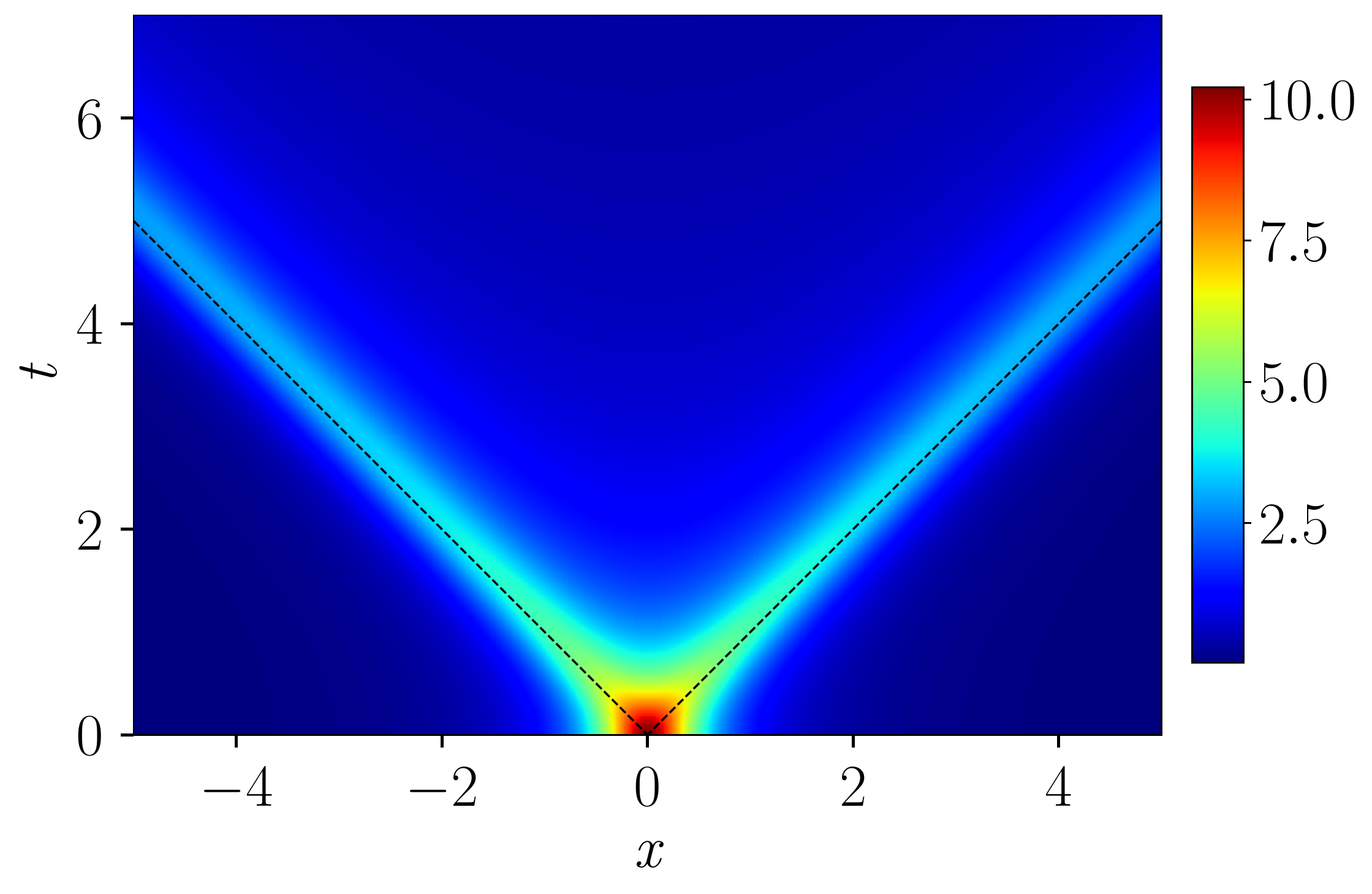}}
	\hspace{0.05\textwidth}
	\subfloat[$m = 5$]{\includegraphics[width=0.44\textwidth]{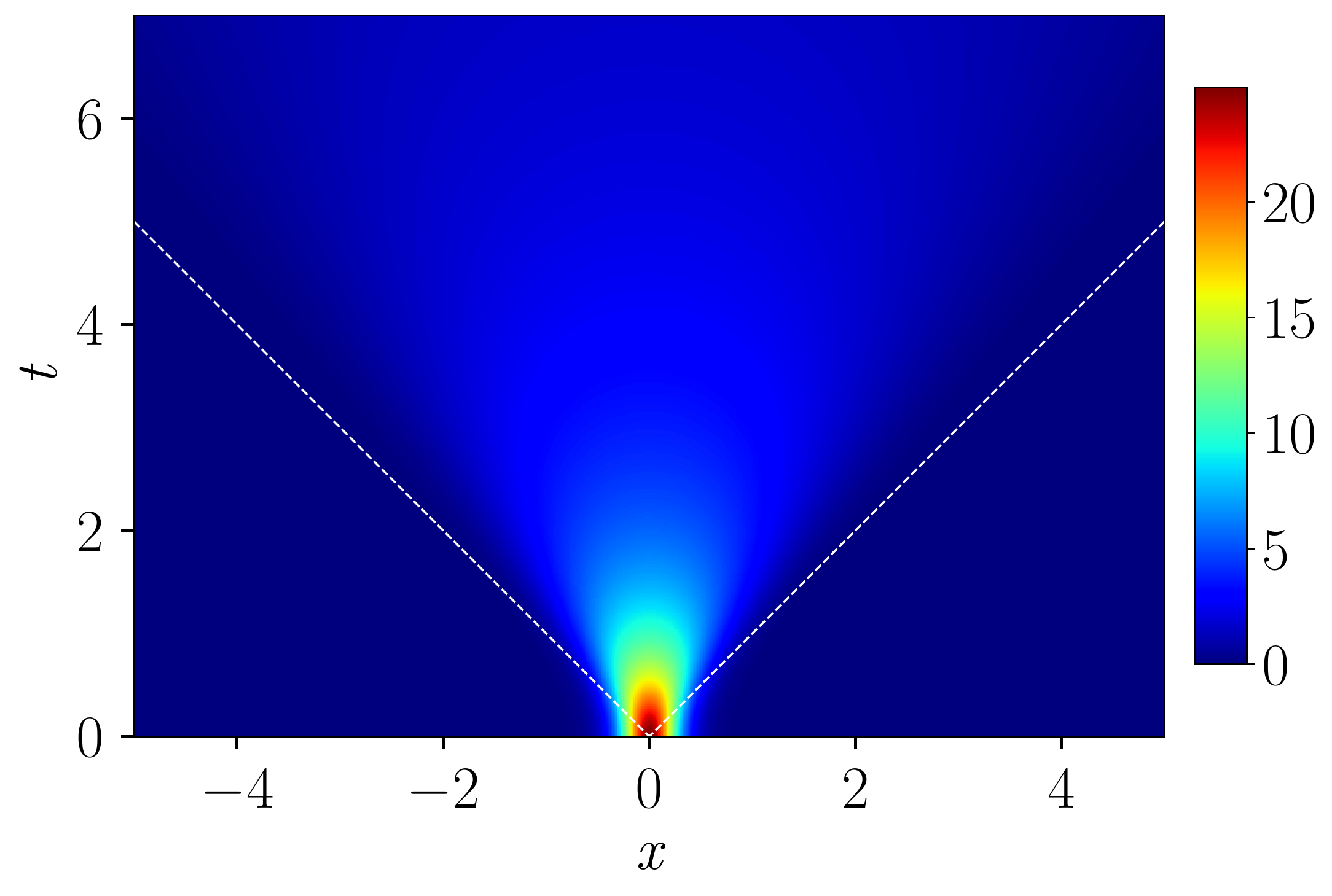}}
	\caption{Charge density evolution following the local quench by the operator $\phi$ in $d = 2$ massive complex scalar field theory~\eqref{eq:charge_quench}. The left figure corresponds to $m = 0.1$ and the right one to $m = 5$, and $\eps = 0.5$ is fixed for both figures. Dotted lines mark the lightcone.}
	\label{fig:charge_quench}
\end{figure}

\begin{figure}[ht]
	\centering
	\subfloat[$d = 2$]{\includegraphics[width=0.32\textwidth]{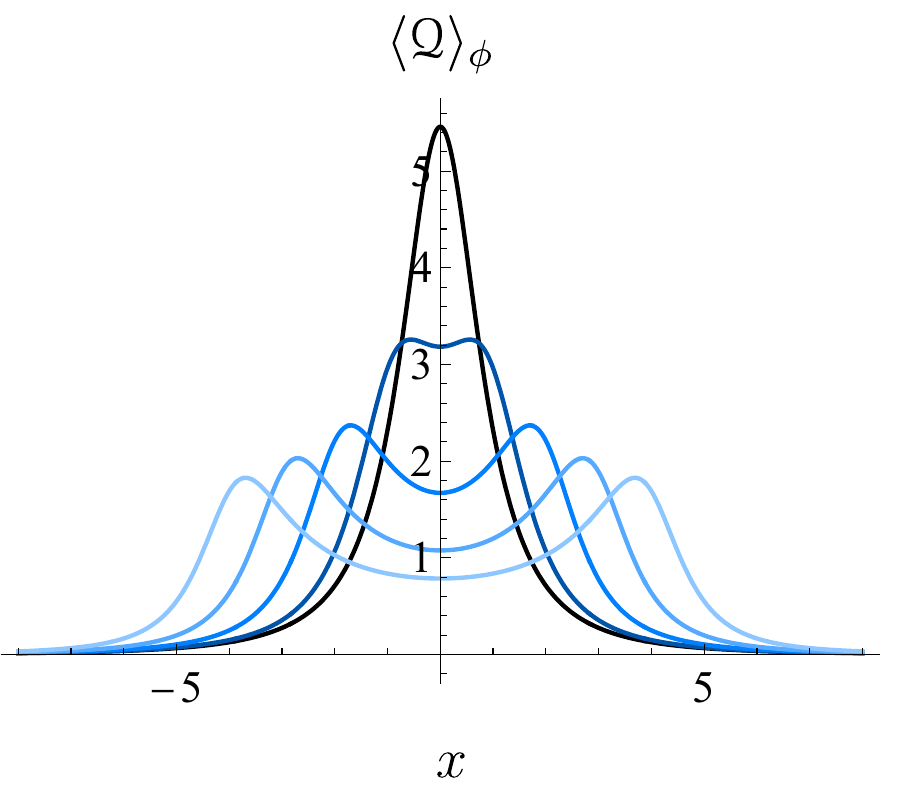}} \hfill
	\subfloat[$d = 3$]{\includegraphics[width=0.32\textwidth]{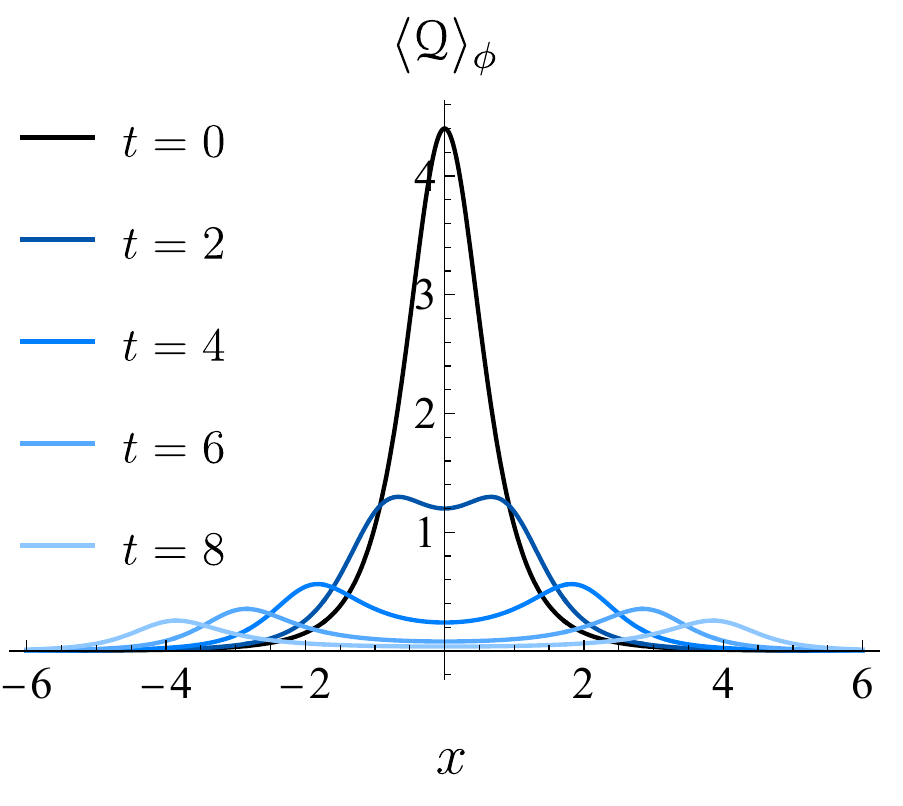}} \hfill
	\subfloat[$d = 4$]{\includegraphics[width=0.32\textwidth]{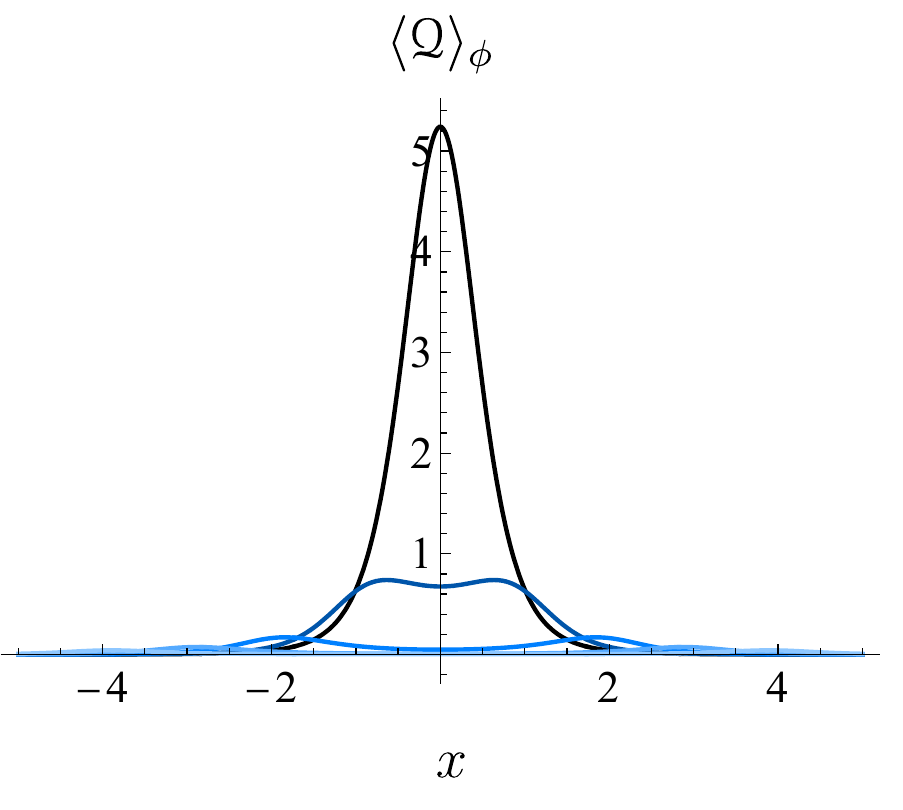}}
	\caption{Spatial charge density dynamics along the $x$-axis after the local $\phi$-quench for fixed time moments and $\eps = 1$, $m = 0.1$ in theories with different number of dimensions, $d = 2$ (left), $d = 3$ (middle) and $d = 4$ (right).}
	\label{fig:charge_quench_timeslices}
\end{figure}

The evolution of the charge density~\eqref{eq:charge_quench} is shown in figures~\ref{fig:charge_quench} and~\ref{fig:charge_quench_timeslices}. It resembles the evolution of the energy density demonstrating three regimes of the propagation: a double-hill configuration for lower masses, a single-maximum configuration for masses higher than some critical value, and the critical configuration propagating in the form of an almost flat plateau. The critical mass in the leading order is $m_{\text{crit}} = \eps^{-1}$, which is obtained by the analysis of the sign of the second spatial derivative at $x = 0$ near the time infinity (see~\eqref{eq:crit_m}).

\skipline

Let us also give the answer for the evolution of the charge density $\Qcal$ after the local quench by the operator~$\partial\phi$
\be
    \begin{aligned}
        & \corrfunc{\Qcal(t, x)}_{\partial\phi} = \frac{m^2(\eps^2 + (t + x)^2)}{2K_2(2m\eps)}\left[\frac{4\eps}{(\eps^2 + (t - x)^2)^2}\left|K_2\left(m\sqrt{(\eps - it)^2 + x^2}\right)\right|^2 + \right. \\
        & \left. + \left(\frac{m(\eps - it)K_1\left(m\sqrt{(\eps - it)^2 + x^2}\right)K_2\left(m\sqrt{(\eps + it)^2 + x^2}\right)}{(\eps^2 + (t - x)^2)\sqrt{(\eps - it)^2 + x^2}} + \text{c.c.}\right)\right].
   \end{aligned}
\ee

This expression has a well-defined massless limit
\be
    \corrfunc{\Qcal(t, x)}_{\partial\phi}\Big|_{m\,=\,0} = \frac{16\eps^3}{(\eps^2 + (t - x)^2)^3}.
\ee

\skipline

The same analysis generalises to the case of an arbitrary number of dimensions, $d > 2$. The Euclidean action for the charged scalar field is
\be
    S = \frac{1}{8\pi}\int d^{\,d}x\left(\partial_\tau\phi^*\partial_\tau\phi + \partial\phi^*_i\partial\phi^i + m^2\phi^*\phi\right),
\ee
and we define the $d$-dimensional version of the charge operator as
\be
    Q = \int d^{\,d - 1}x\,\left(\phi^*{\partial_\tau\phi} - {\partial_\tau\phi}^*\phi\right),
\ee
with the same charge density operator as in the two-dimensional case~\eqref{eq:charge_density}.

In this setup, we obtain the evolution of the charge density after the quench
\be
    \begin{aligned}
        & \corrfunc{\Qcal(t, x^i)}_{\phi,\,d} = \frac{2\eps^{\frac{d}{2} - 1}m^{\frac{d}{2}}}{\pi^{\frac{d}{2} - 1}K_{\frac{d}{2} - 1}(2\eps m)} \times \\
        & \times \left[\frac{(\eps - it)\sqrt{(\eps + it)^2 + \rho^2}}{\left|(\eps - it)^2 + \rho^2\right|^{\frac{d}{2}}} K_{\frac{d}{2} - 1}\left(m\sqrt{(\eps + it)^2 + \rho^2}\right)K_{\frac{d}{2}}\left(m\sqrt{(\eps - it)^2 + \rho^2}\right) + \text{c.c.}\right].
    \end{aligned}
    \label{eq:charge_quench_ddim}
\ee

The series expansion of the divergent term coming from the composite operator after the point-splitting gives
\be
    \mathcal{C}^{\mathcal{Q}}_{\phi,\,d} + \mathcal{D}^{\mathcal{Q}}_{\phi,\,d} = \lim_{\delta\,\to\,0}\left[\frac{4\Gamma\left(\frac{d}{2}\right)}{d^{\frac{d}{2}}\pi^{\frac{d}{2} - 1}} \cdot \frac{1}{\delta^{d - 1}} - \frac{4m^2\Gamma \left(\frac{d}{2} + 1\right)}{d^{\frac{d}{2}}\pi^{\frac{d}{2} - 1}(d - 2)} \cdot \frac{1}{\delta^{d - 3}} + O\left(\frac{1}{\delta^{d - 2n - 1}}\right)\right], \,\,\, n \in \mathbb{Z}.
\ee
For an odd number of dimensions, $d \equiv 2l + 1$, the terms of the orders lower than $d$ are divergent and the $d$-order term contributes as a constant, $m^2/\sqrt{3}$ for $d = 3$, $m^4/(8\sqrt{5}\pi)$ for $d = 5$ etc. For $d \equiv 2l$, there is no constant term: all terms of the orders lower than $d$ contribute as divergences.

The large-time (with spatial distance fixed) and large-distance (with time coordinate fixed) dynamics are given by
\be
    \corrfunc{\Qcal(t, x^i)}_{\phi,\,d} \largetime \frac{2}{m}\cdot\frac{\gamma}{t^{d - 1}} + \left[\frac{(d - 2\eps m)\rho^2}{m} + \frac{(d - 1)(d - 2\eps m - 1)\eps}{2m^2}\right]\frac{\gamma}{t^{d + 1}} + O\left(t^{-d - 3}\right),
    \label{eq:chargeDdim_large-t}
\ee
and
\be
    \corrfunc{\Qcal(t, x^i)}_{\phi,\,d} \underset{\rho\,\to +\infty}{\approx} \frac{2\gamma\eps e^{2\eps m}}{m\rho^{d}}\,e^{-2m\rho}\,e^{\frac{m}{\rho}(t^2 - \eps^2)} + O\left(\rho^{-d - 1}e^{-2m\rho}\,e^{\frac{m}{\rho}(t^2 - \eps^2)}\right),
    \label{eq:chargeDdim_large-x}
\ee
where $\gamma$ is defined by~\eqref{eq:gamma_def}.

The charge density dynamics for the case $d = 3$ is visualized in figure~\ref{fig:charge_quench_timeslices}. As in the case of the energy dynamics, the perturbation propagates in the form of radially symmetric waves.

The charge density dynamics for a massless scalar field ($d > 2$) can be obtained as the $m \to 0$ limit of the general expression~\eqref{eq:charge_quench_ddim}
\be
\begin{aligned}
    & \corrfunc{\Qcal(t, x^i)}_{\phi,\,d}\Big|_{m\,=\,0} = \frac{2\Gamma\left(\frac{d}{2}\right)}{\pi^{\frac{d}{2} - 1}}\left[\frac{(\eps - it)}{\left((\eps - it)^2 + \rho^2\right)^{\frac{d}{2}}} + \text{c.c.}\right].
\end{aligned}
\ee

\section{Local quenches on a cylinder}

We have already studied the simplest setup on a cylinder --- the CFT case, leading to the energy density~\eqref{eq:CFT_quench_cylinder}. In this section, we consider the local quench in massive field theory on a cylinder, where the coordinate $x$ is periodic with the period~$L$. In this case, it is not possible to derive the exact analytical expressions for the coordinate-space two-point function (see~\eqref{eq:massive_2point_cyl}). The two-point function is given as a Fourier transformation, which we perform numerically summing up the series and truncating it at some order. To have an analytical reference, we also consider the vanishing-mass limit, which can be solved exactly up to the leading-order terms.

In what follows, we start with the action of the theory, define the two-point function and calculate the energy density. When the final correlation function is derived, we perform Wick rotation to study the dynamics on a Lorentzian cylinder $S^1\,\times$ Time.

\skipline

To get analytical intuition, we derive the leading-order contribution to the $\phi$-quench on a cylinder in the limit of a vanishingly small mass and then study the energy density evolution following the quench by the operator $\phi$ applying Wick's theorem with the two-point function~\eqref{eq:app:massless_2point_cyl} obtained in appendix~\ref{appendix:cyl}. The result reads
\be
    \begin{aligned}
        \corrfunc{\Ecal(t, x)}_{\phi} & \underset{m\,\to\,0}{\approx} -\frac{\pi^2}{3L^2} + \frac{\pi m}{L}\left[2 + \frac{\sin^2\left(\frac{2\pi x}{L}\right)}{\left|\cosh\left(\frac{2\pi\sqrt{(\eps + it)^2}}{L}\right) - \cos\left(\frac{2\pi x}{L}\right)\right|^2}\right. + \\
        & + \left.\left|\frac{\eps + it}{\sqrt{(\eps + it)^2}}\left(\frac{\sinh\left(\frac{2\pi\sqrt{(\eps + it)^2}}{L}\right)}{\cosh\left(\frac{2\pi\sqrt{(\eps + it)^2}}{L}\right) - \cos\left(\frac{2\pi x}{L}\right)}\right)\right|^2\right] + O(m^2).
    \end{aligned}
    \label{eq:phi_quench_cylinder_mass_deformed}
\ee
The point-splitting of the composite operator leads to the following constant and divergent terms
\be
    \mathcal{C}_{\phi} + \mathcal{D}_{\phi} = \lim_{\delta\,\to\,0}\left[-\frac{\pi^2}{3L^2} + \frac{m\pi}{L} - \frac{m^2}{2}\ln\left(\frac{2\sqrt{2}\,\pi\delta}{L}\right) + O(\delta)\right].
\ee
We should stress that the massless limit of the energy density is not well-defined in the case of the $\phi$-quench because the two-point function of the operator $\phi$ on a cylinder is ill-defined (see appendix~\ref{appendix:cyl} for details).

An analogous expression for the leading-order massive correction to the massless case can be derived for the energy density evolution following the $\partial\phi$-quench. The result is given by
\be
    \begin{aligned}
        \corrfunc{\Ecal(t, x)}_{\partial\phi} & \underset{m\,\to\,0}{\approx} -\frac{\pi^2}{3L^2} + \frac{4\pi^2}{L^2} \cdot \frac{\sinh^2\left(\frac{2\pi\eps}{L}\right)}{\left(\cos\left(\frac{2\pi(x - t)}{L}\right) - \cosh\left(\frac{2\pi\eps}{L}\right)\right)^2} + \\
        & + \frac{\pi m}{L}\cdot\frac{\sinh\left(\frac{2\pi\eps}{L}\right)\sinh\left(\frac{4\pi\eps}{L}\right)}{\cos\left(\frac{2\pi(x - t)}{L}\right) - \cosh\left(\frac{2\pi\eps}{L}\right)} + O(m^2),
    \end{aligned}
    \label{eq:dphi_quench_cylinder_mass_deformed}
\ee
with the constant term and the divergence coming from the point-splitting procedure
\be
    \mathcal{C}_{\partial\phi} + \mathcal{D}_{\partial\phi} = \lim_{\delta\,\to\,0}\left[-\frac{\pi^2}{3L^2} - \frac{1}{\delta^2} + O(\delta)\right].
\ee
Notice that the massless limit reproduces the CFT result~\eqref{eq:CFT_quench_cylinder}.

\skipline

For an arbitrary value of $m$, we are not able to derive an analytical expression for the two-point function. In terms of an infinite series, it follows from~\eqref{eq:massive_2point_cyl} with the choice $A = 1/(4\pi)$ of the normalization constant
\be
    \begin{aligned}
        & \corrfunc{\phi(\tau, x)\phi(\tau_0, x_0)} = \\
        & = \frac{4\pi}{2mL}\,e^{-m\sqrt{(\tau - \tau_0)^2}} + \frac{4\pi}{L}\sum\limits_{n\,>\,0}\frac{e^{-\sqrt{\frac{4\pi^2 n^2}{L^2} + m^2}\sqrt{(\tau - \tau_0)^2}}}{\sqrt{\frac{4\pi^2 n^2}{L^2} + m^2}}\cos\left(\frac{2\pi n(x - x_0)}{L}\right).
    \end{aligned}
    \label{eq:prop_cyl_massive_series}
\ee
This series converges as $O\left(e^{-n}/n\right)$ and therefore, the main contribution comes from the first few orders. This fact allows to truncate the series at some order to get a numerical approximation. To derive the final expression for the energy density evolution after the $\phi$-quench on a cylinder, one should follow the same procedure of the Wick's contractions as in the previous sections. It is not possible to derive the constant and the divergent parts of the correlator explicitly, since we are not able to sum up the whole series analytically. For this reason, we apply the same procedure as in the flat space~\eqref{eq:sub_plain}, which subtracts all contributions from a composite operator, and after that add the constant $-\pi^2/(3L^2)$ from the leading-order contribution~\eqref{eq:phi_quench_cylinder_mass_deformed},~\eqref{eq:dphi_quench_cylinder_mass_deformed}. It is the same constant that follows from the anomalous transformation of the energy-momentum tensor in the CFT$_2$~\eqref{eq:CFT_quench_cylinder} and denotes the vacuum energy density (the Casimir effect).

\begin{figure}
	\subfloat[$m = 0.01$]{\includegraphics[width=1\textwidth]{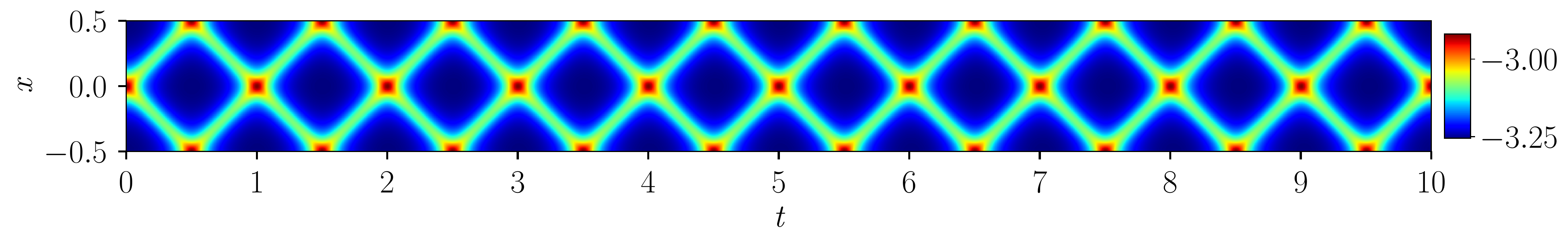}} \\
	\subfloat[$m = 1$]{\includegraphics[width=0.97\textwidth]{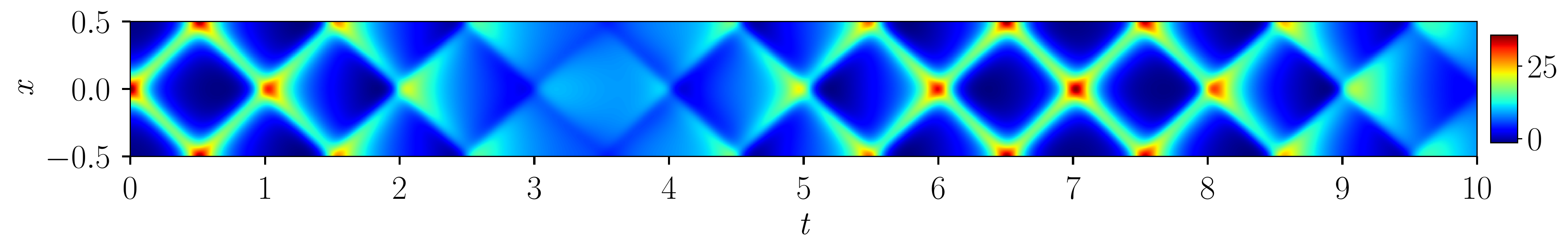}} \\
	\subfloat[$m = 5$]{\includegraphics[width=0.98\textwidth]{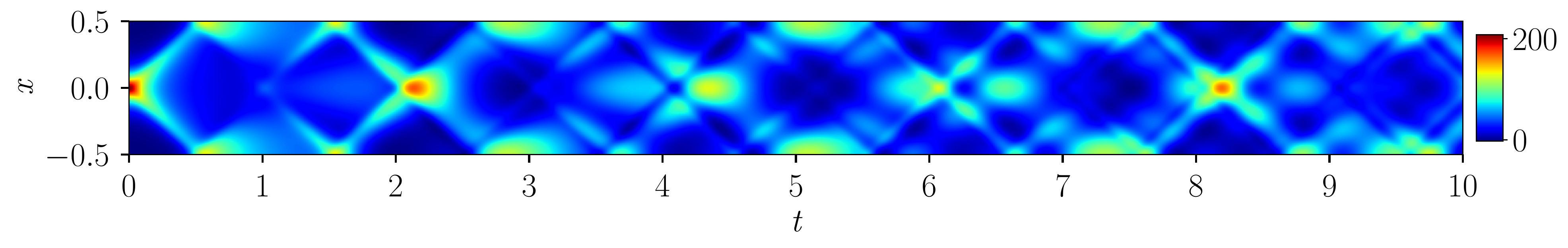}} \\
	\subfloat[$m = 10$]{\includegraphics[width=0.98\textwidth]{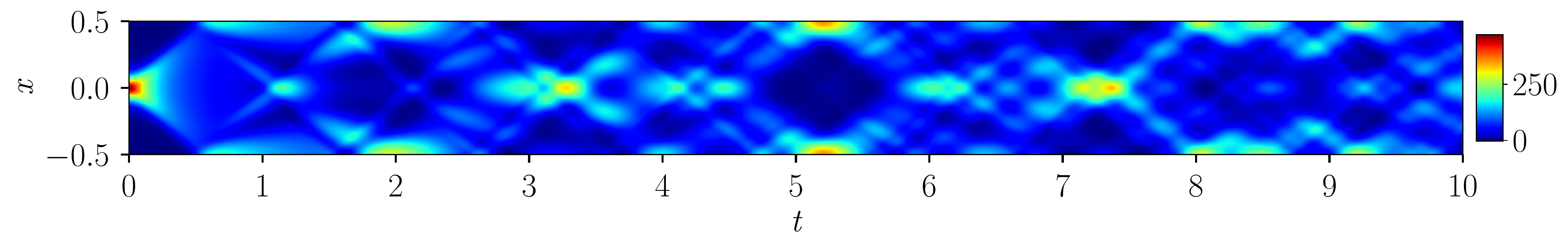}}
    \caption{Energy density evolution after the local $\phi$-quench on a cylinder for masses $m = 0.01$ (top), $m = 1$ (upper middle), $m = 5$ (lower middle), $m = 10$ (bottom). For all the figures, we fix $\eps = 0.1$ and $L = 1$. The result is the sum of $N = 15$ terms in the series.}
	\label{fig:phi_quench_cylinder}
\end{figure}

The plot of the energy density dynamics after the $\phi$-quench calculated in this manner is shown in figure~\ref{fig:phi_quench_cylinder}. The feature of the propagation of the massive perturbation is that it does not decay at large times as one might have expected in analogy to the flat space result~\eqref{eq:massive_quench}. This is explainable because the exponents in the series~\eqref{eq:prop_cyl_massive_series} become oscillating after performing Wick rotation. Oscillations of the first several terms in the series contribute the most, since the amplitudes get suppressed with growing $n$. From a physical point of view, we explain this behaviour as interference of perturbations propagating in the opposite spatial directions and winding around the cylinder. It is worth noticing (and can be checked numerically) that the energy is conserved for this regime of propagation.

\begin{figure}
	\centering
	\subfloat[$m = 0.1$]{\includegraphics[width=0.65\textwidth]{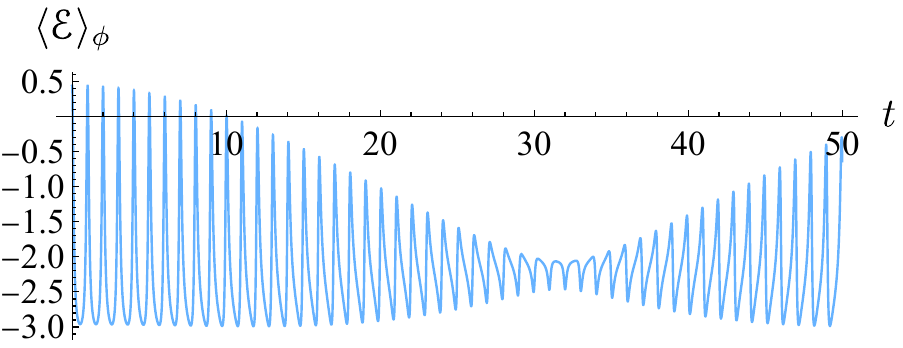}} \\
	\subfloat[$m = 1$]{\includegraphics[width=0.65\textwidth]{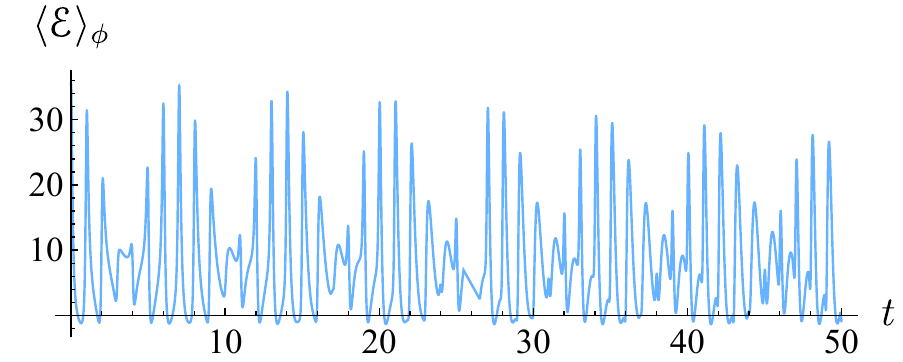}} \\
	\subfloat[$m = 5$]{\includegraphics[width=0.65\textwidth]{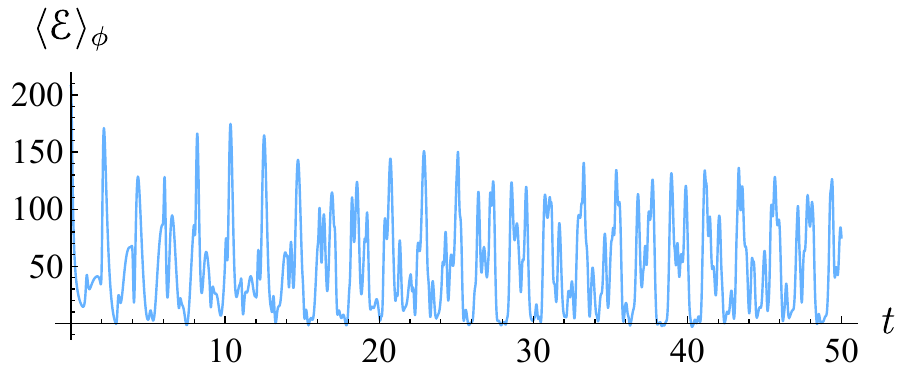}} \\
	\subfloat[$m = 10$]{\includegraphics[width=0.65\textwidth]{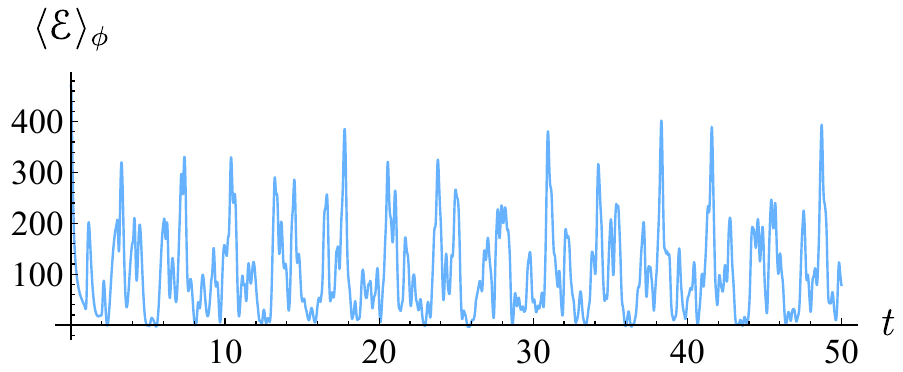}}
	\caption{Time dependence of the energy density in massive theory with cylindrical spacetime geometry after the local $\phi$-quench for $x = 0$. The parameters are fixed as $L = 1$, $\eps = 0.1$. The mass ranges as $m = 0.1$ (a), $m = 1$ (b), $m = 5$ (c), $m = 10$ (d). The result is the sum of $N = 15$ terms in the series.}
	\label{fig:phi_cyl_timeslice}
\end{figure}

From figures~\ref{fig:phi_quench_cylinder} and \ref{fig:phi_cyl_timeslice}, one can note a quite complicated picture of the post-$\phi$-quench energy propagation. The presence of scale parameters (regularization parameter $\eps$, mass $m$ and cylinder circumference $L$) suggests different regimes. The massless perturbation does not decay and freely winds around the cylinder. Relatively small mass increases lead to alternating interchange of free propagation and decaying with subsequent revivals. Large masses, which in the flat-space case lead to configurations, which decay after the initial quench, here manifest themselves in an erratic and seemingly chaotic localization/delocalization pattern.

\FloatBarrier

\begin{figure}
	\centering
	\subfloat[$m = 1$]{\includegraphics[width=1\textwidth]{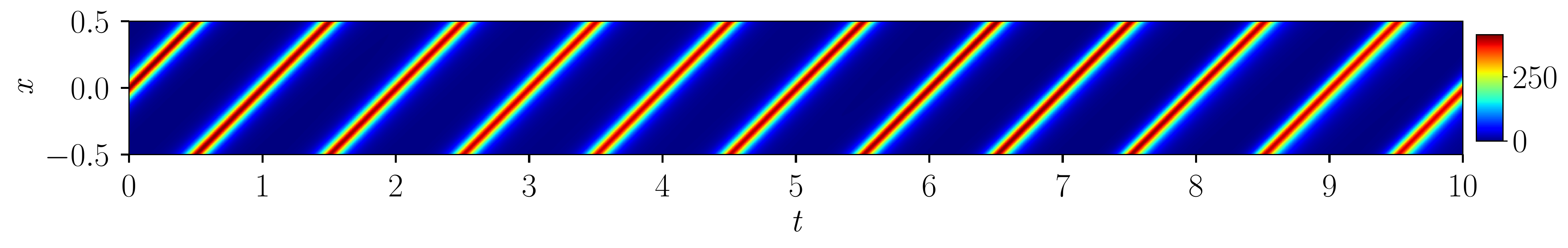}} \\
	\subfloat[$m = 2$]{\includegraphics[width=1\textwidth]{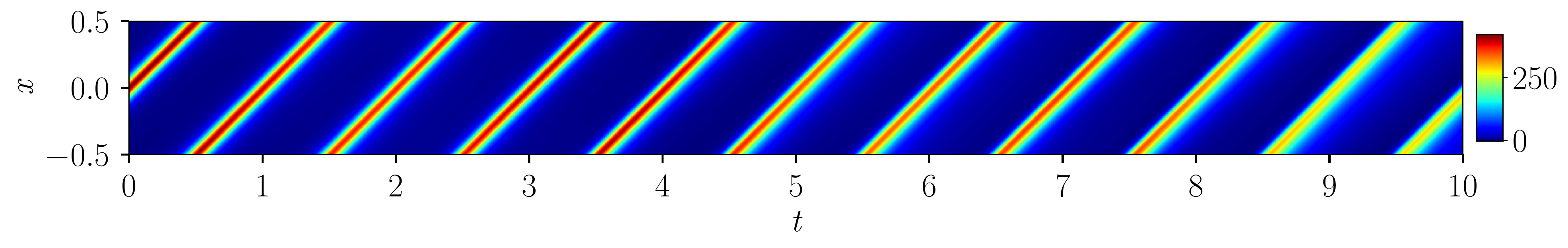}} \\
	\subfloat[$m = 5$]{\includegraphics[width=1\textwidth]{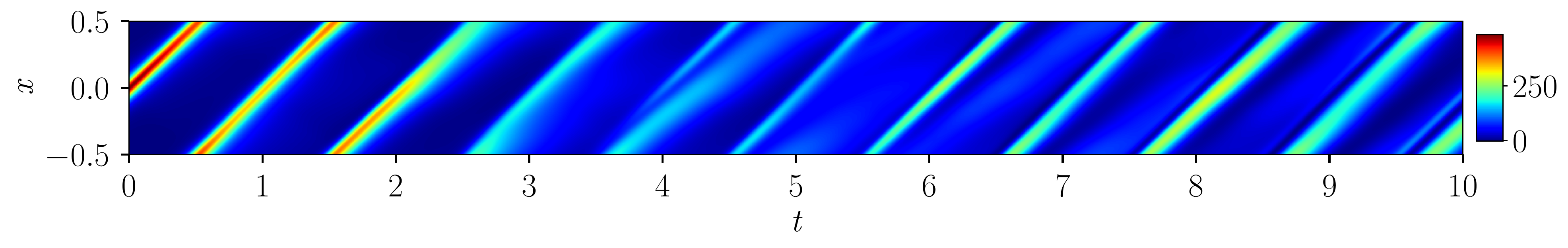}} \\
	\subfloat[$m = 10$]{\includegraphics[width=1\textwidth]{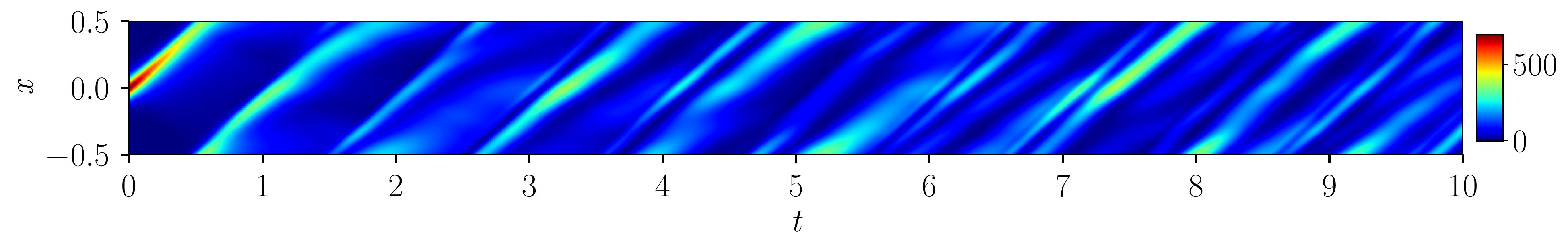}}
	\caption{Energy density evolution after the local $\partial\phi$-quench on a cylinder for masses $m = 1$ (top), $m = 2$ (upper middle), $m = 5$ (lower middle), $m = 10$ (bottom). For all the figures, we fix $\eps = 0.1$ and $L = 1$. The result is the sum of $N = 15$ terms in the series.}
	\label{fig:dphi_quench_cylinder}
\end{figure}

\begin{figure}
	\centering
	\subfloat[$m = 1$]{\includegraphics[width=0.65\textwidth]{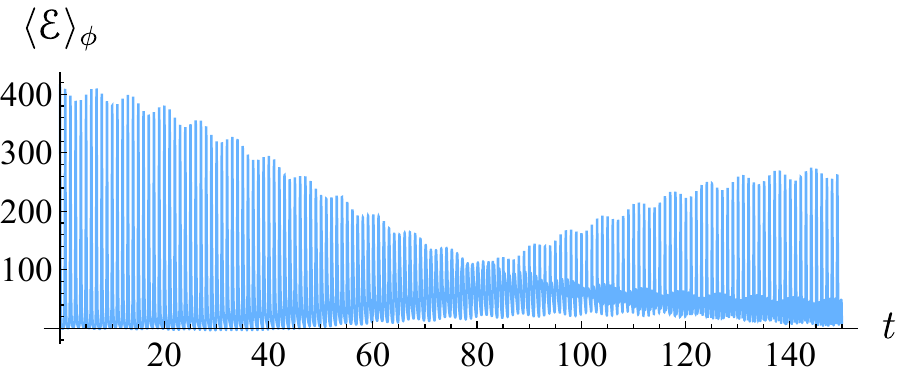}} \\
	\subfloat[$m = 2$]{\includegraphics[width=0.65\textwidth]{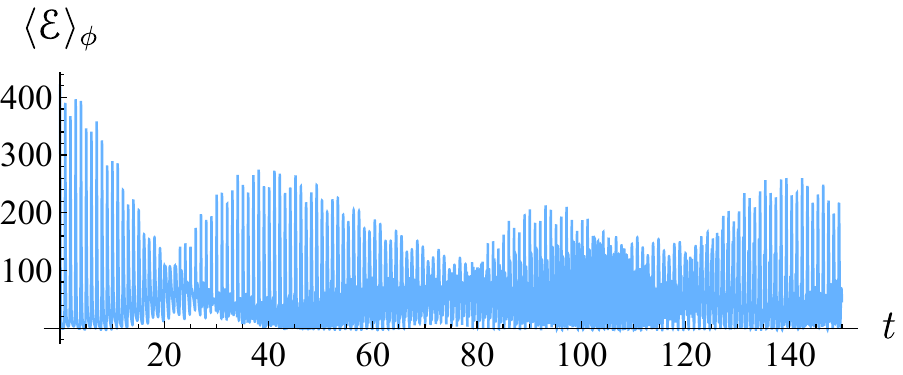}} \\
	\subfloat[$m = 5$]{\includegraphics[width=0.65\textwidth]{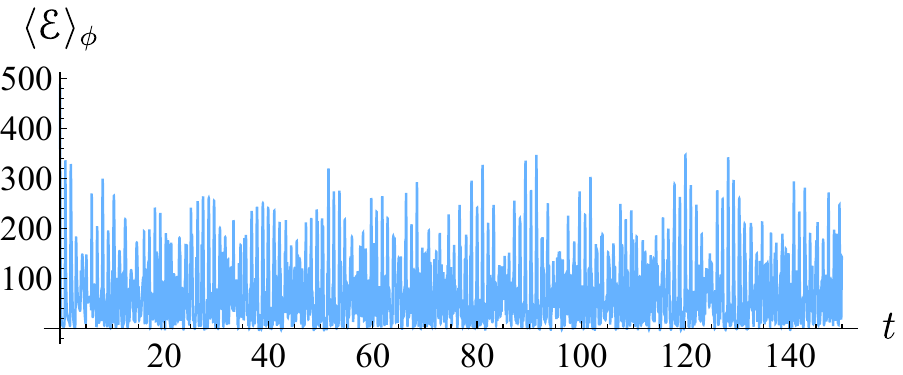}} \\
	\subfloat[$m = 10$]{\includegraphics[width=0.65\textwidth]{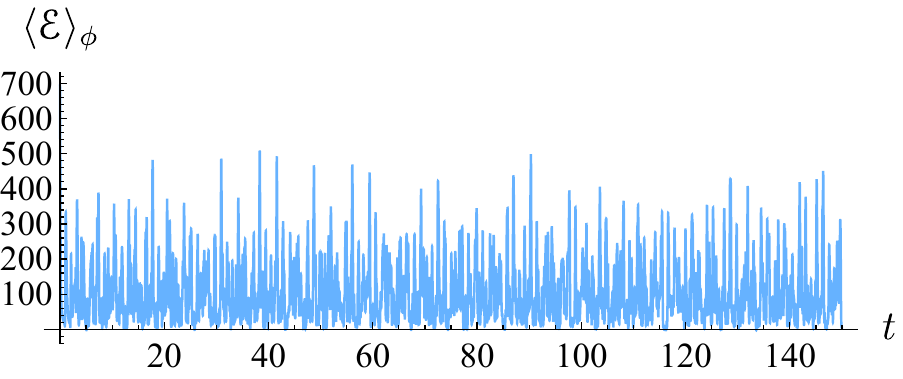}}
	\caption{Time dependence of the energy density in massive theory with cylindrical spacetime geometry after the local $\partial\phi$-quench for $x = 0$. The parameters are fixed as $L = 1$, $\eps = 0.1$. The mass ranges as $m = 1$ (a), $m = 2$ (b), $m = 5$ (c), $m = 10$ (d). The result is the sum of $N = 15$ terms in the series.}
	\label{fig:dphi_cyl_timeslice}
\end{figure}

The complicated behaviour of the post-$\partial\phi$-quench dynamics of the energy density can be read off from the figures~\ref{fig:dphi_quench_cylinder} and \ref{fig:dphi_cyl_timeslice}. Even though in \eqref{eq:phi_quench_cylinder_mass_deformed} and \eqref{eq:dphi_quench_cylinder_mass_deformed} we write down the \mbox{$1^{\text{st}}$-order} corrections in mass to the energy density dynamics on a cylinder, it is analytically easier to estimate the value of the mass, at which the CFT picture breaks down, from the $2^{\text{nd}}$-order expansion of the propagator in momentum space (see appendix~\ref{appendix:cyl})
\be
    \frac{1}{q^2 + m^2 + \om_n^2} \simeq \frac{1}{q^2 + \om_n^2} - \frac{m^2}{\left(q^2 + \om_n^2\right)^2} + O(m^4).
\ee
Therefore, the critical mass $m^{\text{cyl}}_{\text{crit}}$, at which the $m^2$-correction is of order of the massless propagator, is estimated from
\be
    \frac{\left(m^{\text{cyl}}_{\text{crit}}\right)^2}{\left(q^2 + \om_n^2\right)^2} \Bigg/ \frac{1}{q^2 + \om_n^2} \sim 1.
\ee
Numerically, it gives $m^{\text{cyl}}_{\text{crit}} \sim 1/L$ for $q \sim 1/L$. While $m < m^{\text{cyl}}_{\text{crit}}$, almost no effect, which would distinguish a mass-deformed theory from CFT, can be seen. As the mass increases, we observe subsequent dampings and revivals of the amplitude of the energy density. At $m \gg m^{\text{cyl}}_{\text{crit}}$, oscillations of different frequencies overlap, and the structure seems to become chaotic.

\clearpage


\section{Conclusions and future prospects}

In this paper, we have studied operator local quenches in massive scalar field theory. The previous research focused mainly on two-dimensional CFT. We extend the study of localized perturbations to a wider class of systems that do not necessarily have an underlying conformal symmetry. Let us briefly mention possible future research topics.
\begin{itemize}
    \item Free fermionic systems (for example, the effective theory of graphene)  seem to have interesting applications in condensed matter theories.
    
    \item It would be intriguing to consider certain interacting field theories, for example, interacting scalars and vector models, as well as to study the dynamics in gauge theories with applications.
    
    \item An interesting direction would be to extend the higher-dimensional local quenches to the cases of curved spacetimes. For instance, de Sitter space as a background suggests possible applications in cosmology: the cosmological collider and inflation.
    
    \item Finally, we expect a non-trivial and complicated dynamics in different exotic models, like Lifshitz, fracton, ultrametric and fractal field theories.
\end{itemize}

\acknowledgments

D.S.A, A.I.B. and V.V.P. are supported by the Foundation for the Advancement of Theoretical Physics and Mathematics ``BASIS''. The work of D.S.A. was performed at Steklov International Mathematical Center and supported by the
Ministry of Science and Higher Education of the Russian Federation (Agreement No. 075-15-2019-1614).

\appendix
\section{Analytic continuation of Euclidean correlators}
\label{appendix:cont}

In this appendix, we overview the relevant for our setup material of~\cite{Hartman:2015lfa, Asplund:2014coa}, accompanying it with our additional comments. We consider the example of the two-point function, which illustrates the main features of correlators in both Euclidean and Lorentzian signatures. After that, we remind the Osterwalder-Schrader prescription~\cite{Osterwalder:1973dx, Osterwalder:1974tc}, which is equivalent to operator ordering in terms of the branch structure of analytically continued correlators on the complex time plane. At the end, we discuss the operator local quench setup for finite $\eps$.

\subsubsection*{Lightcones, branch cuts and Euclidean and Lorentzian correlators}

Since we perform calculations in Euclidean signature, while observables live in Lorent\-zian, we should clearly understand analytical properties of analytically continued correlation functions, and also be able to relate Euclidean to Lorentzian correlators.

The difference between the signatures is related to the choice of time coordinates --- Euclidean $\tau$ and Lorentzian $t$. Let us introduce a complex variable $\mathfrak{t}$ such that its imaginary part coincides with the Euclidean time, while the real part --- with Lorentzian: $\tau = \re\mathfrak{t}$ and $t = \im\mathfrak{t}$. To obtain a Lorentzian correlator given a Euclidean one, we first analytically continue the Euclidean correlator to the complex time $\mathfrak{t}$. Then, dealing with the analytical structure of the analytically continued correlator, we reduce its domain to $\im\mathfrak{t}$ in a way that gives a real-time $n$-point function with a uniquely defined chronological ordering of the operators.

Analytically continued $n$-point correlators possess $n - 1$ branch points and branch cuts. For instance, the CFT two-point function of primary operators can be rewritten in the following form
\be
    \frac{1}{\big[(\mathfrak{t_i} - \mathfrak{t_j})^2 + (x_i - x_j)^2\big]^\Delta} = \exp\left[-\Delta\ln\left((\mathfrak{t_i} - \mathfrak{t_j})^2 + (x_i - x_j)^2\right)\right],
\ee
hence, there is a branch cut due to the complex logarithm.

In \textit{Euclidean signature}, there is no notion of timelike or null separation, hence, operators inserted in non-coinciding points are spacelike separated and, therefore, commute. On the complex time plane, there are no branch points lying on the real axis. This means that a correlator as a function of $\tau = \re\mathfrak{t}$ is single-valued. Any swap of operators does not change its value and hence, there is no notion of operator ordering.

Reducing from the complex time to \textit{Lorentzian time}, we should choose a particular branch of the analytically continued correlator in order to unambiguously define its Lorentzian version. There may be number of ways to do so, which result in different \textit{operator orderings}. Note that zero spacetime interval in Lorentzian signature denotes both coincident and null-separated points. The first correspond to a pole of the analytically continued correlator in the complex $\mathfrak{t}$-plane, while the second --- to a branch point. For a pair of operators, the branch point denotes when one operator hits the lightcone of the other. The branch cut starting from this point separates two causal domains, and different values of the correlator on the adjacent branches correspond to different orderings. Thus, the branch cut generates a non-trivial commutator of these operators across the lightcone.


\begin{figure}[ht]\centering
    \includegraphics[width=0.45\textwidth]{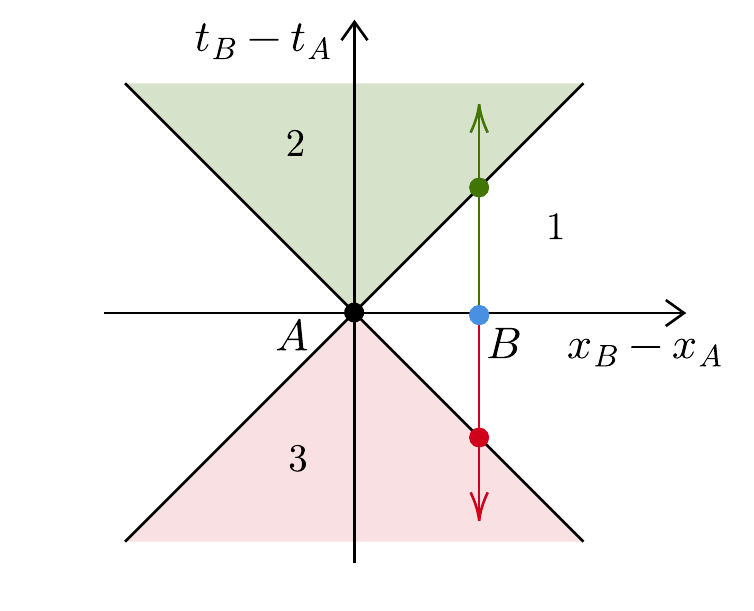}
    \includegraphics[width=0.45\textwidth]{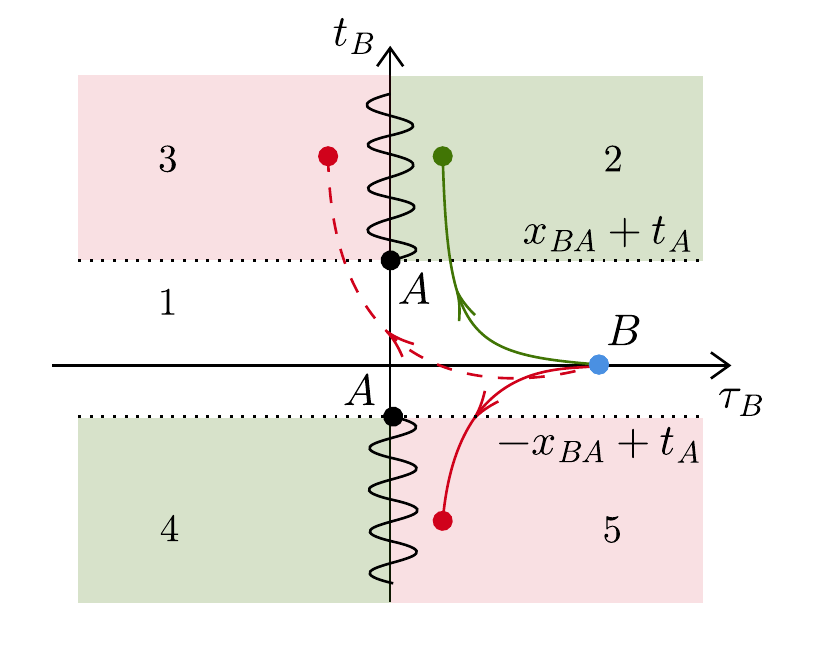}
	\caption{\textit{Left:} Operator $B$ (blue dot) hits the future lightcone (region 2) of the operator $A$ (black dot). In the region 1, there is no notion of time ordering. In the region 2, the operators are time-ordered. Inside the past lightcone (region 3), the operators are anti-time-ordered. The green and red dots show the moments when the operator $B$ crosses the future and past lightcones, respectively. \textit{Right:} Positions of branch cuts on the complex time plane. Real axis corresponds to Euclidean time, imaginary --- to Lorentzian. Analytic continuation along the green contour leads to time ordering of operators, while along the red curve --- to anti-time ordering. Since the regions 3 and 5 lie on the same branch, it is equivalent to continue along the dashed red curve to obtain anti-time ordered operators.}
    \label{fig:app:HittingLightCone}
\end{figure}

\textbf{Example: two-point correlator.} Let us consider a Lorentzian two-point function $\corrfunc{A(t_A, x_A) B(t_B, x_B)}_L$. One has to specify a particular ordering of the operators based on the branch structure of this correlator in complex time $\mathfrak{t}$. Let us assume that the operator $A$ is placed at a fixed spacetime point~$(t_A, x_A)$, while $B$ is at arbitrary point $(t_B, x_B)$. As time evolves, the operator $B$ eventually hits the future lightcone of $A$, see figure~\ref{fig:app:HittingLightCone}, left. When the operators are spacelike-separated, $t_{BA} < x_{BA}$ ($t_{BA} \equiv t_B - t_A$ and $x_{BA} \equiv x_B - x_A$), there is no notion of the operator ordering, hence $[A, B] = 0$ or $\corrfunc{A B} = \corrfunc{B A}$. When $B$ is inside the future lightcone of~$A$, $t_{BA} > x_{BA}$, and time ordering implies that $\corrfunc{T A B} = \corrfunc{B A} = \corrfunc{A B} + \corrfunc{[B, A]}$. Conversely, the evolution backward in time after hitting the past lightcone of $A$ by the operator $B$ would correspond to the following anti-time ordering:
$\corrfunc{\Tbar A B} = \corrfunc{A B} = \corrfunc{B A} + \corrfunc{[A, B]}$.

In order to define a Lorentzian correlator given a Euclidean one, we need to perform analytic continuation from Euclidean time $\re\mathfrak{t}_B$ to Lorentzian $\im\mathfrak{t}_B$ along some continuous contour on the complex $\mathfrak{t}_B$-plane. In the domain $-x_{BA} + t_A < \im\mathfrak{t}_B < x_{BA} + t_A$, the two-point function is analytic and single-valued. From a physical point of view, this fact reflects that spacelike-separated operators are causally disconnected, and it makes no difference how we order them. As $\im\mathfrak{t}_B > x_{BA} + t_A$ ($\im\mathfrak{t}_B < -x_{BA} + t_A$), the correlator as a function of the complex time has a branch cut along the imaginary axis, with a branch point at $\im\mathfrak{t}_{BA} = x_{BA} + t_A$ ($\im\mathfrak{t}_{BA} = -x_{BA} + t_A$). This cut divides the function into two branches. Each branch corresponds to a particular ordering of the operators. The green path in figure~\ref{fig:app:HittingLightCone}, right, corresponds to time ordering, while the red one --- to anti-time ordering. Alternatively, we can continue along the dashed red contour, because the regions 2, 4 and 3, 5 are equivalent. Since $\corrfunc{B A}$ and $\corrfunc{A B}$ lie on different branches, a non-zero value of the commutator $[A, B]$ is generated as we go around the branch point. 

\subsubsection*{Osterwalder-Schrader prescription}

\begin{figure}[ht]\centering
    \includegraphics[width=0.45\textwidth]{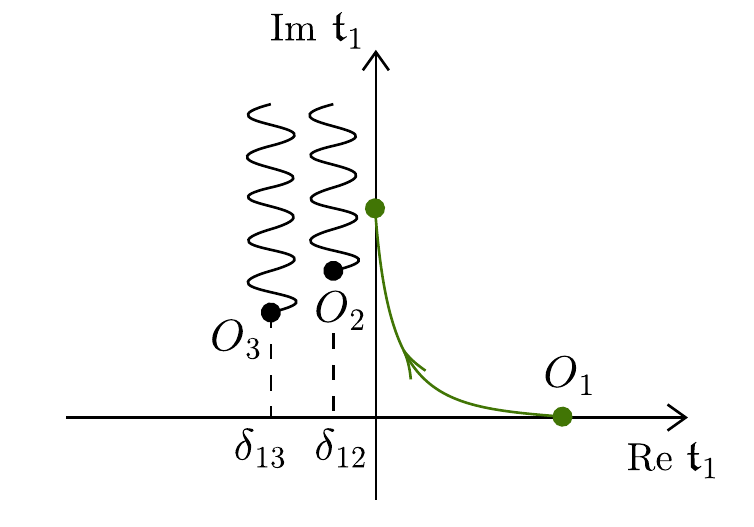}
    \caption{In the Osterwalder-Shrader $i\delta$-prescription for the three-point function $\corrfunc{O_1 O_2 O_3}_E$, the branch cuts are shifted by $i\delta$'s to the left of the imaginary axis $\im\mathfrak{t}_1$, which ensures time-ordering, $\corrfunc{T O_1 O_2 O_3}_L$. Real axis corresponds to Euclidean time and imaginary axis to Lorentzian time. Here, $\delta_{ij} \equiv \delta_i - \delta_j$.}
    \label{fig:app:Osterwalder}
\end{figure}

A common way to time order operators is the Osterwalder-Schrader $i\delta$-prescription~\cite{Osterwalder:1973dx, Osterwalder:1974tc}. For a time-ordered three-point function, one proceeds with the following Wick rotation
\be
    \begin{aligned}
        & \corrfunc{T O_1(t_1, x_1) O_2(t_2, x_2) O_3(t_3, x_3)}_L = \\
        & = \lim\limits_{\begin{subarray}{c} \delta_i\,\to\,0^{+} \\ \delta_1\,>\,\delta_2\,>\,\delta_3 \end{subarray}} \left\langle O_1\big(i(t_1 - i\delta_1), x_1\big) O_2\big(i(t_2 - i\delta_2), x_2\big) O_3\big(i(t_3 - i\delta_3), x_3\big)\right\rangle_E.
    \end{aligned}
    \label{eq:app:Osterwalder}
\ee
From the point of view described above, this prescription means the following. The operator $O_1$ hits the lightcone of $O_2$ at $\mathfrak{t}_{1,O_{2}} = it_2 - \delta_{12} \pm i(x_1 - x_2)$ and the lightcone of $O_3$ at $\mathfrak{t}_{1,O_{3}} = it_3 - \delta_{13} \pm i(x_1 - x_3)$ on the complex $\mathfrak{t}_1$-plane, $\delta_{12} \equiv \delta_1 - \delta_2 > 0$, $\delta_{13} \equiv \delta_1 - \delta_3 > 0$, $\delta_{12} < \delta_{13}$. Therefore, the addition of $i\delta$'s shifts the branch cuts to the right or left of the imaginary axis and away from each other. We can imply by considering pairs of operators $O_1, O_2$ and $O_1, O_3$ that this leads to time ordering of the operators in the Lorentzian correlator, see figure~\ref{fig:app:Osterwalder}. 

\subsection*{Operator local quench setup}

We are interested to define a specific Lorentzian correlator from the following Euclidean
\be
    \corrfunc{\Odag(\eps, 0)\Ocal(\tau, x)O(-\eps, 0)}_E \to \corrfunc{\Odag(i\eps, 0) \Ocal(t, x) O(-i\eps, 0)}_L.
    \label{eq:app:setup}
\ee

\begin{figure}[ht]\centering
    \includegraphics[width=0.55\textwidth]{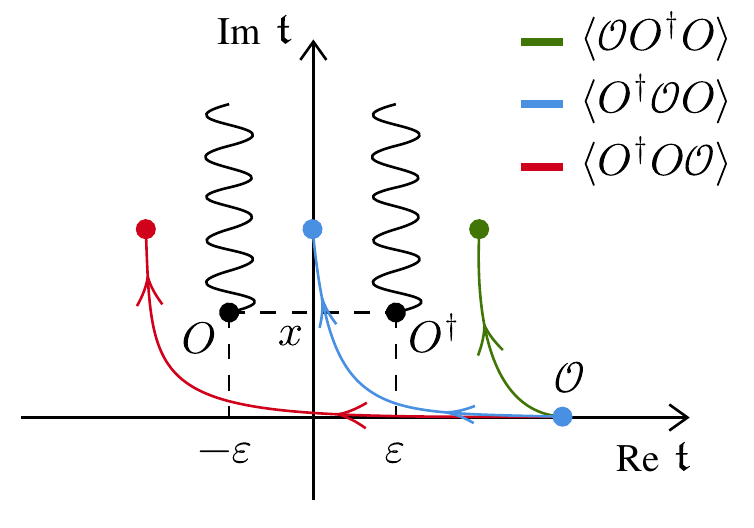}
    \caption{Analytical structure of the three-point function~\eqref{eq:app:setup}. Real axis corresponds to Euclidean time and imaginary --- to Lorentzian. The branch cuts are drawn in black. The only possible Wick rotation, consistent with the setup~\eqref{eq:our_setting}, which implies finite $\eps$, is along the blue curve. However, if we are to consider a vanishingly small $\eps$, then the continuation along the green curve would correspond to a time-ordered correlator, while along the red curve --- to an anti-time-ordered.}
    \label{fig:app:QuenchBCuts}
\end{figure}

The branch points for the pair of the operators $\Ocal$ and $\Odag$ ($\Ocal$ and $O$) are at $\mathfrak{t} = \pm ix + \eps$ ($\mathfrak{t} = \pm ix - \eps$), see figure~\ref{fig:app:QuenchBCuts}. The finite parameter $\eps$ shifts the branch cuts along the real axis, hence, automatically defining operator ordering. Indeed, continuing along the blue curve ($\tau \to it$), we obtain the operators $\Odag$ and $\Ocal$ anti-time-ordered, while $\Ocal$ and $O$ time-ordered. Other orderings can be obtained only for \textit{vanishingly small} values of $\eps$, since in this case, the branch cuts fall onto the axis $\im\mathfrak{t}$ (see the green and the red curves in figure~\ref{fig:app:QuenchBCuts}). To sum up, in our setup with a finite $\eps$, we perform Wick rotation as $\tau \to it$ (as in~\cite{Caputa:2014eta, Caputa:2014vaa, Shimaji:2018czt, Caputa:2019avh, Bhattacharyya:2019ifi}).

\section{Derivation of the two-point function on a cylinder}
\label{appendix:cyl}

Let us derive the two-point function of a free scalar on a cylinder wrapped in the \mbox{$x$-direction}. To find the Euclidean Green's function of this theory, we should solve the following equation with periodic boundary conditions
\be
    \left\{
    \begin{aligned}
        & A\left(-\Delta + m^2\right)K(\vec{x}_1 - \vec{x}_2) = \delta^{(2)}(\vec{x}_1 - \vec{x}_2), \\
        & K(\tau, x + L) = K(\tau, x),
    \end{aligned}
    \right.
    \label{eq:app:equation_for_cylinder}
\ee
where we denoted $\Delta = \partial_{\tau}^2 + \partial_x^2$; $A$ is the normalization factor of the corresponding action (throughout the article we use the convention $A = 1/(4\pi)$), and $\vec{x}_1 = \{\tau_1, x_1\}$, $\vec{x}_2 = \{\tau_2, x_2\}$ are Euclidean 2-vectors.

\skipline

For non-vanishing mass, the solution to the equation~\eqref{eq:app:equation_for_cylinder} does not have a simple analytical expression, and it can be written only as a formal sum. In order to get some intuition, let us consider the solution to the equation~\eqref{eq:app:equation_for_cylinder} for a \textit{massless} scalar field $\phi(\tau, x)$, which should correspond to the \textit{formal} expression for the two-point correlator $\corrfunc{\phi(\tau, x)\phi(0, 0)}\Big|_{m\,\to\,0}$. We consider the series
\be
    \corrfunc{\phi(\tau, x)\phi(0, 0)}\Big|_{m\,\to\,0} = K(\tau, x)\Big|_{m\,\to\,0} = \frac{1}{A L}\sum\limits_{n}\int\limits_q\frac{dq}{2\pi}\,\frac{e^{i\om_n x + iq\tau}}{q^2 + \om_n^2}\Bigg|_{\om_n = \frac{2\pi n}{L}},
\ee
leading to a divergent integral at $n = 0$. Here $\om_n = 2 \pi n/L$ is the Matsubara-like frequency. Regularizing this expression by restoring mass in the $n = 0$ term\footnote{This result with the restored mass term is a helpful analytical test for numerical calculations with a vanishingly small mass.}, we get the following series
\be
    \corrfunc{\phi(\tau, x)\phi(0, 0)}\Big|_{m\,\to\,0} = \frac{1}{AL}\lim_{m\,\to\,0}\left[\int\limits_q\frac{dq}{2\pi}\,\frac{e^{iq\tau}}{q^2 + m^2}\right] + \frac{1}{AL}\sum\limits_{n\,\neq\,0}\int\limits_q\frac{dq}{2\pi}\,\frac{e^{i\om_n x + iq\tau}}{q^2 + \om_n^2}\Bigg|_{\om_n = \frac{2\pi n}{L}}, 
    \label{eq:app:massless_cyl_not_prim}
\ee
which by virtue of the Fourier transformation 
\be
    \int\frac{dq}{2\pi}\,\frac{e^{iq\tau}}{q^2 + \om_n^2} = \frac{1}{2|\om_n|}e^{-|\om_n|\cdot|\tau|},
    \label{eq:how_to_continue}
\ee
summation over $n \neq 0$
\be
    \begin{aligned}
        & \frac{1}{2}\sum\limits_{n\,>\,0}\frac{e^{i\om_n x - |\om_n|\cdot|\tau|}}{|\om_n|}\Bigg|_{\om_n = \frac{2\pi n}{L}} = \frac{L}{4\pi}\sum\limits_{n\,>\,0}\frac{1}{|n|}\,e^{\frac{2\pi}{L}\left(ix - |\tau|\right)|n|} = -\frac{L}{4\pi}\ln\left(1 - e^{\frac{2\pi}{L}\left(ix - |\tau|\right)}\right), \\
        & \frac{1}{2}\sum\limits_{n\,<\,0}\frac{e^{i\om_n x - |\om_n|\cdot|\tau|}}{|\om_n|}\Bigg|_{\om_n = -\frac{2\pi |n|}{L}} = \frac{L}{4\pi}\sum\limits_{n\,<\,0}\frac{1}{|n|}\,e^{-\frac{2\pi}{L}\left(ix + |\tau|\right)|n|} = -\frac{L}{4\pi}\ln\left(1 - e^{-\frac{2\pi}{L}\left(ix + |\tau|\right)}\right),
    \end{aligned}
\ee
and substitution $|\tau| \to \sqrt{\tau^2}$ leads to a formal expression of the vanishingly-small mass limit of the two-point function of the field $\phi$
\be
    \begin{aligned}
        & \corrfunc{\phi(\tau, x)\phi(0, 0)}\Big|_{m\,\to\,0} = \\
        & = \frac{1}{AL}\lim_{m\,\to\,0}\left[\frac{e^{-m\sqrt{\tau^2}}}{2m}\right] - \frac{\ln 2}{4\pi A} + \frac{\sqrt{\tau^2}}{2 A L} - \frac{1}{4\pi A}\ln\left[\cosh\left(\frac{2\pi\sqrt{\tau^2}}{L}\right) - \cos\left(\frac{2\pi x}{L}\right)\right],
    \end{aligned}
    \label{eq:app:massless_2point_cyl}
\ee
where we pick out the divergence arising in the limit $m \to 0$.

Now several comments are in order. We need to establish how to perform Wick rotation to real-time dynamics. However, the function $|\tau|$, $\tau \in \mathbb{R}$, is not unambiguously defined for complex times. To use $\tau \to it$ as before, we extend $|\tau|$ for $\tau \in \mathbb{R}$ to $\sqrt{\tau^2}$ for $\tau \in \mathbb{C}$. This is reasonable, because $\tau$ coincides with the principal branch of~$\sqrt{\tau^2}$ in the left-half complex plane, while $(-\tau)$ --- with that in the right-half plane. The branch cut of $\sqrt{\tau^2}$ should be treated as in appendix~\ref{appendix:cont}. This method is also verified by the fact that calculating so, we reproduce the well-known CFT answer for the local quench by primary operator on a cylinder~\eqref{eq:CFT_quench_cylinder} obtained by use of conformal symmetries. In~\eqref{eq:app:massless_2point_cyl}, we have already written explicitly time dependence following this extension of the absolute value of a function.

\skipline

In the massive case, the propagator reads
\be
    \corrfunc{\phi(\tau, x)\phi(0, 0)} = \frac{1}{AL}\sum\limits_{n}\int\limits_q\frac{dq}{2\pi}\,\frac{e^{i\om_n x + iq\tau}}{q^2 + \om_n^2 + m^2}\Bigg|_{\om_n = \frac{2\pi n}{L}}.
\ee
It is only possible to evaluate the integral corresponding to the time Fourier transformation, but the Fourier series corresponding to the spatial transformation cannot be summed over and should be left as an infinite sum
\be
    \corrfunc{\phi(\tau, x)\phi(0, 0)} = \frac{1}{AL}\cdot\frac{e^{-m\sqrt{\tau^2}}}{2m} + \frac{1}{AL}\sum\limits_{n\,\neq\,0}\frac{e^{i\om_n x - |\widetilde{\om}_n|\sqrt{\tau^2}}}{2|\widetilde{\om}_n|}\Bigg|_{\begin{subarray}{l} \om_n = \frac{2\pi n}{L} \\ \widetilde{\om}_n = \sqrt{\om_n^2 + m^2} \end{subarray}}.
    \label{eq:massive_2point_cyl}
\ee

An analogous expression for the two-point function of the field operator $\partial\phi$ in holomorphic coordinates $\sigma = x + i\tau$ and $\sbar = x - i\tau$ is given by
\be
    \begin{aligned}
        \corrfunc{\partial\phi(\sigma, \sbar)\partial\phi(0, 0)} & = -\frac{m}{8AL}e^{-\frac{im}{2}\sqrt{s^2}} + \\
        & + \sum\limits_{n\,\neq\,0}\frac{4\pi n\left(2\pi ns - \sqrt{\left(L^2 m^2 + 4\pi^2 n^2\right)s^2}\right) + L^2 m^2 s}{8AL^2\sqrt{L^2 m^2 + 4\pi^2 n^2}\,s} \times \\
        & \times \exp\left(-\frac{i\sqrt{\left(L^2 m^2 + 4\pi^2 n^2\right)s^2} - 2\pi np}{2L}\right),
    \end{aligned}
\ee
where $s \equiv \sigma - \sbar$ and $p \equiv \sigma + \sbar$, where one can notice that adding mass leads to mixing of the lightcone coordinates. The limit $m \to 0$ is well-defined in contrast to~\eqref{eq:app:massless_2point_cyl} and yields
\be
    \corrfunc{\partial\phi(\sigma, \sbar)\partial\phi(0, 0)} = - \frac{1}{4\pi A}\left(\frac{\pi}{L}\right)^2\sin^{-2}\left(\frac{\pi\sigma}{L}\right).
    \label{eq:app:massless_dphi_2point_cyl}
\ee

\newpage
\bibliography{bosonic_local_quench}
\bibliographystyle{JHEP}


\end{document}